\documentclass[pre,twocolumn,amsmath,amssymb,superscriptaddress]{revtex4-1}

\usepackage{xr}
\usepackage{url}
\usepackage{hyperref}
\hypersetup{
    colorlinks,
    citecolor=blue,
    filecolor=blue,
    linkcolor=blue,
    urlcolor=blue
}

\usepackage{tabularx}

\usepackage{mathrsfs}
\usepackage{verbatim}
\usepackage{graphicx}	
\usepackage{dcolumn}	
\usepackage{bm}		
\usepackage{bbm,upgreek}
\usepackage{mathtools}
\usepackage{physics}

\usepackage{xr}
\usepackage{color}
\usepackage[cal=boondoxo]{mathalfa}
\usepackage{palatino}

\makeatletter
\def\maketitle{
\@author@finish
\title@column\titleblock@produce
\suppressfloats[t]}
\makeatother

\usepackage{dsfont}
\usepackage{cancel}
\usepackage{xcolor}
\usepackage{harpoon}
\usepackage{accents}

\newcommand*{\idea}[1]{\medskip \noindent \textbf{#1}}

\begin{document}

\title{The high-resolution \textit{in vivo} measurement of replication fork velocity and pausing by lag-time analysis}

\author{Dean Huang}
\affiliation{Department of Physics, University of Washington, Seattle, Washington 98195, USA}
\author{Anna E.~Johnson}
\affiliation{Department of Biochemistry, Vanderbilt University, Nashville, Tennessee 37205, USA}
\affiliation{Department of Pathology, Microbiology, and Immunology, Vanderbilt University Medical Center, Nashville, Tennessee 37232, USA}
\author{Brandon S.~Sim}
\affiliation{Department of Physics, University of Washington, Seattle, Washington 98195, USA}
\author{Teresa Lo}
\affiliation{Department of Physics, University of Washington, Seattle, Washington 98195, USA}
\author{Houra Merrikh}
\email{houra.merrikh@vanderbilt.edu}
\affiliation{Department of Biochemistry, Vanderbilt University, Nashville, Tennessee 37205, USA}
\affiliation{Department of Pathology, Microbiology, and Immunology, Vanderbilt University Medical Center, Nashville, Tennessee 37232, USA}
\author{Paul A. Wiggins}
\email{pwiggins@uw.edu}\homepage{http://mtshasta.phys.washington.edu/}
\affiliation{Department of Physics, University of Washington, Seattle, Washington 98195, USA}
\affiliation{Department of Bioengineering, University of Washington, Seattle, Washington 98195, USA}
\affiliation{ Department of Microbiology, University of Washington, Seattle, Washington 98195, USA}

\begin{abstract}

An important step towards understanding the mechanistic basis of the central dogma is the quantitative characterization of the dynamics of nucleic-acid-bound molecular motors in the context of the living cell, where a crowded cytoplasm as well as competing and potentially antagonistic processes may significantly affect their  rapidity and reliability. To capture these dynamics, we develop a novel method, lag-time analysis, for measuring \textit{in vivo} dynamics. The approach uses exponential growth as the stopwatch to resolve dynamics in an asynchronous culture and therefore circumvents the difficulties and potential artifacts associated with synchronization or fluorescent labeling. Although lag-time analysis has the potential to be widely applicable to the quantitative analysis of \textit{in vivo} dynamics,  we focus on an important application: characterizing replication dynamics.    To benchmark the approach, we analyze replication dynamics in three different species and a collection of mutants. We provide the first quantitative locus-specific measurements of fork velocity, in units of kb per second, as well as replisome-pause durations, some with the precision of seconds. The measured fork velocity is observed to be both locus and time dependent, even in wild-type cells. In addition to quantitatively characterizing known phenomena, we detect brief, locus-specific pauses at rDNA in wild-type cells for the first time. We also observe temporal fork velocity oscillations in three highly-divergent bacterial species.
Lag-time analysis not only has great potential to offer new insights into replication, as demonstrated in the paper, but also has potential to provide quantitative insights into other important processes.



\end{abstract}



\maketitle

\section{Introduction}

At a single-molecule scale, all cellular processes are both highly stochastic as well as subject to a crowded cellular environment where they typically compete with a large number of  potentially-antagonistic processes that share the same substrate \cite{phillips2013physical,Tinoco:2011jv}. In spite of these challenges, essential processes must be robust at a cellular scale to facilitate efficient cellular proliferation. Understanding how these processes are regulated to achieve robustness remains an important and outstanding biological question \cite{Elias-Arnanz:1999rz,Deshpande:1996ng,Liu:1995yd,Rocha:2008fl,Mirkin:2007pt,Yao:2009hb,Pomerantz:2008ye}. However, a central challenge in investigating these questions is the quantitative characterization of the activity of enzymes in the context of the living cell. For instance, although single-molecule assays can resolve the pausing of molecular motors on nucleic-acid substrates in the context of  \textit{in vitro} measurements \cite{Adelman:2012uc,Davenport:2000lr}, performing analogous measurements in the physiologically-relevant environment of the cell, where these processes are subject to antagonism, poses a severe challenge to the existing methodologies \cite{Mangiameli:2017oc}.

In this paper, we develop an approach, \textit{lag-time analysis}, that facilitates the quantitative characterization of dynamics, with resolution of seconds, in the context of the living cell. The approach exploits exponential growth as the stopwatch to capture dynamics in exponentially proliferating cellular cultures  \cite{Huang2022} and unlike competing approaches, it can circumvent the difficulties and potential artifacts introduced by cell synchronization \cite{Bates:2005jc} or fluorescent labeling. Although in principle the approach is broadly applicable to all central dogma processes (\textit{i.e.}~replication, transcription, and translation) as well as other dynamics, we will focus on replication dynamics exclusively for concreteness.

In this paper, we demonstrate the first locus-specific genome-wide measurement of the fork velocity, in units of kilobases per second, and the duration of replisome pauses.  It facilitates detailed comparisons to be made, not just between different loci in a single cell, but between wild-type and mutant cells as well as between bacterial species. We apply this approach to analyze three model bacterial systems: \textit{Bacillus subtilis},  \textit{Vibrio cholerae}, and \textit{Escherichia coli}.  In \textit{B.~subtilis}, we analyze transcription-induced replication antagonism which is the main determinant of replisome dynamics in a set of mutants with retrograde (reverse-oriented) fork motion.
An analysis of  \textit{V.~cholerae} provides evidence that fork number is an important determinant of fork velocity, but also provides clear evidence that fork velocity is time dependent. To explore this time-dependence, we analyze the fork-velocity in \textit{E.~coli} which provides strong evidence for temporally-oscillating fork velocity, consistent with a recent report \cite{Bhat:2022wf}. Finally, we demonstrate that these oscillations are observed in all three organisms. In summary, the observed phenomena demonstrate the central importance of characterizing central dogma processes in the context of the living cell, where their activity is regulated and modulated by the cellular environment.

    \begin{figure}[t]
\centering
\includegraphics[width=1\linewidth]{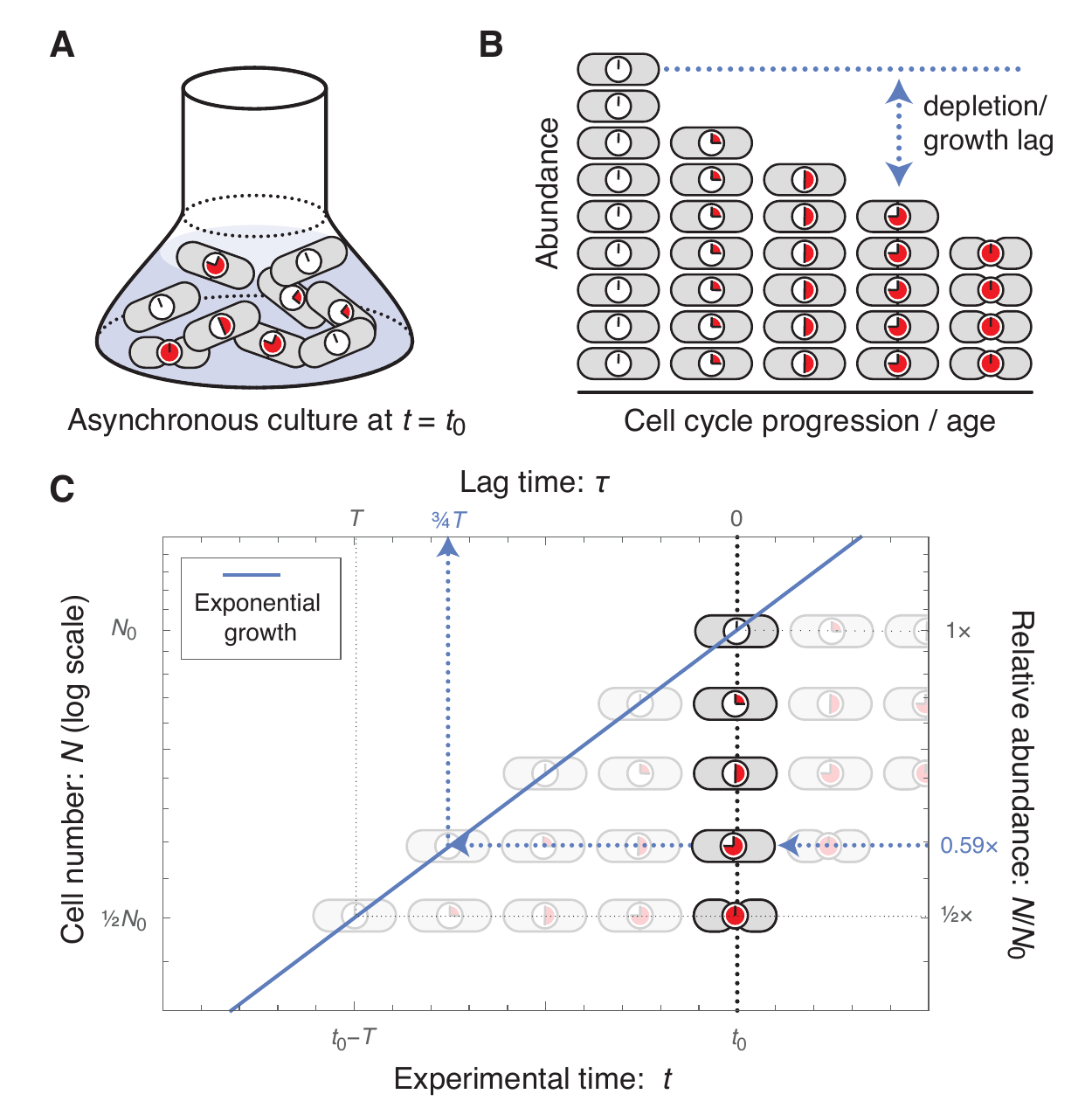}
\caption{
\textbf{Lag-time analysis.}
\textbf{Panel A: Sample preparation.} An asynchronous culture in steady-state exponential growth is harvested at time $t = t_0$. \textbf{Panel B: Quantitation of demographics.} Cell abundance is quantified. For analyzing replication dynamics, cell quantitation is performed by next-generation sequencing.  
\textbf{Panel C: Measurement of lag time.} The dotted black line represents the culture at $t = t_0$. Cells with greater cell cycle progression (\textit{i.e.}~age) are depleted relative to newborn cells. For each cell age, the relative abundance determines the \textit{lag time}. Their abundance is equivalent to an exponentially proliferating species that lags newborn cells by a time equal to its age. For instance, the nine-o'clock cell is at a relative abundance of 0.59 with a lag time of 3/4ths the mass-doubling time $T$. Schematically, start from the observed number of nine-o'clock cells, and follow that lineage horizontally (back in time) until you reach the newborn cell, born at $t = t_0-\tau$ (blue dotted line). For a stochastic cell cycle, lag time measures the exponential mean of the stochastic time (Eq.~\ref{eqn:expmean}). } 
\label{fig:LagTime}
\end{figure}


\begin{figure*}[t]
\centering
\includegraphics[width=0.9\linewidth]{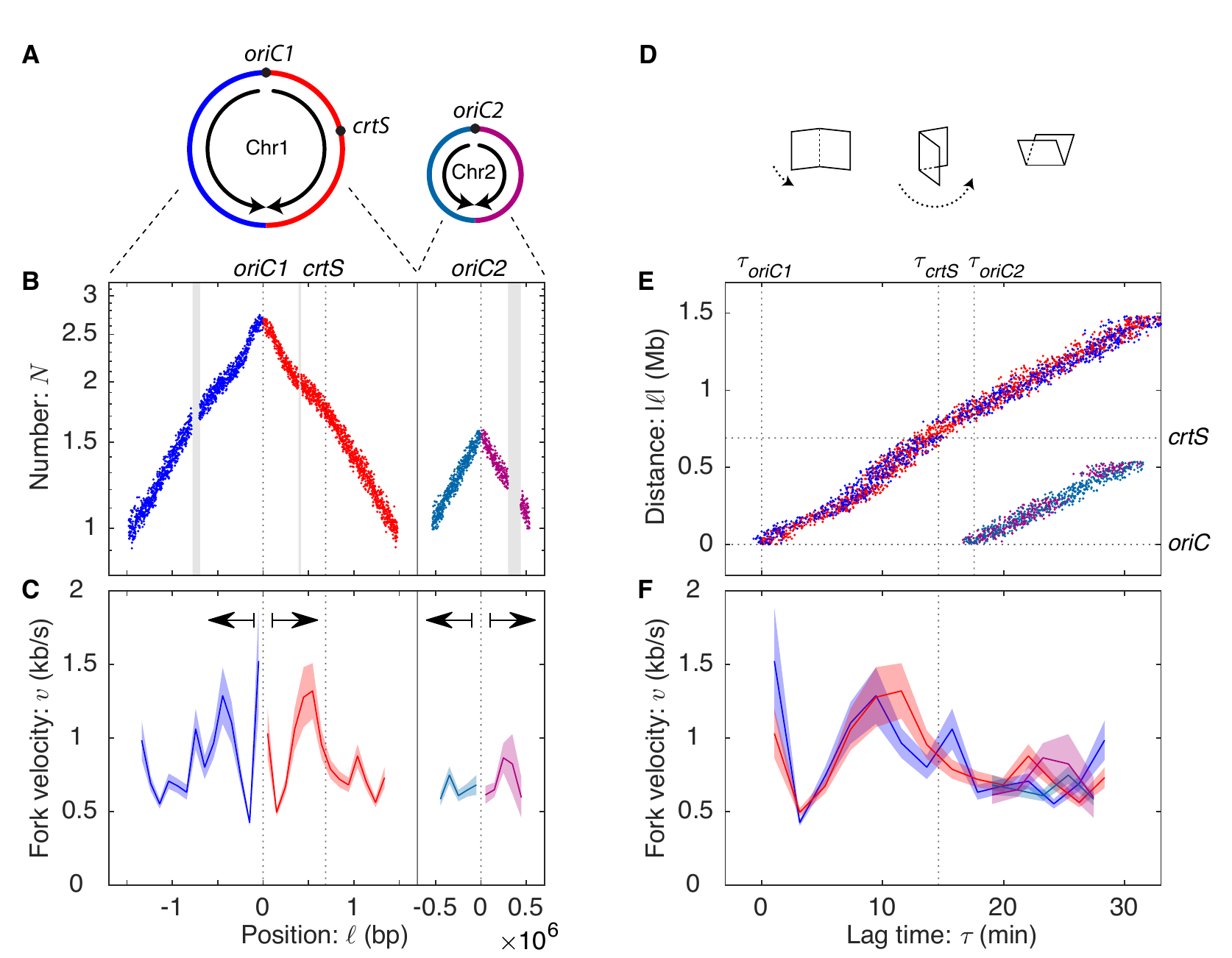}
\caption{ \textbf{Panel A: Chromosome organization in \textit{V.~cholerae.}} \textit{V.~cholerae} harbors two chromosomes Chr1 and Chr2. \textit{oriC2} initiates shortly after the \textit{crtS} sequence is replicated on the right arm of Chr1.
\textbf{Panel B: Marker frequency for \textit{V.~cholerae} grown on LB.} 
Repetitive sequences that cannot be mapped result in gaps.
\textbf{Panel C: Fork velocity is locus-dependent.} The fork velocity is shown as a function of genomic position with an error region. Statistically significant differences in the fork velocity are observed between loci. There is significant bilateral (\textit{i.e.}~mirror) symmetry around the origin.
\textbf{Panel D:} A visual representation of the relation between the log-marker-frequency and lag-time plots: Fold at the origin and rotate.
\textbf{Panel E: Lag-time analysis.} The replication forks start at the origin at lag time zero and then accelerate and decelerate synchronously, as the forks move away from the origin. The consistency in arm position is a manifestation of bilateral symmetry. 
\textbf{Panel F: Fork velocity as a function of lag time.} 
In addition to bilateral symmetry, after Chr2 initiates, all four forks show roughly consistent velocities.
}
\label{fig:VcCombofig}
\end{figure*}

\section{Results}

\idea{The bacterial cell cycle.} The bacterial cell cycle is divided into three periods \cite{Wang:2009oj,Willis:2017ye}: The B period is analogous to the G$_1$ phase of the eukaryotic cell cycle, corresponding to the period between cell birth and replication initiation.  The C period is analogous to the S phase (and early M phase) in which the genome is replicated and simultaneously and sequentially segregated \cite{Reyes-Lamothe:2019fd}. The D period is analogous to a combination of phases G$_2$ and late-M, corresponding to a period of time between replication termination and cell division, including the process of septation (\textit{i.e.}~cytokinesis). 

The demographics of cell-cycle periods of exponentially-growing bacterial cells were first quantitatively modeled by Cooper and Helmstetter in an influential paper \cite{Cooper:1968gd} and then refined by multiple authors \cite{Bremer:1977fv,Bird:1972ux,Pritchard:1975do}. In the Methods Section, we generalize these models to demonstrate that the measured marker-frequency analysis quantitatively measures the cell-cycle replication dynamics. The key results are summarized below.

\idea{Lag-time analysis.} Our strategy will be \textit{to use exponential growth as the stopwatch with which we resolve cell-cycle dynamics}. In short, cells with greater cell-cycle progression (\textit{i.e.}~age) are depleted in the population, equivalent to an independent, exponentially-proliferating species that \textit{lags} newborn cells by a time equal to its age \cite{Huang2022}. (See Fig.~\ref{fig:LagTime} for a schematic illustration of the approach.) \textit{Lag-time analysis} is the measurement of this time lag. 
In principle, this approach can be applied to characterize the dynamics of any biological molecules or complexes; however, for concreteness, we will focus on replication dynamics since the replication process is of great biological interest and next-generation sequencing provides a powerful tool for digital, as well as genome-wide, quantitation of the number of genomic loci.

In 
marker-frequency experiments, the number of each sequence $N(\ell)$ in a steady-state, asynchronously growing population is determined by mapping next-generation-sequencing reads to the reference genome.  This marker frequency can be reinterpreted as a measurement of the \textit{lag time} $\tau(\ell)$:
\begin{equation}
\tau(\ell) = \textstyle\frac{1}{k_G}\ln   {\textstyle\frac{N_0}{N(\ell)}}, 
\label{eqn:sol}
\end{equation}
where $N(\ell)$ is the observed number of the locus at genomic position $\ell$ and $N_0$ is the observed number of the origin in the culture and  $k_G$ is the growth rate.
This relation can  be understood as a consequence of the exponential growth law \cite{Huang2022}.

In a deterministically-timed model, the measured lag time would be equal to the replication time relative to initiation.
In reality, the timing of all processes in the cell cycle is stochastic. We previously showed that the measured lag time is related to the distribution of durations in single cells by the exponential mean \cite{Huang2022}:
\begin{equation}
{\tau_i} \equiv  -\textstyle\frac{1}{k_G}\ln\mathbb{E}_{t} \exp( -k_Gt ), \label{eqn:expmean}
\end{equation}
where $\mathbb{E}_{t}$ is the expectation over stochastic time $t$ with distribution $t \sim p_{i}(\cdot)$.  

\idea{Determination of replisome pause durations.} Replisome pause durations or the lag time difference between the replication of any two loci can be computed using the difference of lag times between the two loci:
\begin{equation}
\Delta \tau_{ij} \equiv \tau_j- \tau_i = \textstyle\frac{1}{k_G}\ln   {\textstyle\frac{N(\ell_i)}{N(\ell_j)}}. \label{eqn:pauseeqn}
\end{equation}
We emphasize that the observed  lag-time difference is the exponential mean of the stochastic time difference, which has important consequences for slow processes.

\idea{Determination of the fork velocity.} For fast processes, like single nucleotide incorporation, the exponential mean leads to a negligible correction (see Methods); therefore,  the fork velocity has a simple interpretation: it is the slope of the genomic position versus lag-time curve:
\begin{equation}
v(\ell) \equiv \textstyle\frac{{\rm d}\ell}{{\rm d} \tau} = \frac{k_G}{\alpha(\ell)}, \label{eqn:vel}
\end{equation} 
or equivalently it is the ratio of the growth rate to  the log-slope:
\begin{equation}
\alpha(\ell) \equiv -\textstyle \frac{\rm d}{\rm d \ell }\ln N(\ell), \label{eqn:slope}
\end{equation}
which can be directly determined from the marker frequency. 

\begin{table*}
\resizebox{\textwidth}{!}{\begin{tabularx}{1.1 \textwidth}{ X | X | X | X | X | X | X | X | X }
\hline
                         &    &    &         \multicolumn{5}{c|}{Fork statistics}          &    \centering{Statistical significance}   \cr
 Organism                &    \centering{Growth condition}                   &   \centering{Doubling time}:                  &   \centering{C period}:            &  \centering{Fork number:}      &   \centering{Velocity mean:}         &   \centering{std:}              & \centering{Symmetry: } & \centering{p-value:} \cr
                         &   &     \centering{$T$ (min)}              &  \centering{$C$ (min)}             &   \centering{$\overline{N}_F$}  &   \centering{$\overline{v}$ (kb/s)}  &  \centering{$\sigma_v$ (kb/s)}  & \centering{$f_S$ }     & \centering{$p$} \cr
              
\hline
\hline
 \textit{E.~coli}     & \raggedleft  LB & \raggedleft 19 & \raggedleft 30 & \raggedleft 3.8 & \raggedleft 1.3 & \raggedleft 0.19 & \raggedleft 84\% & \raggedleft $\ll 10^{-30}$ \cr 
                     & \raggedleft   M9 & \raggedleft 69 & \raggedleft 46 & \raggedleft 1.2 & \raggedleft 0.85 & \raggedleft 0.12 & \raggedleft 59\% & \raggedleft $6\times 10^{-12}$\cr 
                     \hline
 \textit{V.~cholerae} & \raggedleft LB & \raggedleft 22 & \raggedleft 31 & \raggedleft 4.3 & \raggedleft 0.82 & \raggedleft 0.27 & \raggedleft 76\% & \raggedleft $\ll 10^{-30}$ \cr 
                     & \raggedleft  M9  & \raggedleft 50 & \raggedleft 32 & \raggedleft 1.5 & \raggedleft 0.84 & \raggedleft 0.28 & \raggedleft 70\% & \raggedleft $\ll 10^{-30}$ \cr 
                                          \hline
 \textit{B.~subtilis}    & \raggedleft  S7 & \raggedleft 64  & \raggedleft 42  & \raggedleft 0.82 & \raggedleft 1.1 & \raggedleft 0.68 & \raggedleft 50\% & \raggedleft $\ll10^{-30}$ \cr            
     \ \  \textit{rrnIHG} &\raggedleft MOPS+CA & \raggedleft 44  & \raggedleft 40  & \raggedleft 1.2 & \raggedleft 0.86 & \raggedleft 0.45 & \raggedleft 45\% & \raggedleft $\ll10^{-30}$ \cr 
                                &  \raggedleft     MOPS &  \raggedleft 50       & \raggedleft 41  & \raggedleft 1.1 & \raggedleft 0.99 & \raggedleft 0.52 & \raggedleft 57\% & \raggedleft $\ll10^{-30}$ \cr                 
                     \hline
\end{tabularx}}
\caption{ \textbf{Fork number and velocities under different growth conditions.} Increasing fork number by increasing cell metabolism does not consistently reduce fork velocity. Fork velocities in fast growth are higher in \textit{E.~coli} and lower in \textit{V.~cholerae}. The statistical significance column shows the p-value for the null hypothesis of constant fork velocity. For more details on how these values are calculated, see Supplemental Material Sec.~IV. \label{tab1} }
\end{table*}




\idea{Lag-time analysis reveals \textit{V.~cholerae} replication dynamics.} To explore the application of lag-time analysis to characterize replication dynamics, we begin our analysis in the bacterial model system \textit{Vibrio cholerae}, which harbors two chromosomes: Chromosome 1 (Chr1) is 2.9 Mb and Chromosome 2 (Chr2)  is 1.1 Mb. The origin of Chr1, \textit{oriC1}, fires first and roughly the first half of replication is completed before the replication-initiator-RctB-binding-site \textit{crtS} is replicated, triggering Chr2 initiation at \textit{oriC2} \cite{Val:2014cb,Srivastava:2006ze,Rasmussen:2007le}. Chr1 and Chr2 then replicate concurrently for the rest of the C period. (See Fig.~\ref{fig:VcCombofig}A.) 

To demonstrate the power of lag-time analysis, we compute the marker frequency, lag-time, and fork velocities. To measure pause times and replication velocities, we generate a piecewise linear model with a resolution set by the Akaike Information Criterion (AIC). The AIC-optimal model for fast growth (in LB) had 39 knots, spaced by 100 kb, generating 38 measurements of locus velocity across the two chromosomes. The replication dynamics for growth in LB is shown in Fig.~\ref{fig:VcCombofig}.

\idea{Lag-time analysis enables the measurement of the duration of fast processes.} 
We focus first on the duration of time between \textit{crtS} replication and the initiation of \textit{oriC2}. Fluorescence microscopy imaging reveals that this wait time is very short \cite{Val:2016qu}, but it is very difficult to quantify since the precise timing of the replication of the \textit{crtS} sequence is difficult to determine by fluorescence imaging; however, this is a natural application for lag-time analysis. To measure the lag-time difference between \textit{crtS} replication and  \textit{oriC2} replication, we use Eq.~\ref{eqn:pauseeqn} to compute the replication time difference from the relative copy numbers. For this analysis, we generate a piecewise linear model with  knots at  the \textit{crtS} and \textit{oriC2} loci. The measured lag time is 
\begin{equation}
\Delta \tau_{\rm pause} =  3.5 \pm 0.1\text{\ min},
\end{equation}
a pause duration which is clearly resolved in the lag-time plot shown in Fig.~\ref{fig:VcCombofig}E. 



\idea{The fork velocity is locus dependent.} It is qualitatively clear from the fork-distance-versus-time plot (Fig.~\ref{fig:VcCombofig}E) that the fork velocity is locus dependent since the trajectory is not straight. To test this question statistically, we compare the 39-knot model to the null hypothesis (constant fork velocity), which is rejected with a p-value of $p \ll 10^{-30}$ and therefore the data cannot be described by a single fork velocity.  (See Tab.~\ref{tab1}.)  The resulting velocity profiles are shown in Fig.~\ref{fig:VcCombofig}CF.  

\idea{Bilateral symmetry supports a time-dependent mechanism.} Our understanding of the replication process motivated two general classes of mechanisms: (i) \textit{time-dependent} and (ii) \textit{locus-dependent} mechanisms. 
Time-dependent mechanisms, like a dNTP-limited replication rate,  affect all forks uniformly and therefore loci equidistant from the origin should have identical fork velocities: 
\begin{equation}
v(\ell) = v(-\ell),
\end{equation}
where $\ell$ is the genetic position relative to the origin. In contrast, in a locus-dependent mechanism, like replication-conflict-induced slowdowns, the slow regions are expected to be randomly distributed over the chromosome. In this scenario we expect to see no bilateral symmetry between arms (\textit{e.g.}~Fig.~\ref{fig:lagtime}B).

A bilateral symmetry between the arms is clearly evident in the data (the mirror symmetry about the origin in  Figs.~\ref{fig:VcCombofig}BC and is manifest in the lag-time analysis as the coincidence between the left and right arm trajectories and velocities in Figs.~\ref{fig:VcCombofig}EF. 
To quantitate this symmetry, we divide the variance of the fork velocity into symmetric and antisymmetric contributions. (See the Supplemental Material Sec.~IV~A.) A time-dependent mechanism would generate a $f_S = 100\%$ symmetric variance, whereas a locus-dependent mechanism would be expected to generate equal symmetric and antisymmetric variance contributions ($f_S = 50\%$). \textit{V.~cholerae} Chr1 and Chr2 have $f_S = 76\%$ symmetry, consistent with a time-dependent mechanism playing a dominant but not exclusive role in determining the fork velocity.  (See Tab.~\ref{tab1}.) 

%

\begin{figure*}[t]
\centering
\includegraphics[width=1\linewidth]{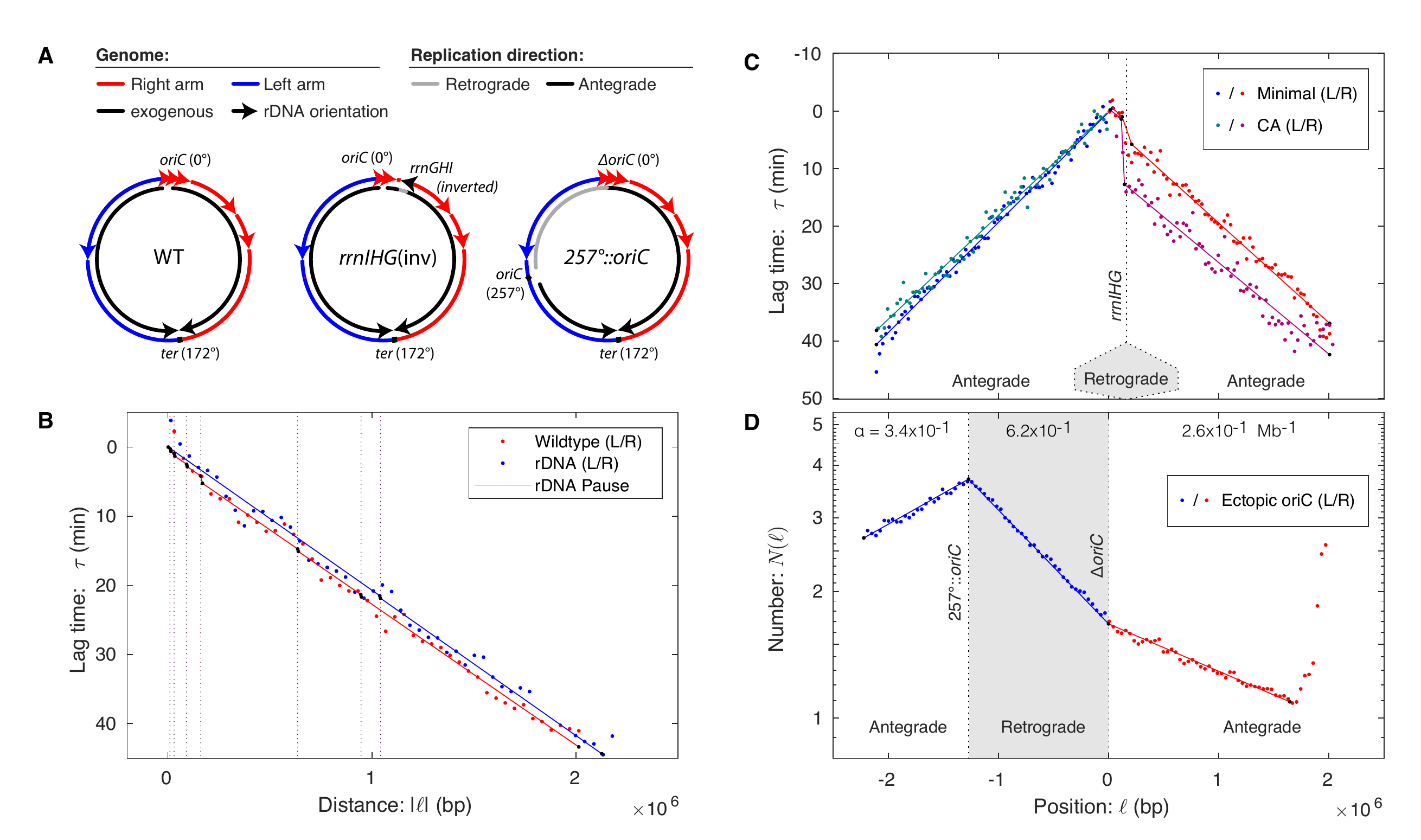}
\caption{
\textbf{\textit{B.~subtilis} fork dynamics and transcriptional conflicts.}
\textbf{Panel A: Chromosomal structure for wild-type and mutant \textit{B.~subtilis} strains.} The \textit{ter} region in wild-type \textit{B.~subtilis} is positioned at 172$^\circ$, rather than 180$^\circ$, making the right arm shorter than the left. In \textit{rrnIHG}(inv), the \textit{rrnIHG} locus is inverted so that it is transcribed in a head-on orientation with respect to replication.  In 257$^\circ$::\textit{oriC}, the origin is moved to 257$^\circ$, resulting in a short left arm that terminates at the terminus and a long right arm that replicates initially in the retrograde direction, before replicating the residuum of the right arm in the antegrade orientation.
\textbf{Panel B: Lag time in wild-type cells.} Replication on the right arm  (red)  is delayed relative to the left arm (blue) by multiple endogenous co-directional rDNA loci.
\textbf{Panel C: Head-on conflicts lead to pausing.} The \textit{rrnIHG} genes are inverted so that transcription of the rDNA locus is in the head-on direction. A longer lag-time pause is observed at intermediate growth rates (CA, purple) than slow growth (minimal, red). Fork velocities elsewhere are roughly consistent. 
\textbf{Panel D: Retrograde fork motion is slow.} 
The retrograde fork motion in R is slow compared to antegrade replication in A1. Late antegrade motion in A2 is faster than early antegrade motion in A1.  
}
\label{fig:Bsfig}
\end{figure*}

\idea{The replisome pauses briefly at rDNA in \textit{B.~subtilis}.} 
To explore the possibility that locus-dependent mechanisms could play a dominant role in determining the fork velocity profile, we next characterized the fork dynamics in the context of replication conflicts, where the antagonism between active transcription and replication, have been reported to stall the replisome by a locus-specific mechanism \cite{Pomerantz:2008ye,French:1992ir}.  In \textit{B.~subtilis}, there are seven highly-transcribed rDNA loci on the right arm and only a single locus of the left arm. Consistent with the notion of rDNA-induced pausing, the \textit{ter} locus is positioned asymmetrically  on the genome, at 172$^\circ$ rather than 180$^\circ$, leading the right arm of the chromosome to be shorter than the left arm. (See Fig.~\ref{fig:Bsfig}A.) In spite of the difference in length, both arms terminate roughly synchronously, implying that the average fork velocity is lower on the right arm, consistent with putative fork pausing at the rDNA loci. Are these conflict-induced pauses present in wild-type cells where the replication and transcription are co-directional? We have previously reported evidence based on single-molecule imaging that they are \cite{Mangiameli:2017oc}, but there is as of yet no other unambiguous supporting evidence. 

To detect putative  short pauses at the rDNA loci in wild-type \textit{B.~subtilis}, a low-noise dataset was essential. We therefore examined a number of different datasets, including our own, to search for a dataset with the lowest statistical and systematic noise. A marker-frequency dataset for a nearly wild-type strain growing on minimal media was identified for which the noise level was extremely low. (See Supplemental Material Sec.~III~C2.)  The lag-time analysis is shown in Fig.~\ref{fig:Bsfig}B.
Replication pauses should result in discrete steps in the lag time (\textit{e.g.}~Fig.~\ref{fig:lagtime}B); however, no clearly-defined steps are visible in the lag-time plot. The pauses are either absent or too small to be clearly visible without statistical analysis. 

To achieve optimal statistical resolution, we used the AIC model-selection framework \cite{BurnhamBook,akaike1773} on four competing hypotheses: In Model 1, fork velocities are constant and equal on both arms with no pauses. In Model 2, fork velocities are constant but unequal on the left and right arms with no pauses. In Model 3, fork velocities are constant and equal on the left and right arms with equal-duration pauses at each rDNA locus. In Model 4, fork velocities are constant and unequal on the left and right arms with equal-duration pauses at each rDNA locus. AIC selected Model 3 (equal arm velocities with rDNA pauses) and a pause duration of:
\begin{equation}
\Delta \tau_{\rm pause} = 17 \pm 8\text{ s},
\end{equation}
is observed.  The pause models were strongly supported over the non-pause models ($\Delta {\rm AIC}_{23} = 4.3$ and $\Delta {\rm AIC}_{43} =  9.4$). Therefore, statistical analysis supports the existence of short slowdowns (\textit{i.e.}~pauses) at the rDNA, even if these features are not directly observable without statistical analysis. In higher-noise datasets, the statistical inference was ambiguous.


\idea{Strong, head-on conflicts lead to long pauses.} Although we have just demonstrated that endogenous co-directional conflicts are detected statistically, they do not lead to a clear unambiguous signature. In contrast, strong, exogenous head-on conflicts in which the replisome and transcriptional machinery move in opposite directions can lead to particularly potent conflicts and even cell death \cite{Elias-Arnanz:1999rz,Deshpande:1996ng,Liu:1995yd,Rocha:2008fl,Mirkin:2007pt,Yao:2009hb,Pomerantz:2008ye,Dutta:2011fu}.  The ability to engineer conflicts at specific loci facilitates the use of  lag-time analysis for measuring the duration of the replication pauses.

To measure the pause durations due to head-on conflicts, we analyze the marker frequency from a strain, \textit{rrnIHG}(inv), generated by Srivatsan and coworkers with three rDNA genes (\textit{rrnIHG}) inverted so that they are transcribed in the head-on orientation.
Marker-frequency datasets were reported for this strain in two growth conditions: minimal supplemented with casamino acids, in which the strain grows at an intermediate growth rate, and unsupplemented minimal media \cite{Srivatsan:2010ag}. (Mutant cells cannot proliferate in rich media, presumably because the transcription conflicts are so severe \cite{Srivatsan:2010ag}.) In both slow and intermediate growth conditions, a clearly resolved step at the head-on locus is observed in the marker-frequency and lag-time analysis (Fig.~\ref{fig:Bsfig}B), exactly analogous to the simulated pause (Fig.~\ref{fig:lagtime}B).

To determine the pause durations in the two growth conditions, we again consider a model with an unknown pause duration (at the inverted rDNA locus) and constant but unequal fork velocities on the left and right arms. The observed lag-time pauses are
\begin{equation}
 \Delta \tau_{\rm pause} = \begin{cases} 3.3\pm 0.7\text{ min (slow)}\\
 9.7\pm 0.9\text{ min (intermediate)} \end{cases}, 
 \end{equation}
 for the slow and intermediate growth rates, respectively. 
 
 Although lag-time analysis  reports a precise pause duration, it is important to remember that the observed lag time corresponds to the exponential mean of the stochastic state lifetime (Eq.~\ref{eqn:expmean}), including cells that arrest and therefore never complete the replication process. Eq.~\ref{eqn:pause} accounts for the pause generated by this arrested cell fraction. Srivatsan and coworkers report that 10\% of the cells are arrested in intermediate growth, which accounts for 8.3 min of the lag time, leaving an estimated pause time of $\Delta \tau_{\rm pause} = 1.4\pm0.9$ min for non-arrested cells, which is roughly consistent with the pause time observed in slow growth conditions. 

\idea{Slow retrograde replication in \textit{B.~subtilis}.} Are all conflict-induced slowdowns consistent with long pauses at a small number of rDNA loci? 
 Wang et al. have previously engineered a head-on strain, 257$^\circ$::\textit{oriC}, with less severe conflicts by moving  \textit{oriC} down the left arm of the chromosome to 257$^{\circ}$ \cite{Wang:2007ip}. (See Fig.~\ref{fig:Bsfig}A.) The resulting strain has a very short left arm and a very long right arm, the first third of which is replicated in the \textit{retrograde} (\textit{i.e.}~reverse to wild-type) orientation. This retrograde region contains only a single rDNA locus. All other regions are replicated in the \textit{antegrade} (\textit{i.e.}~endogenous) orientation.  

Consistent with the analysis of Wang et al., we position knots to divide the chromosome into three regions with three distinct slopes: an early antegrade region $A1$ (the short left arm) with log-slope $\alpha_{A1}= 0.34 \pm 0.01\ {\rm Mb}^{-1}$, a retrograde region $R$ with log-slope $\alpha_{R}=0.63 \pm 0.01\ {\rm Mb}^{-1}$  and a late antegrade region $A2$ with log-slope $\alpha_{A2}=0.26\pm 0.01 \ {\rm Mb}^{-1}$, that replicates after the left arm terminates. (See Fig.~\ref{fig:Bsfig}C.) Due to the higher percentage of head-on genes in the $R$ region compared with the $A1$ region, the conflict model predicts more rapid replication in region $A1$ versus $R$. Consistent with this prediction, the  ratio of replication velocities is:
\begin{equation}
v_{A1}/v_{R} = 1.84\pm 0.4, \label{eqn:dirdep}
\end{equation}
revealing a strong replication-direction dependence. The slope appears relatively constant, consistent with a model of uniformly-distributed slow regions rather than a small number of long pauses as observed in the reversal of the rDNA locus \textit{rrnIHG}. Our quantitative analysis is consistent with the interpretation of Wang et al. \cite{Wang:2007ip}.

\idea{Rapid late replication due to genomic asymmetry.} This dataset has a striking feature that is not emphasized in previous reports: Late antegrade fork velocity is faster than early antegrade velocity:  
\begin{equation}
v_{A2}/v_{A1}=1.29\pm 0.05.\label{eqn:forknumeffect}
\end{equation}
Although this effect is weaker than the replication-direction dependence discussed above (Eq.~\ref{eqn:dirdep}), its size is still comparable.
An analogous late-time speedup is seen in two other ectopic origin strains. (See the Supplemental Material Secs.~VI~H and VI~J.)

One potential  hypothesis is that a locus-dependent mechanism slows the fork in the $A1$ region relative to the $A2$ region; however, no velocity difference is evident in these regions in the wild-type cells (Fig.~\ref{fig:Bsfig}B).
Alternatively, one could hypothesize that there is some form of communication between forks that leads to a slowdown in region $A1$ due to the slowdown in region $R$; however, no coincident slowdown is observed in \textit{rrnIHG}(inv) at a position opposite the \textit{rrnIHG} locus, inconsistent with this hypothesis. Another possible hypothesis is that late-time replication is always rapid; however, no significant speedup is observed in either wild-type \textit{B.~subtilis} (Fig.~\ref{fig:Bsfig}A)  or \textit{V.~cholerae} cells at the end of the replication process (Fig.~\ref{fig:Bsfig}B and Fig.~\ref{fig:VcCombofig}E). However, there is one extremely important difference between 257$^\circ$::\textit{oriC} and the wild-type strains: Due to the asymmetric positioning of the origin and replication traps at the terminus (Fig.~\ref{fig:Bsfig}A), there is only a single active replication fork as the A2 region is replicated. We therefore hypothesize that the fork velocity is inversely related to active fork number.

\idea{Fork number determines velocities in  \textit{V.~cholerae}.} 
To explore the effects of changes in the fork number on fork velocity, it is convenient to return to \textit{V.~cholerae}. In slow growth conditions, the cells start the C period with a pair of replication forks, for which the fork-number model  predicts faster fork velocity, and  finish the replication cycle with two pairs of forks, predicting slower fork velocity. 

Although the structure of the velocity profile is more complex than predicted by the fork-number model alone, the observed fork velocity is broadly consistent with its predictions. If a mean fork velocity is computed before and after \textit{oriC2} initiates, the ratio is:
\begin{equation}
v_{\rm before}/v_{\rm after} = 1.46\pm 0.02, \end{equation}
which is quantitatively consistent with the hypothesis that more forks lead to a slowdown in replication and the size of the effect is comparable to what is observed in \textit{B.~subtilis} (Eq.~\ref{eqn:forknumeffect}).


  \begin{figure}[t]
\centering
\includegraphics[width=.9\linewidth]{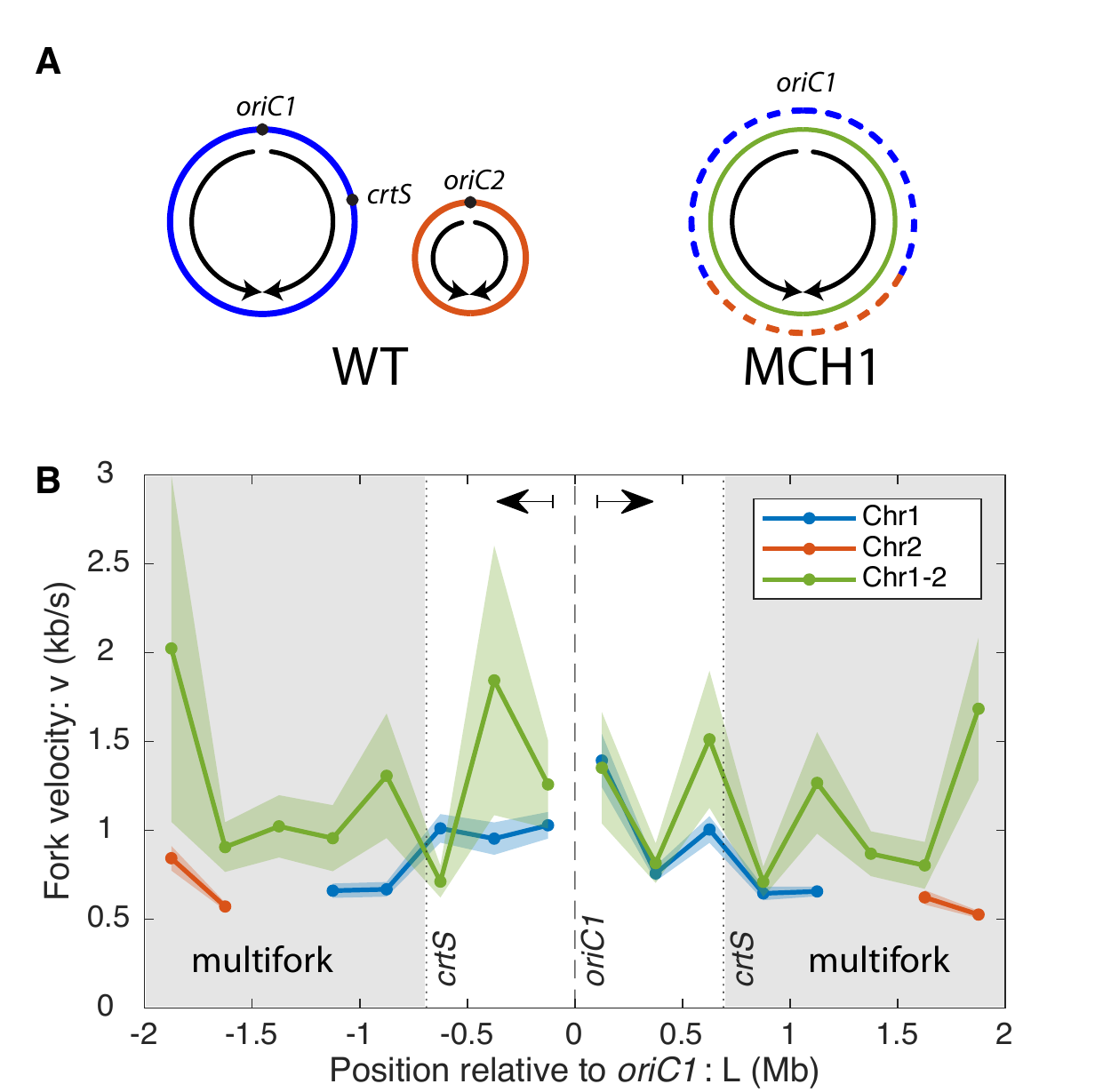}
\caption{\textbf{Reducing fork number increases fork velocity.} \textbf{Panel A:} The monochromosomal strain MCH1 has a single chromosome (green) which was constructed by recombining Chr2 (orange) into the terminus of Chr1 (blue) \cite{Val2012}. Under slow growth conditions the first part of the chromosome in both strains is replicated by a single pair of forks. When the fork reaches the \textit{crtS} sequence on the right arm, Chr2 is initiated at \textit{oriC2} in the wild-type cells. All of Chr2 and the residuum of Chr1 replicate simultaneously, resulting in two pairs of active forks. In contrast, all sequences in MCH1 are replicated using a single pair of forks. \textbf{Panel B:} In MCH1, where all sequences are replicated by a single pair of forks, the fork velocity is faster than is observed in WT cells during the multifork region (grey--sequences replicated after \textit{crtS}).}
\label{fig:Mono_vel}
\end{figure}

A mutant \textit{V.~cholerae} strain has been constructed that facilitates a non-trivial test of the fork-number model: In the monochromosomal strain MCH1, Chr2 is recombined into Chr1 at the terminus of Chr1, resulting in a single monochromosome (Chr 1-2). (See Fig.~\ref{fig:Mono_vel}A.) Both the wild-type and MCH1 strains have essentially identical sequence content, implying the locus-dependent model would predict identical replication velocities; however, all replication in MCH1 occurs with only a single set of forks whereas the wild-type strain replicates the latter half of the C period with two pairs of forks, one pair on each chromosome.   

The measured fork velocities are shown in Fig.~\ref{fig:Mono_vel}B and support the fork-number model: 
MCH1 replicates the sequences after \textit{crtS} at roughly $1.6$ times the fork velocity of the wild-type cells, consistent with the fork-number model.
Alternatively, we can consider the same quantitation of fork velocity we considered above: The ratio of fork velocities of loci replicated before \textit{crtS} to those replicated afterwards:
\begin{equation}
v_{\rm before}/v_{\rm after} = 1.11\pm0.03,
\end{equation}
therefore only a very small slowdown is observed after \textit{crtS} is replicated in MCH1, even though exactly the same sequences are replicated, again consistent with the fork-number model.




\begin{figure}[t]
\centering
\includegraphics[width=.9\linewidth]{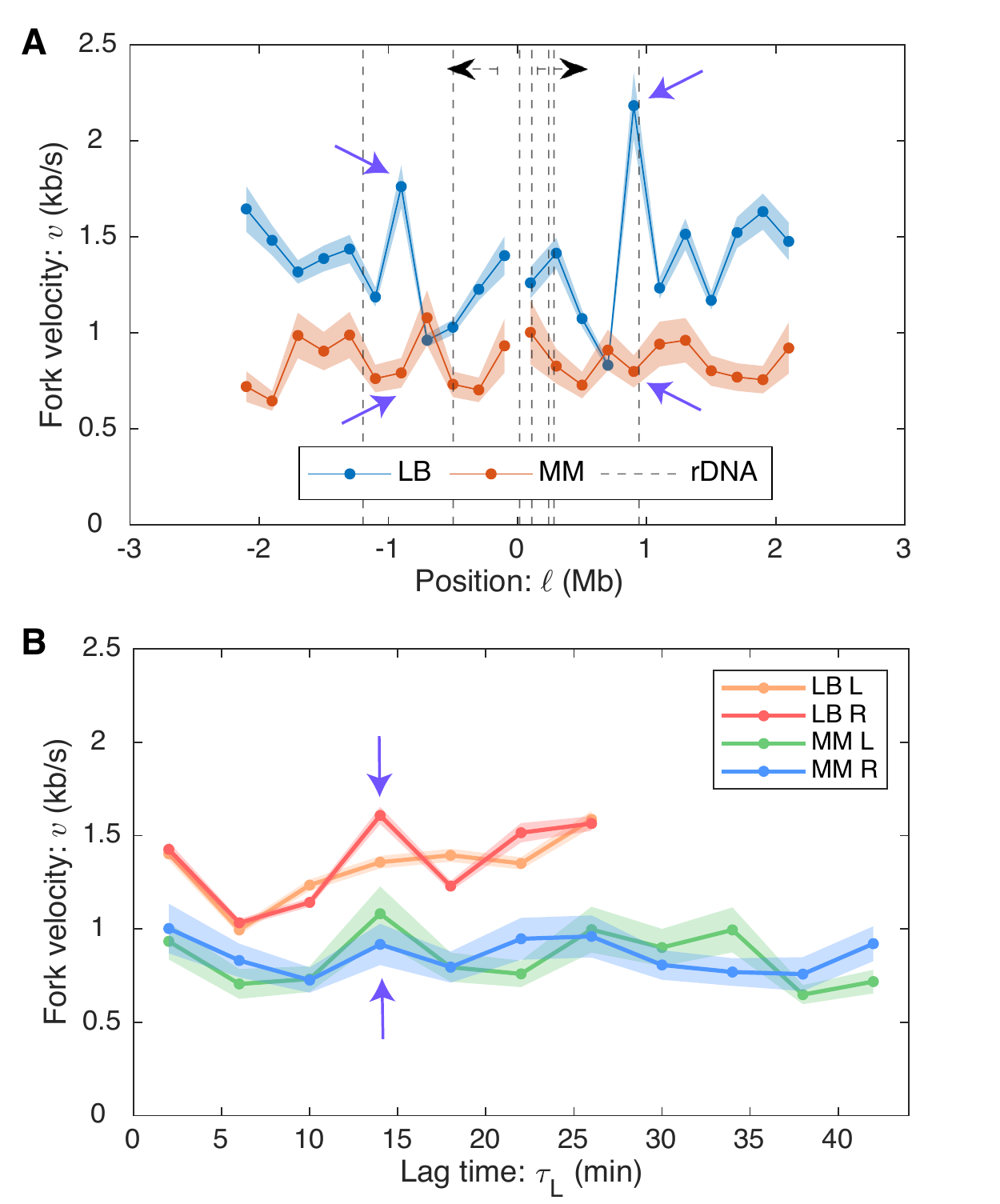}
\caption{ \textbf{Panel A: Velocity oscillations with respect to position in \textit{E.~coli}.}  We compare fork velocities as a function of genomic position (with respect to \textit{oriC}) under rapid (LB) and slow (minimal media--MM) growth conditions. Motivated by conflict-induced pauses, we have annotated the rDNA positions; however, slow velocities are not consistently coincident with rDNA loci. 
Regions with high fork velocities are not consistent between rapid and slow growth.  \textit{E.g.}~see the purple arrows. 
\textbf{Panel B: Velocity oscillations with respect to lag time in \textit{E.~coli}.} The velocity profiles have significant bilateral symmetry: the right and left arm velocities oscillate up and down together. Furthermore, not only are the oscillations consistent between left and right arms, they are also consistent between rapid (LB) and slow growth (minimal media--MM). \textit{E.g.}~see the purple arrows.  }

\label{fig:Ecoli_vel}
\end{figure}


\idea{The fork velocity oscillates in \textit{E.~coli}.} Although experiments in \textit{V.~cholerae} clearly support the fork-number model, there is significant variability that cannot be explained by this model alone. Are time-dependent variations in fork velocity also observed in organisms that replicate a single chromosome? To answer this question, we worked in the gram-negative model bacterium \textit{Escherichia~coli}, which harbors a single 4.6~Mb chromosome. 
A large collection of marker-frequency datasets have already been generated for both rapid and slow growth conditions by the Rudolph lab \cite{Rudolph:2013ho}. As with the \textit{B.~subtilis} marker-frequency datasets, we selected those that had the lowest statistical and systematic noise. (See the Supplemental Material Sec.~III~C2.) 

The fork velocities are shown in Fig.~\ref{fig:Ecoli_vel}. As before, statistically significant variation is observed in the fork velocity as a function of position. (See Tab.~\ref{tab1} and Supplemental Material Sec.~IV~C.) 
As discussed above in the context of \textit{V.~cholerae}, we had initially hypothesized that this variation might be a consequence of rDNA position or some other locus-dependent mechanism;
however, there are three  arguments against this hypothesis: (i) The slow-velocity regions  are not coincident with rDNA locations (Fig.~\ref{fig:Ecoli_vel}A) or relative GC content (Supplemental Material Sec.~VI~A). (ii) Consistent with the time-dependent model, 84\% (and 59\%) of the observed variation in the fork velocity is symmetric for fast  (and slow) growth. (iii) We would expect that a locus-dependent model would predict slow regions that are consistent between fast and slow growth, which is not observed. (See the purple arrows in Fig.~\ref{fig:Ecoli_vel}A.) We therefore conclude that the dominant mechanism for determining the fork velocity is a time-dependent mechanism, consistent with our observations for \textit{V.~cholerae}.

The lag-time analysis is particularly informative in the context of the \textit{E.~coli} data  (Fig.~\ref{fig:Ecoli_vel}B): Although there is no alignment in the velocity with respect to locus position, there is clear alignment of the fork velocity variation with respect to lag time, not only between the left and right arms of the chromosome, but between slow and fast growth conditions. The observed oscillations appear roughly sinusoidal in character.

\idea{Fork velocity oscillations are observed in three organisms.} Temporal oscillations in the fork velocity are an unexpected phenomenon. Are these features a systematic error with a single dataset? First we note that these oscillations are present in two \textit{E.~coli} growth conditions (LB and minimal).
This phenomenon would be on sounder footing if similar oscillations are observed in two evolutionarily distant species: the gram-negative \textit{V.~cholerae} and gram-positive  \textit{B.~subtilis}. If this phenomenon is observed, to what extent are the oscillations of similar character (\textit{e.g.}~phase, amplitude, and period)?

\begin{figure}[t]
\centering
\includegraphics[width=1\linewidth]{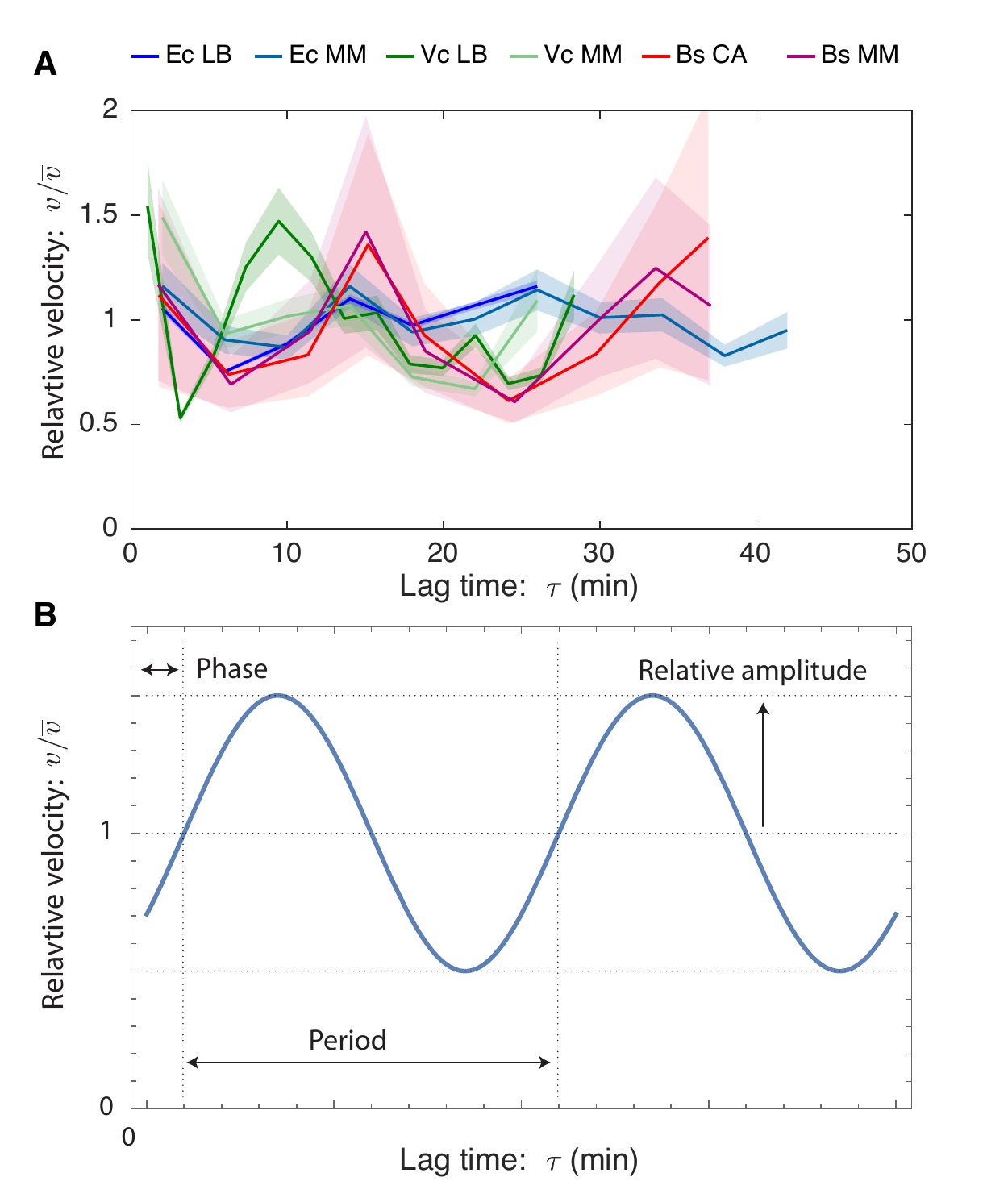}
\resizebox{.8\linewidth}{!}{\begin{tabularx}{.48 \textwidth}{ X | X | X | X | X}
\hline
 Organism        & \centering{Growth}    &  \centering{Period} &  \centering{Phase}           &  Relative  \cr
                 & \centering{condition} &  \centering{(min)}  &   \centering{(degrees)}      &   amplitude\cr
\hline
\hline
 \textit{E.~coli}     & \raggedleft Fast (LB) & \raggedleft 15 & \raggedleft $-78^\circ$ & \raggedleft $18\%$ \cr
                     & \raggedleft Slow  (M9) & \raggedleft 12 & \raggedleft $-45^\circ$ & \raggedleft $18\%$ \cr
 \textit{V.~cholerae} & \raggedleft Fast (LB) & \raggedleft 12 & \raggedleft $-81^\circ$ & \raggedleft $31\%$ \cr
                     & \raggedleft  Slow  (M9) & \raggedleft 10 & \raggedleft $-39^\circ$ & \raggedleft $36\%$ \cr
\textit{B.~subtilis} & \raggedleft M9+CA      & \raggedleft $17$ &  \raggedleft $-110^\circ$ & \raggedleft $26\%$ \cr
                     & \raggedleft  M9 & \raggedleft 15 &  \raggedleft $-150^\circ$ & \raggedleft $30\%$ \cr
                     \hline
\end{tabularx}}
\caption{\textbf{Panel A: Temporal velocity oscillations are observed in three bacterial species:} \textit{E.~coli} (Ec), \textit{B.~subtilis} (Bs),  and \textit{V.~cholerae} (Vc). The fork velocity starts high before decaying rapidly and then recovering. \textbf{Panel B: Oscillation characteristics.}  The oscillatory characteristics  are broadly consistent  both between conditions and species.  }
\label{fig:Mixed_vel}
\end{figure}

We compared the lag-time-dependent fork velocity for all three species.  
In \textit{B.~subtilis}, we have already discussed a rDNA-induced pausing on the right arm, which could complicate the interpretation of the data. We therefore consider the fork velocity on just the left arm.  For \textit{E.~coli} and \textit{V.~cholerae}, 
we compute the average velocity as a function of lag time between the two arms. Since the different organisms and growth conditions have different mean fork velocities, we compare the fork velocity relative to the overall mean. The results are shown in Fig.~\ref{fig:Mixed_vel}. 

All three organisms show oscillations with the same qualitative features: Each fork velocity has roughly the same phase: The velocity begins high, before decaying. The relative amplitudes, roughly 0.5 peak-to-peak, are all comparable with  the largest-amplitude oscillations observed in \textit{V.~cholerae} and the smallest in \textit{E.~coli}. When the relative velocities are compared, it is striking how much consistency there  is between growth conditions in \textit{E.~coli} and \textit{B.~subtilis}. Finally, the period of oscillation is comparable but distinct in all three organisms, ranging from 10 to 15 minutes. The oscillation characteristics are summarized in a table in Fig.~\ref{fig:Mixed_vel}.

%
%
%
%

%

\section{Discussion}

The focus of this paper is on the development of lag-time analysis, which uses exponential growth as the timer to characterize cellular dynamics. Although the approach is in principle widely applicable to characterize the dynamics of any molecule of complex, we focus on replication dynamics to explore this novel approach.
Although the approach has been understood at a conceptual level since the pioneering work of Cooper and Helmstetter \cite{Cooper:1968gd}, our recent exploration of stochastic models and the introduction of the exponential mean have clarified its interpretation \cite{Huang2022} and, from an experimental perspective, the introduction of next-generation sequencing greatly expanded the potential of the approach for characterizing the dynamics of nucleic acids and in particular replication dynamics, where it has the potential to make precise measurements of replication timing. In  multiple applications, we have used this approach to quantitatively measure time durations as short as seconds, a time resolution that is challenging, if not impossible, to achieve using other methods.

 To appreciate the power of our approach, it is useful to compare our results to recent results of Nieduszynski and coworkers who have used experimental methods, cell synchronization (sync-seq), to directly resolve replication dynamics \cite{Batrakou:2020rt,Muller:2014fh}. In this case, the time resolution is limited by a combination of the precision of cell synchronization, which is an imperfect tool \cite{Withers:1998zt}, and the frequency of fraction collection (every 5 minutes). Although it would be interesting to compare the relative precision of our approach to this competing method, the authors do not report fork velocities, pause durations, or provide an error analysis of their reported replication times, questioning to what extent the approach is truly quantitative. Since the fractions are collected on five-minute intervals, this time resolution is the floor of the direct time resolution achieved by this approach. 
In contrast, we report on a range of pause durations that are shorter than 5 minutes. 

A significant experimental shortcoming of the sync-seq approach is the necessity of cell synchronization. In most systems,  synchronization requires cell-cycle arrest, which introduces 
a significant potential for artifactual results \cite{Bates:2005jc}; whereas lag-time analysis probes dynamics in steady-state  growth. Our own preliminary analysis suggests that the timing of initiation at a subset of loci in \textit{Saccharomyces cerevisiae} is changed by the cell synchronization procedure relative to steady-state growth \cite{Huang2023}. 

In addition to these quantitative and high-time-resolution applications, we have also demonstrated the approach in a more conceptual context: using lag-time analysis to argue that the observed oscillations in fork velocity were temporal rather than locus dependent. 
Together, these examples demonstrate both conceptual and concrete advantages to lag-time analysis over the traditional marker-frequency analysis.

\idea{The significance of the fork velocity.}  Previous marker-frequency analyses have often reported a log-slope (\textit{e.g.}~\cite{ref:Vibrio_Galli_data,Wang:2007ip}), which is closely related to the fork velocity. What new insights does the measurement of the fork velocity offer over this closely related approach? The fork velocity approach has two  important advantages: (i) The first advantage is a conceptual one. The underlying quantity of interest is velocity (or rate per base pair). This is the quantity that is measured \textit{in vitro} and is relevant in a mechanistic model. In contrast, the log-slope is an emergent quantity that is only relevant in the context of exponential growth. (ii) The second advantage is concrete: Although log-slope measurements allow ratiometric comparisons between  fork velocity at different loci in the same dataset, \textit{they cannot be used to make comparisons across datasets.} Any comparison of the log-slope between cells with different growth rate (\textit{e.g.}~due to changes in growth conditions, mutations, species, \textit{etc.})~are meaningless. 
For instance, the log-slopes of the wild-type and MCH1 \textit{V.~cholerae} strains are very different even though the changes in the fork velocity are small. Our wide-ranging comparisons between growth conditions, mutants, and organisms demonstrate the power of reporting fork velocity over the log-slope.

\idea{Applications to eukaryotic cells.} Although our focus has been on replication in bacterial cells, an important question is to what extent our approach could be adapted to eukaryotic cells.
First, we emphasize that the lag-time analysis is directly applicable without modification to the eukaryotic context. As such, the timing of the replication of loci can be analyzed; however,  since the S phase is typically a smaller fraction of the cell cycle and the genomes of eukaryotic cells are larger, deeper sequencing will be required to achieve the same resolution we demonstrate in the context of bacterial cells. One significant potential refinement to this approach is the use of cell sorting (sort-seq) to enrich for replicating cells which can greatly increase the signal-to-noise ratio \cite{Batrakou:2020rt,Muller:2014fh}; however, this approach appears to lead to significant flattening near early-firing origins, as we have observed in other contexts (Supplemental Material Sec.~III~C2), and therefore increasing sequencing depth is probably the most promising approach for eukaryotic systems when quantitative characterization is a priority. (See  Methods  Eqs.~\ref{eqn:timeres} and \ref{eqn:errorVel} for an estimate of resolution.)

Although  lag-time analysis can easily be extended to the eukaryotic context, the measurement of the fork velocity will require some care. A critical assumption in our analysis is that  replication forks move unidirectionally at any particular locus, \textit{i.e.}~it can be either rightward or leftward moving but not both. (See Supplemental Material Sec.~III~C9.) Fork traps prevent this bidirectionality in many bacterial cells, but this does not apply to eukaryotic cells. However, if regions of the chromosome can be found where fork movement is unidirectional, \textit{e.g.}~sufficiently close to early-firing origins, fork velocity measurements could be made in eukaryotic cells. For instance, these conditions appear to be met for a significant fraction  of the \textit{Saccharomyces cerevisiae} genome \cite{Muller:2014fh}. With significant increases in sequencing depth, we expect analogous replication phenomenology, including pausing and locus- and time-dependent fork velocities,  will be observed in eukaryotic systems using lag-time analysis.

\idea{Importance of a model-independent approach.} As we prepared this manuscript, we became aware of a competing group which also uses marker-frequency analysis to test a specific hypothesis: the fork velocity is oscillatory in \textit{E.~coli} \cite{Bhat:2022wf}, consistent with our own observations. Although our reports share some conclusions, this competing approach requires detailed models for the cell cycle and  the fork velocity, along with explicit stochastic simulations. 
We demonstrate an approach to measure fork velocities \textit{independent of model assumptions} or \textit{detailed hypotheses for the fork velocity}, without the need for numerical simulation and complete with the ability to perform an explicit and tractable error analysis.
We therefore expect our analysis will be both widely applicable as well as accessible to other investigators without specialized tools or modeling expertise. 



\idea{Systematic error in datasets.} It may seem perplexing that we have not pooled many existing datasets from multiple independent experiments. This would na\"ively increase the statistical resolution and sensitivity from an analytical perspective. However, it is important to emphasize that not all marker-frequency experiments are of equal quality and that many datasets we have analyzed have clear signatures of systematic error. (See the Supplemental Material Sec.~III~C2.) 
In our analysis, we have prioritized the selection of artifact-free datasets over the indiscriminate pooling of data. We emphasize that to date, datasets have not been generated with quantitative replication dynamics analysis as a goal and we are confident that experimental protocols can be optimized to improve the data. Ref.~\cite{Knoppel2021} describes a promising approach, including harvesting populations earlier in exponential phase. We too are developing new protocols to increase data quality.

%

\idea{Multiple factors determine replisome dynamics.} Our measurements of the replication velocity reveal that there are multiple important determinants that results in complex velocity profiles.

\idea{dNTP pools regulate the fork velocity.}  Previous work had already demonstrated that increases (or decreases) in dNTP pool levels
lead to concomitant decreases (or increases) in the C period duration,  consistent with a dNTP-limited model of the replication velocity \cite{Churchward:1977ab,Zaritsky:1973yw,Odsbu:2009rs,Zhu:2017kg}.
%
Our data is broadly consistent with these previous results, but in a subcellular context:  (i) The fork-number model, in which fork velocities decrease as the number of active forks increase, is clearly consistent with a mechanism in which the nucleotide pool levels, although highly-regulated \cite{Gon:2006oa}, cannot completely compensate for the increased incorporation rate associated with multiple forks. 
(ii) The observation of the fork velocity oscillations is also consistent with an analogous failure of the regulatory response to compensate, this time temporally. The initial fall in the fork velocity is consistent with a model in which dNTP levels initially fall as replication initiates and nucleotides begin to be incorporated into the genome.  Reduction in the dNTP levels causes a regulatory response to increase dNTP synthesis by ribonucleotide reductase \cite{Gon:2006oa}, but the finite response time of the network could lead to dynamic overshoot in the regulatory feedback, leading to oscillations \cite{Arfken:379118}. Ref.~\cite{Bhat:2022wf} has also argued that this oscillating-dNTP-level model would lead to time-dependent oscillations in the mutation rate which are consistent with the origin-mirror-symmetric distribution of the mutation observed in \textit{E.~coli}. However, this interesting phenomenon and this hypothesized mechanism will require further investigation. 

\idea{Retrograde fork motion leads to slow replication velocities.}
Retrograde fork motion, where the fork moves in the opposite direction from wild-type cells, lead to the largest changes in fork velocity observed. 
 To what extent is the observed slowdown a consequence of a few long-duration pauses versus a region-wide slowdown? In regions which exclude the rDNA, the effect appears well distributed. 
However, it is important to note that the genomic resolution of lag-time analysis is still much too low to resolve individual transcriptional units. We anticipate that with increased sequencing depth as well as improvements in sample preparation, this approach could detect genomic structure in the fork velocity at the resolution of individual transcriptional units. 
Although we did analyze a number of mutants with retrograde fork movement in  \textit{V.~cholerae} and \textit{E.~coli} (analysis not shown), the competing effect of increased fork number as well as the genomic instability of these strains  made these experiments difficult to interpret quantitatively, since fork number and direction were both affected in these strains \cite{Wang:2007ip,Dimude:2018fj}. We concluded qualitatively that retrograde replication direction appears not to play as large a role in these gram-negative bacteria as it does in gram-positive \textit{B.~subtilits}, consistent with previous evidence \cite{Mirkin:2005qs,Mirkin:2006od,Wang:2007ip,Srivatsan:2010ag,Lang:2018eo}. However, we expect lag-time analysis could be used to characterize even small effects of the retrograde fork orientation in more-carefully engineered strains, analogous to those that we analyzed in the context of \textit{B.~subtilis} \cite{Wang:2007ip,Srivatsan:2010ag}. 

Previous reports \cite{Wang:2007ip,Srivatsan:2010ag}, including our own \cite{Lang:2017ce,Merrikh:2011cq,Mangiameli:2017oc,Merrikh:2015cz,Million-Weaver:2015hk}, had reported long-duration replication-conflict induced pauses, especially in mutant strains where the orientation of rDNA \cite{Srivatsan:2010ag} or other highly transcribed genes \cite{Lang:2017ce} are inverted to give rise to a head-on conflict between replication and transcription. The contribution of lag-time analysis  in this context is multifold: First, we provide a quantitative number in the context of the very-short-duration pauses for co-directional transcription in wild-type cells. 
This analysis supports a long-standing hypothesis that the right arm of the \textit{B.~subtilis} chromosome is shorter than the left arm to compensate for pausing at the rDNA loci that arm predominately located on this arm.  
 
We also report quantitative measurements for the longer pauses that results from head-on conflicts in mutants where highly transcribed genes are inverted.  Our analysis gives us the ability to quantitatively differentiate the contributions of fork pausing and  arrest in the analysis of the marker frequency, which was previously impossible. 
Our measurement of a timescale of minutes is consistent with our previous \textit{in vivo} single-molecule measurements in which we report transcription-dependent disassembly of the core replisome \cite{Mangiameli:2017oc}. Could the observed fork-velocity oscillations be misinterpreted as pauses? The observed lag-time offset between the two arms (\textit{e.g.}~Fig.~\ref{fig:Bsfig}B) is not predicted by fork-velocity oscillations.

\idea{Conclusion.} In this paper, we introduce a novel method for quantitively characterizing cellular dynamics by lag-time analysis. Although more broadly applicable, we focus our analysis on the characterization of replication dynamics using next-generation sequencing to quantitate DNA  locus copy number genome-wide.
%
The approach has the ability to make precise, even at the resolution of seconds, measurements of time differences and pause durations, as well as the ability to quantitatively measure fork velocities \textit{in vivo} in physiological units of kb/s, at genomic resolutions of roughly 100 kb, for the first time. Importantly, unlike marker-frequency analysis, our approach allows direct quantitative comparisons to be made between growth conditions, mutant strains, and even different organisms. The resulting measurements of replication dynamics reveal complex phenomenology, including temporal oscillations in the fork velocity as well as  evidence for multiple mechanisms that determine the fork velocity. The lag-time analysis has great potential for application  beyond bacterial systems as well as the potential to significantly increase in  resolution and sensitivity  as sequencing depth  and  sample preparation improve.


\section{Methods}

\idea{Introduction to marker-frequency analysis.}
Our focus will be on marker-frequency analysis, which measures the total number of a genetic locus in an asynchronous population.  The model was generalized to predict the marker frequency $N(\ell)$ of a locus a genomic distance $\ell$ away from the origin \cite{Bremer:1977fv,Bird:1972ux,Pritchard:1975do}: 
\begin{eqnarray}
N(\ell) &=& N_{0}\ {\rm e}^{-\alpha|\ell|}, \label{eqn:CP1} 
\end{eqnarray}
where $N_0$ is the number of origins, which grows exponentially in time with the rate of mass doubling of the culture, $k_G$. Since the origin is replicated first, the number of origins is always largest compared to the numbers of other loci. Quantitatively, the copy number is  predicted to decay exponentially with log-slope: 
\begin{eqnarray}
\alpha &=& -\textstyle \frac{\rm d}{\rm d \ell }\ln N(\ell)  = k_G/v, \label{eqn:CP2} 
\end{eqnarray}
where $k_G$ is the population growth rate and $v$ is the fork velocity, typically expressed in units of kilobases per second. To derive this result, two critical assumptions were made: (i) the timing of the cell cycle is deterministic and (ii) the fork velocity is constant \cite{Cooper:1968gd,Bremer:1977fv}. 

Initially, our na\"ive expectation was  that the interplay between the significant stochasticity  of the cell-cycle timing with the asynchronicity of the culture would prevent marker-frequency analysis from being used as a quantitative tool for characterizing cell-cycle dynamics. For instance, significant stochasticity is observed in the duration of the B period \cite{Si:2019ty} (\textit{i.e.}~the duration of time between cell birth and the initiation of replication). Does this stochasticity lead to a failure of the log-slope relation 
(Eq.~\ref{eqn:CP2})?

\begin{figure}[t]
\centering
\includegraphics[width=1\linewidth]{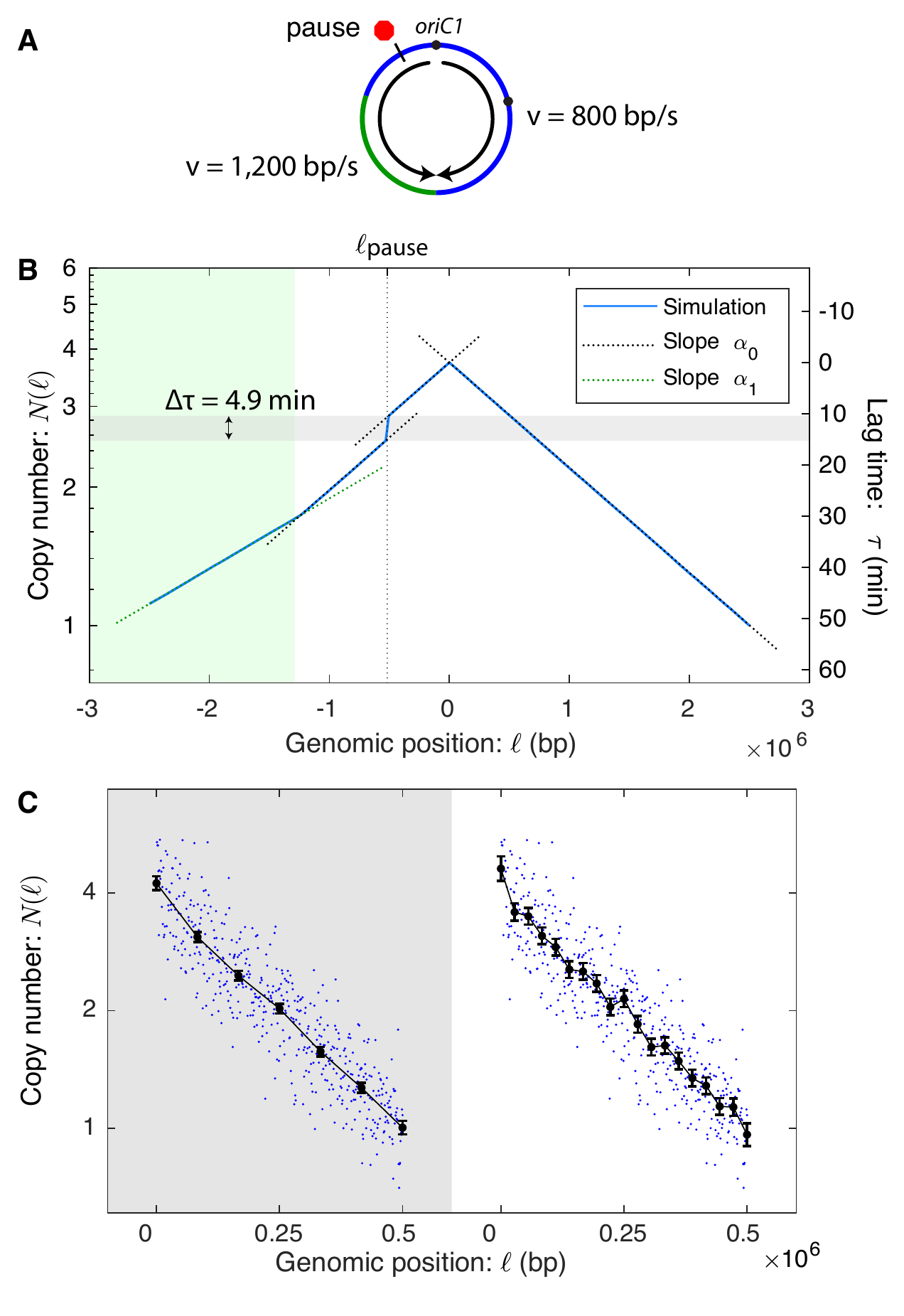}
\caption{\textbf{Panel A: A schematic of the simulated chromosome.} Replication initiates at the origin, pauses at a locus (red octagon) on the left arm and the velocity is increased on the lower left arm (green). \textbf{Panel B: Simulated marker frequency obeys the log-slope law.} The stochastic simulation generates a marker-frequency curve (blue). The model is stochastic in the timing of replication initiation as well as the fork dynamics and it includes two regions (blue and green) with different fork velocities as well as a pause with a stochastic lifetime. (See the \textit{Terminus 4} model in the Supplemental Material Sec.~V~A.) In spite of the stochasticity, it obeys the log-slope law locally (Eq.~\ref{eqn:slope}). Furthermore, the inferred lag-time pause (4.9 min) is predicted by the exponential mean (Eq.~\ref{eqn:expmean}).
\textbf{Panel C: Tradeoff between genomic resolution and velocity precision.}  As the spacing between knots decreases, increasing the genomic resolution, the error in the velocity measurement increases. 
}
\label{fig:lagtime}
\end{figure}

\idea{Stochastic simulations support the universality of the log-slope relation.}  To explore the role of stochasticity and a locus-dependent fork velocity  in shaping the marker frequency, we simulated the cell cycle using a stochastic simulation. Our aim was not to perform a simulation whose mechanistic details were correct, but rather to study how strong violations of the Cooper-Helmstetter assumptions, in particular how stochasticity, as strong or stronger than that observed, influenced the observed marker frequency and the log-slope relation (Eq.~\ref{eqn:CP2}).  
In short, we used a Gillespie simulation  \cite{Gillespie1977} where the B period duration and the lifetime of replisome nucleotide incorporation steps are exponentially distributed, and we added regions of the genome where the incorporation rate was fast as well as a single slow step on one arm. See Fig.~\ref{fig:lagtime}A and the Supplemental Material Sec.~V~A for a detailed description of the model, as well as movies of the marker frequency approaching steady-state growth, starting from a single-cell progenitor.

To our initial surprise, the stochasticity of the model had no effect on the predicted log-slope of the locus copy number. (See Fig.~\ref{fig:lagtime}B.) In spite of the stochastic duration of the B period and the locus-dependence, the marker frequency still decays exponentially with the same decay length \textit{locally}, \textit{i.e.}:
\begin{equation}
\alpha(\ell) \equiv -\textstyle \frac{\rm d}{\rm d \ell }\ln N(\ell) =  k_G/v(\ell), \label{eqn:slope}
\end{equation}
where $k_G$ was the empirically determined growth rate in the simulation and $v(\ell)$ was the local fork velocity at the locus with position $\ell$. 

We  hypothesized that this result might be a special case of choosing an exponential lifetime distribution, since this is consistent with a stochastic realization of chemical kinetics. To test this hypothesis, we simulated several different distributions, including a uniform distribution, for the duration of the B period and the stepping lifetime for the replisome (as well as simulating multifork replication). 
In each case, the local log-slope relation (Eq.~\ref{eqn:slope}) held, even as the growth rate and fork velocities changed with the changes in the underlying simulated growth dynamics. We therefore hypothesized that Eq.~\ref{eqn:slope} was a universal law of cell-cycle dynamics and  independent of Cooper and Helmstetter's original assumptions.

\idea{The exponential-mean duration.}
Motivated by this empirical evidence, we exactly computed the population demography in a class of stochastically-timed cell models. We described these results elsewhere ~\cite{Huang2022}. In short, we showed that there is an exact correspondence between these stochastically-timed models and deterministically-timed models in exponential growth. 
The relationship between the corresponding deterministic lifetime ${\tau}_i$ of a state $i$ and the underlying distribution $p_{i}$ in the stochastic model is the exponential mean (Eq.~\ref{eqn:expmean}) \cite{Huang2022}. The exponential mean biases the mean towards short times, the growth rate $k_G$ determines the strength of this bias, and the biological mechanism for this bias is due to the enrichment of young cells relative to old cells in an exponentially growing culture \cite{Huang2022}. 

To understand the consequences of this result, we consider two special cases of this exponential mean. For processes with lifetimes short compared to the doubling time, Eq.~\ref{eqn:expmean} can be Taylor expanded to show that the exponential mean is:
\begin{equation}
\tau \approx  \mu_t - \textstyle\frac{1}{2}k_G\sigma^2_t+...,
\end{equation}
the regular arithmetic mean $\mu_t$ with a leading-order correction proportional to the product of the growth rate and variance $\sigma_t^2$. In the context of single-nucleotide incorporation, this correction is on order one-part-in-a-million and therefore can be ignored. As a consequence, Eq.~\ref{eqn:slope}, corresponding to the transitions between states with short-lifetimes, is unaffected by the stochasticity, exactly as we observed in our simulations.

Another important case to consider is the strong disorder limit, in which a small fraction of the population $\epsilon$ stochastically arrests, \textit{i.e.}~with lifetime $\infty$, while the other individuals have exponential-mean lifetime $\tau_0$.  Using the definition in Eq.~\ref{eqn:expmean}, it is straightforward to show that the deterministic lifetime is:
\begin{equation}
\tau = \tau_0-T\log_2 (1-\epsilon) \approx \tau_0+ \textstyle\frac{\epsilon}{\ln 2}T,\label{eqn:pause}
\end{equation}
where $T$ is the population doubling time and the second equality is an approximation for small $\epsilon$. The exponential mean duration is extended by the arrest, but remains finite. Therefore, an arrest of a subpopulation is indistinguishable from a longer duration pause in an exponentially proliferating population. (See Ref.~\cite{Huang2022}.)

\idea{Marker-frequency demography.} For a stochastic model with locus-dependent fork velocity, we showed that Eqs.~\ref{eqn:CP1} and \ref{eqn:CP2} generalize to
\begin{equation}
N(\ell) = N_{0}\ {\rm e}^{-k_G\tau(\ell)}, \label{eqn:dem}
\end{equation}
where we will call $\tau(\ell)$ the lag time of a locus at position $\ell$, which is equal to the sum of the differential lag times for each sequential step:
\begin{equation}
\tau_{j} =\sum_{i=0}^{j-1} \delta \tau_i,
\label{eqn:emt0}
\end{equation}
where $\delta \tau_i$ is the differential lag time for state $i$ or the exponential mean of the state lifetime \cite{Huang2022}. In the continuum limit, it is more convenient to represent this sum as an integral: 
\begin{equation}
\tau(\ell_i) = \int^{\ell_i}_{0}\textstyle \!\!\!\! {\rm d} \ell \frac{1}{v(\ell)}, \label{eq:int}
\end{equation}
where the fork velocity is defined: $v(\ell_i) \equiv 1\ $bp$ / \delta \tau_i$. To demonstrate that the generalized stochastic model predicts the log-slope relation (Eq.~\ref{eqn:slope}), the log-slope can be derived by substituting Eq.~\ref{eq:int} into Eq.~\ref{eqn:dem}, as was observed in the stochastic simulations, demonstrating the universality of Eq.~\ref{eqn:vel}. We note that Wang and coworkers had previously derived an equivalent expression using the deterministic framework of the Cooper-Helmstetter model in the Material and Methods Section of Ref.~\cite{Srivatsan:2010ag}.

\idea{Stochasticity has a minimal effect on the marker frequency.} We initially had hypothesized that stochasticity should affect the marker frequency. As explained above, it is the rapidity of base incorporation that explains why stochasticity is dispensable in this context. The same argument does not apply to the B period  which is comparable to the duration of the cell cycle. However, for the marker frequency, it is  lag-time differences between the replication times of  loci that is determinative, and therefore the lag time of the B period cancels from these lag-time differences.
Although it is mostly irrelevant for understanding wild-type cell dynamics,  stochasticity and an arrested subpopulation will play an important role in one phenomenon we analyze: replication-conflict induced pauses.



\idea{Time resolution.} Due to the large number of reads achievable in next generation sequencing, the time resolution will be high in carefully designed analyses. The number of reads is subject to counting or Poisson noise. It is therefore straightforward to estimate the experimental uncertainty in the lag time due to finite read number:
\begin{eqnarray}
\sigma_{\tau_j} &=& k_G^{-1}\textstyle \frac{1}{\sqrt{N_j}} = 1\ {\rm s}\cdot(\textstyle\frac{6\times 10^6}{N_j})^{1/2}, \label{eqn:timeres}
\end{eqnarray}
where we have used a read number inspired by the  replication-conflict pausing example. This estimate suggests that under standard conditions, time measurements with an uncertainty of seconds are possible using this approach.



\idea{Fork-velocity resolution.} To compute the slope in Eq.~\ref{eqn:vel}, the log-marker-frequency is fit to a piecewise linear function with equal spacing between knots. See Fig.~\ref{fig:lagtime}B. 
There is an important tradeoff between genomic resolution (\textit{i.e.}~the genomic distance between knots) and fork velocity precision (\textit{i.e.}~the uncertainty in velocity measurement): Increasing the genomic distance between knots reduces the genomic resolution  but also reduces the uncertainty in the velocity measurement.  
 We therefore consider a series of models with increasing genomic resolution and use the Akaike Information Criterion (AIC) to select the optimal model \cite{BurnhamBook,akaike1773}. See Supplemental Material Sec.~III~C10. This approach balances the desire to resolve features by increasing the genomic resolution with the loss of velocity precision.
 
Given a knot spacing, it is straightforward to estimate the relative error:
\begin{equation}
\textstyle\frac{\sigma_v}{v} = \sqrt{\frac{2}{n\,(\Delta \ell)^3}} \frac{v}{k_G} \approx 0.1 \cdot \left(\frac{1.5}{n}\right)^{1/2} \left(\frac{100\ {\rm kb}}{\Delta \ell}\right)^{3/2}, \label{eqn:errorVel}
\end{equation}
where $n$ is the read depth in reads per base and $\Delta \ell$ is the spacing between knots in basepairs. Therefore, for a canonical next-generation-sequencing experiment, we can expect to achieve roughly 10\% error in the fork velocity for 100 kb genomic resolution.
Note that in our error analysis, we have included only the error from cell number $N$, not the error from the uncertainty in the cell-cycle duration, which covaries between loci in a particular experiment.



\idea{Acknowledgments.} The authors would like to thank B.~Traxler, A.~Nourmohammad, J.~Mougous, and J.~Mittler for many useful conversations.  We would like to thank P.~Levin, J.~Wang and L.~Simmons for advice on our manuscript. We thank S.~B.~Peterson and A.~Schaefer for help with \textit{V.~cholerae}. We would like to thank J.~Wang, C.~Possoz, F.-X.~Barre, C.~Rudolph for detailed conversations about their data. This work was supported by NIH grant R01-GM128191.

\idea{Data availability statement.} 
The sequencing datasets generated during the current study are currently available on reasonable request, but will be uploaded to a public repository prior to publication. Relevant sequencing data from other studies can be found at the accession numbers given in Sec.~II of the Supplemental Material. Digitized data are available on reasonable request.

\idea{Code availability statement.}
All MATLAB scripts written for this study are available on reasonable request.

\bibliographystyle{apsrev4-1}
\bibliography{forkvel}
 
\end{document}


\title{Supplemental Material: \\The high-resolution \textit{in vivo} measurement of replication fork velocity and pausing by lag-time analysis}

\author{Dean Huang}
\affiliation{Department of Physics, University of Washington, Seattle, Washington 98195, USA}
\author{Anna E.~Johnson}
\affiliation{Department of Biochemistry, Vanderbilt University, Nashville, Tennessee 37205, USA}
\affiliation{Department of Pathology, Microbiology, and Immunology, Vanderbilt University Medical Center, Nashville, Tennessee 37232, USA}
\author{Brandon S.~Sim}
\affiliation{Department of Physics, University of Washington, Seattle, Washington 98195, USA}
\author{Teresa Lo}
\affiliation{Department of Physics, University of Washington, Seattle, Washington 98195, USA}
\author{Houra Merrikh}
\email{houra.merrikh@vanderbilt.edu}
\affiliation{Department of Biochemistry, Vanderbilt University, Nashville, Tennessee 37205, USA}
\affiliation{Department of Pathology, Microbiology, and Immunology, Vanderbilt University Medical Center, Nashville, Tennessee 37232, USA}
\author{Paul A. Wiggins}
\email{pwiggins@uw.edu}\homepage{http://mtshasta.phys.washington.edu/}
\affiliation{Department of Physics, University of Washington, Seattle, Washington 98195, USA}
\affiliation{Department of Bioengineering, University of Washington, Seattle, Washington 98195, USA}
\affiliation{ Department of Microbiology, University of Washington, Seattle, Washington 98195, USA}

\onecolumngrid

\pagebreak

\renewcommand{\thepage}{S\arabic{page}}

\setcounter{page}{1}

\setcounter{section}{0}

\bigskip

\pagebreak

\maketitle

\onecolumngrid

\tableofcontents

\section{Strains used in this study}

\begin{table}[h]
{
{\centering

\begin{tabularx}{0.95\textwidth}{ | l || l | X | l | }
\hline
 Short name: & Strain name: & Description: & Source:                      \cr
\hline
\hline 
 \textit{V.~cholerae} WT & O1 biovar El Tor N16961 & \textit{ChapR} $\Delta$\textit{lacZ} gm$^\mathrm{R}$ & \cite{ref:Vibrio_Galli_data} \\ \hline
 \textit{V.~cholerae} MCH1 & MCH1 & \raggedright Integration of N16961 Chr2, without \textit{oriC2} and partition machinery region, in place of the \textit{dif1} site in N16961 Chr1. & \cite{ref:Vibrio_Galli_data} \\ \hline
 \textit{V.~cholerae} \textit{oriR4} & EGV111 & \raggedright N16961 \textit{ChapR} $\Delta$\textit{lacZ oriC1} @ R4 (1,898Mb) gm$^\mathrm{R}$ & \cite{ref:Vibrio_Galli_data} \\ \hline
 \textit{B.~subtilis} WT & JH642 & \textit{trpC2 pheA1} & \cite{Wang:2007ip} \\ \hline
 \textit{B.~subtilis} $257^\circ$::\textit{oriC} & MMB703 & \raggedright \textit{trpC2 pheA1 argG}(257°)::(\textit{oriC}/\textit{dnaAN kan}) $\Delta$(\textit{oriC-L})::\textit{spc} & \cite{Wang:2007ip} \\ \hline
 \textit{B.~subtilis} $94^\circ$::\textit{oriC} & JDW258 & \raggedright \textit{trpC2 pheA1 aprE}(94°)::(\textit{oriC}/\textit{dnaAN kan}) $\Delta$(\textit{oriC-L})::\textit{spc dnaB134ts-zhb83}::Tn\textit{917}(\textit{cat}) & \cite{Wang:2007ip} \\ \hline
 \textit{B.~subtilis} \textit{oriN} & MMB208 & \raggedright \textit{pheA1} (\textit{ypjG-hepT})\textit{122 spoIIIJ}(359°)::(\textit{oriN kan tet}) $\Delta$\textit{oriC-S} & \cite{Wang:2007ip} \\ \hline
 \textit{B.~subtilis} $257^\circ$::\textit{oriN} & MMB700 & \raggedright \textit{pheA1} (\textit{ypjG-hepT})\textit{122 argG}(257°)::(\textit{oriN kan}) $\Delta$\textit{oriC-S} & \cite{Wang:2007ip} \\ \hline
 \textit{B.~subtilis} YB886 & YB886 & \textit{trpC2 metB5 sigB amyE sp$\beta$-ICEBs$^\circ$ xin-} & \cite{Srivatsan:2010ag} \\ \hline
 \textit{B.~subtilis} \textit{rrnIHG}(pre-inv) & JDW858 & \raggedright YB886 \textit{kbaA'::neo' cat::'ybaN rrnG-5S'::erm 'neo::'ybaR} & \cite{Srivatsan:2010ag} \\ \hline
 \textit{B.~subtilis} \textit{rrnIHG}(inv) & JDW860 & \raggedright YB886 \textit{kbaA'::neo erm} inv(\textit{ybaN::rrnG-5S}) \textit{cat::'ybaR} & \cite{Srivatsan:2010ag} \\ \hline
 \textit{E.~coli} WT & K-12 MG1655 & \raggedright F- lambda- \textit{ilvG}- \textit{rfb}-50 \textit{rph}-1 & \cite{MidgleySmith2018} \\ \hline
 \end{tabularx} 

}}
\end{table}

\newpage

\section{Datasets used in this study}

\label{sec:datainfo}

All datasets were obtained from cells grown at 37~$^\circ$C. Next-generation-sequencing FASTQ files and marker frequencies were either generated for this study, obtained from Wang et al. \cite{Wang:2007ip,Srivatsan:2010ag}, or obtained from the European Nucleotide Archive (ENA). The corresponding project accessions are included if applicable. In the following table, Exp is shorthand for exponential phase growth and Stat is shorthand for stationary phase. Growth media are described in more detail in Supplemental Material Sec.~\ref{sec:GrowthMedia}.

{
{\centering

\begin{tabularx}{0.95\textwidth}{ | X || p{2.5cm} | l | l | l | p{3cm} | }
\hline
 Short name: & Growth media: & Doubling time: & Source: & Project accession: & Run accession: \\ \hline
\hline
 \textit{V.~cholerae} WT & LB & $22\pm1$ min & This study & TBA & Exp: 8217-AJ-013 \newline Stat: 8217-AJ-001 \\ \cline{2-6}
 & M9 fructose & $50\pm4$ min & From study \cite{ref:Vibrio_Galli_data} & ENA: PRJEB28538 & Exp: ERX2796386 \newline Stat: ERX2796387 \\ \hline
 \textit{V.~cholerae} MCH1 & LB & $90\pm6$ min & This study & TBA & Exp: 8217-AJ-016\newline Stat: 8217-AJ-004 \\ \cline{2-6}
 & M9 fructose & $50\pm4$ min & From study \cite{ref:Vibrio_Galli_data} & ENA: PRJEB28538 & Exp: ERX2796384\newline Stat: ERX2796385 \\ \hline
 \textit{V.~cholerae} \textit{oriR4} & M9 fructose & $55\pm5$ min & From study \cite{ref:Vibrio_Galli_data} & ENA: PRJEB28538 & Exp: ERX2796379\newline Stat: ERX2796380 \\ \hline
 \textit{B.~subtilis} WT & S7 fumarate & $61.0\pm1.2$ min & From study \cite{Wang:2007ip} & Digitized & Digitized \\ \hline
 \textit{B.~subtilis} $257^\circ$::\textit{oriC} & S7 fumarate & Not reported & From study \cite{Wang:2007ip} & Digitized & Digitized \\ \hline
 \textit{B.~subtilis} $94^\circ$::\textit{oriC} & S7 fumarate & Not reported & From study \cite{Wang:2007ip} & Digitized & Digitized \\ \hline
 \textit{B.~subtilis} \textit{oriN} & S7 fumarate & $64.3\pm2.9$ min & From study \cite{Wang:2007ip} & Digitized & Digitized \\ \hline
 \textit{B.~subtilis} $257^\circ$::\textit{oriN} & S7 fumarate & Not reported & From study \cite{Wang:2007ip} & Digitized & Digitized \\ \hline
 \raggedright\textit{B.~subtilis} \textit{rrnIHG}(pre-inv) & LB & $20\pm1$ min & From study \cite{Srivatsan:2010ag} & Digitized & Digitized \\ \cline{2-6}
 & \raggedright MOPS glucose CA & $28\pm1$ min & From study \cite{Srivatsan:2010ag} & Digitized & Digitized \\ \cline{2-6}
 & \raggedright MOPS glucose MM & $42\pm1$ min & From study \cite{Srivatsan:2010ag} & Digitized & Digitized \\ \hline
 \raggedright\textit{B.~subtilis} \textit{rrnIHG}(inv) & LB & $>160$ min & From study \cite{Srivatsan:2010ag} & Digitized & Digitized \\ \cline{2-6}
 & \raggedright MOPS glucose CA & $44\pm5$ min & From study \cite{Srivatsan:2010ag} & Digitized & Digitized \\ \cline{2-6}
 & \raggedright MOPS glucose MM & $44\pm1$ min & From study \cite{Srivatsan:2010ag} & Digitized & Digitized \\ \hline
 \textit{E.~coli} WT & LB & $19.3\pm1.7$ min & From study \cite{MidgleySmith2018} & ENA: PRJEB25595 & Exp: ERS2298483 \newline Stat: ERS2298484 \\ \cline{2-6}
 & M9 glucose & $68.8\pm6.2$ min & From study \cite{MidgleySmith2018} & ENA: PRJEB25595 & Exp: ERS2298504 \newline Stat: ERS2298505 \\ \hline
 \end{tabularx}

}}

\section{Experimental methods}

\subsection{Growth media and determination of growth phase} \label{sec:GrowthMedia} 
As we have used data from multiple sources, the minimal media and the determination of population growth phase are not consistent across all studies. All studies use the standard recipe for Luria-Bertani (LB) rich medium (1\% tryptone, 0.5\% yeast extract, and 1\% NaCl in H2O), with the exception of Midgley-Smith et al. (2018), where 0.05\% NaCl is used instead. All studies using M9 minimal media have the same base recipe (1X M9 salts, 2~mM MgSO4, and 0.1~mM CaCl2), with different supplements and carbon sources. In our study and in Galli et al. (2019) \cite{ref:Vibrio_Galli_data}, M9 was supplemented with 10~$\mu$g/mL thiamine HCl for the \textit{V. cholerae} strains. The carbon source for M9 is 0.4\% glucose for our study, 0.4\% fructose for Galli et al. \cite{ref:Vibrio_Galli_data}, and 0.2\% glucose for Midgley-Smith et al. \cite{MidgleySmith2018}. In Wang et al. (2007) \cite{Wang:2007ip}, the relevant datasets exclusively use S7 minimal medium (50~mM MOPS, 10~mM (NH4)2SO4, 5~mM potassium phosphate, 2~mM MgCl2, 0.9~mM CaCl2, 50~$\mu$M MnCl2, 5~$\mu$M FeCl3, 10~$\mu$M ZnCl2, and 2~$\mu$M thiamine hydrochloride), supplemented with 1\% sodium fumarate as the carbon source, 0.1\% glutamate, 40~$\mu$g/mL tryptophan, and 40~$\mu$g/mL phenylalanine. For Srivatsan et al. (2010) \cite{Srivatsan:2010ag}, the minimal medium consists of 50~mM MOPS with 1\% glucose, along with different supplements for the CA (0.5\% casamino acids) and MM (40~$\mu$g/mL tryptophan, 40~$\mu$g/mL methionine, 40~$\mu$g/mL phenylalanine, and 100~$\mu$g/mL arginine) growth media.

All populations were grown at 37~$^\circ$C. In our study, exponential phase cultures were grown to OD$_{600}$ of 0.60-0.80. Stationary phase samples were grown to OD$_{600}$ of 1.50 or greater. In Galli et al. \cite{ref:Vibrio_Galli_data}, exponential phase cultures were grown to an OD$_{650}$ of 0.05 and 0.2 in M9 and LB, respectively. In Midgley-Smith et al. \cite{MidgleySmith2018}, exponential phase cultures were grown to an OD$_{600}$ of 0.48. In Srivatsan et al. \cite{Srivatsan:2010ag}, exponential phase cultures were grown to an OD$_{600}$ of 0.2-0.6.

\subsection{Generation of marker frequency data for this study}
For detailed protocols of marker-frequency generation for each dataset, see the corresponding references in Sec.~\ref{sec:datainfo}. In this study, strains were struck on solid agar and grown overnight at 37~$^\circ$C. Three individual colonies were selected from each strain and used to inoculate 10~mL of LB media, which was grown overnight at 37~$^\circ$C w/ 260~rpm shaking. The following morning, these overnight pre-cultures were back-diluted to OD$_{600}$ of 0.05 in 5~mL of either LB or M9 media. The same biological replicate was used to inoculate both growth conditions. Exponential phase cultures were grown to OD$_{600}$ of 0.60-0.80. Stationary phase samples were grown to OD$_{600}$ of 1.50 or greater. To harvest, cultures were centrifuged at 10,000~rpm for 5~mins at 25~$^\circ$C, and the supernatant was removed. gDNA was prepared immediately following harvest using GeneJet Genomic DNA Purification Kit (Thermo Scientific, K0721). Purified gDNA was quantified using Qubit dsDNA HS Assay Kit (Invitrogen, Q32851) and quality was assayed using a Nanodrop 2000 Spectrophotometer (Thermo Scientific, ND-2000). Whole-genome libraries were prepared from purified gDNA using Twist 96-Plex Library Prep Kit (Twist Bioscience, 104950) with dual adapters. Libraries were subsequently pooled and sequenced on Illumina NovaSeq6000 (S4) to a depth of 6,000,000 reads per sample.

\subsection{Marker frequency analysis}

\subsubsection{Sequence alignment}
To align deep sequencing read outputs to the bacterial reference genomes for MFA, we used the read alignment tool Bowtie 2 via the MATLAB function \texttt{bowtie2} \cite{ref:bowtie2}. We then extracted the position information from the resulting SAM file using an in-lab written MATLAB script and the command-line tool SAMtools \cite{ref:SAMtools}.

\subsubsection{Data selection}

\label{sec:datasel}

Not all marker-frequency experiments are of the same quality and many datasets we have analyzed have clear signatures of systematic error. We use two key criteria to determine the quality of the data. The first is a qualitative measure involving the shape of the marker-frequency profile and the second is a quantitative measure based on resolution analysis. 

\begin{figure}
\includegraphics[width=0.5\linewidth]{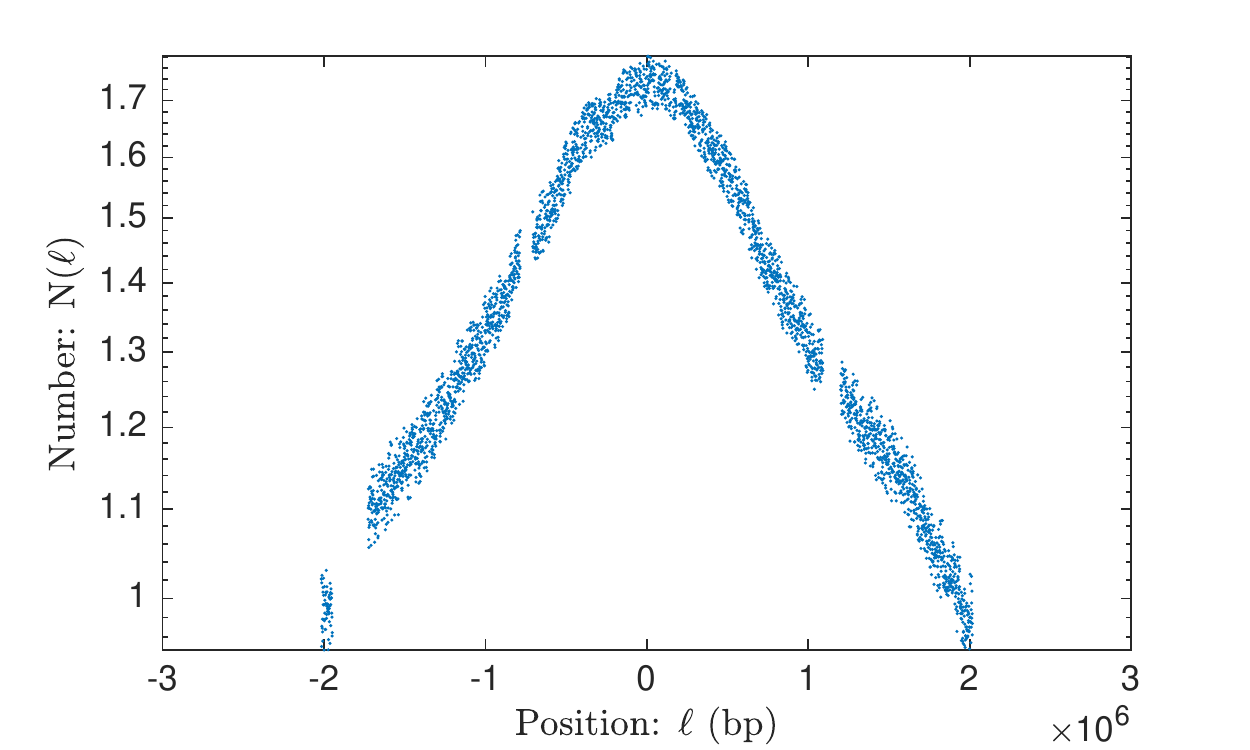}
\caption{Marker frequency profile of \textit{V. cholerae} MCH1 in LB rich media. Flattening of the profile near the origin of replication suggests a slowdown in the rate of replication initiation, consistent with the population entering stationary phase.}
\label{fig:LB_MCH1_bad}
\end{figure}

\idea{Marker profile near origin.} Our analysis uses exponential growth as the stop-watch for resolving dynamics, so it is essential that the population is harvested at the right time for sequencing. We have found that the marker-frequency profile has a signature flattening near the origin of replication if the population is harvested too late in exponential phase, as shown in Fig. \ref{fig:LB_MCH1_bad}. The population is entering stationary phase, where nutrient depletion begins to cause slower growth. Entering stationary phase also causes a decrease in the population-wide rate of replication initiation, which leads to the flattening of the profile near the origin. For the rest of the chromosome, replication continues unchanged for the majority of the population. Since exponential growth is crucial to our analysis, we have chosen to only use datasets with cusp-like behavior near the origin(s), as opposed to a rounded concave-down shape. 

\idea{Resolution analysis.} In the case where there is a single origin of replication for each chromosome, we expect only a single maximum in the marker frequency profile, corresponding to the origin. In an ideal noiseless situation, any local maxima distinct from the origin would correspond to other points of replication initiation and the fit would predict a retrograde fork velocity. However, due to the stochastic nature of the sequencing data, there can be spurious local maxima if the resolution is chosen to be too high. Fluctuations due to noise can cause the knots in the piecewise-linear fit along each arm of the chromosome to be non-monotonic, as shown in Panel C of Fig.~7 in the main paper. Thus, given two datasets for the same organism and growth condition, data quality was determined based on the best resolution that can be achieved without introducing spurious maxima.

\subsubsection{To divide or not to divide by stationary phase data}
To further remove any sequencing induced bias, we divide the copy number data obtained during exponential growth by the data obtained during stationary phase. This normalization step is done with the 1~kb binned values. Since the division of two noisy uncorrelated data sets typically results in a decreased signal-to-noise ratio (SNR), we used a comparison of the variance divided by the mean before and after division by stationary phase.

We expect the relative variance to decrease by normalization if the reduction in sequencing bias resulting from division has a larger effect than the increase in noise from dividing two noisy data sets. We first divide the noisy stationary phase data by its expected mean value. This ensures that it is of unit order and will not introduce a global scale factor during division, which would affect only the normalized data and not the pre-normalization data. We find that after division, the variance is roughly halved, suggesting that the normalization by stationary phase is successful in reducing sequencing bias without resulting in excessive noise. We thus chose to divide all exponential phase data with the corresponding stationary phase data. For some datasets, the stationary phase results were not provided for some of the ectopic origin mutants. The only difference in these mutants from their original strains is the addition of a short segment of DNA for the ectopic origin, which does not significantly change the locations of sequencing bias. Therefore, we chose to divide the exponential phase data for the ectopic origin strains by the stationary phase data from the original strains. 

\subsubsection{Filtering the data}
To remove outliers from the marker-frequency data, we used a centered 200~kb median filter, such that at each position $\ell$ along the chromosome, we set the median marker-frequency from $\ell-100$~kb to $\ell+100$~kb as a baseline. The data was given periodic boundary conditions such that the centered medians still had 200~kb regions at the boundaries of the data set. We then discarded the 10\% of points that most deviated from this centered median baseline, sorted by the absolute value of the difference. We also removed data points from regions known to be associated with mobile elements of low GC content, such as the Super Integron on Chr2 of \textit{V. cholerae}, which is subject to anomalous sequencing bias.

\subsubsection{Nearest-neighbor variance estimator}
We used a nearest-neighbor estimate of variance:
\begin{equation}
\hat{\sigma}^2 \approx \frac{1}{N}\sum_{i=1}^{N}\frac{1}{2}(x_{i+1}-x_{i})^2,
\end{equation}
where ${x_{N+1}=x_1}$, in agreement with periodic boundary conditions. This approach allows us to obtain a variance measurement without assuming the underlying distribution of the data and the corresponding expected value of each bin. 


\subsubsection{Fitting slopes to the log of the copy number}
After taking the natural logarithm of the copy number, we expect the data to follow a roughly linear trend along each arm of the chromosome. Since our goal is to obtain higher resolution measurements of fork velocity along the chromosome, it is necessary to subdivide each arm into smaller segments with variable slopes. The fit to the data should be continuous, since discontinuities would create copy number ambiguity at segment junctions. Therefore, we use a continuous piecewise linear least squares fit to obtain slope measurements from the data. We call the MATLAB function lsqnonlin as part of an in-lab written fitter function. The fitter obtains vertical control point values at the junctions between segments. Each control point is used in the measurement of the slope both to the left and to the right of it.

\subsubsection{Estimating errors for the control points}
\label{sec:errorest}
To obtain error estimates for the control points of the least squares fit, we use the Jacobian that lsqnonlin returns, which is the Jacobian of the difference between the fit data and the experimental data. If we have $k$ fit parameters $\theta_\alpha$, where $\alpha=1,...,k$, and $N$ data points $y_i$, where $i=1,...,N$, then the Jacobian takes the form:
\begin{equation}
J_{i\alpha}=\pdv{}{\theta^\alpha}\qty(y_i-\mu_i(\theta)),
\end{equation}
where $\mu_i(\theta)$ is the fitter estimate of $y_i$, based on the model with the set of parameters $\theta$. We use the matrix form of the Fisher information:
\begin{equation}
\label{eq:Fisher}
[I(\theta)]_{\alpha,\beta}=\mathbb{E}\qty[\qty(\pdv{}{\theta^\alpha}\log f(Y;\theta))\qty(\pdv{}{\theta^\beta}\log f(Y;\theta))\eval\theta],
\end{equation}
where $f(Y;\theta)$ is the probability density function (PDF) of the random variable $Y$, given parameters $\theta$. In our case, we expect the noise to be Gaussian distributed around the mean, so we have the PDF:
\begin{equation}
f(Y;\theta)=\frac{1}{(\sigma\sqrt{2\pi})^N}\prod_{i=1}^N\exp(-\frac{(y_i-\mu_i(\theta))^2}{2\sigma^2}),
\end{equation}
where $\sigma$ is the standard deviation. Now we take the derivative of the logarithm:
\begin{align}
\pdv{}{\theta^\alpha}\log f(Y;\theta)&=\pdv{}{\theta^\alpha}\qty(-\frac{\sum_i(y_i-\mu_i(\theta))^2}{2\sigma^2}-N\log(\sigma\sqrt{2\pi})),\\
&=-\frac{1}{\sigma^2}\sum_{i=1}^N(y_i-\mu_i(\theta))\pdv{}{\theta^\alpha}\qty(y_i-\mu_i(\theta)),\\
&=-\frac{1}{\sigma^2}\sum_{i=1}^N(y_i-\mu_i(\theta))J_{i\alpha}.
\end{align}
Plugging this into Eq.~\ref{eq:Fisher} and noting that the matrix multiplication contracts over the $i$ index, we are left with:
\begin{equation}
I(\theta)=\frac{J^TJ}{\sigma^2}.
\end{equation}
The Cram\'{e}r-Rao bound states that the inverse of the Fisher information provides a lower bound for the covariance matrix of unbiased estimators. More explicitly, the relation is as follows:
\begin{equation}
\mathrm{cov}_\theta\qty(\hat{\theta})\geq I(\theta)^{-1}.
\end{equation}
We take the equality for our error estimates. The variances of the parameter estimates are given by the diagonal elements of the covariance matrix.

\subsubsection{Transformation matrices for obtaining the fork velocity}
\label{sec:transformmatrices}
To convert the control points of the piecewise linear fit into slopes and fork velocities, we use coordinate transformation matrices, which have the added benefit of also transforming the variance estimates from the Fisher information. The coordinate transformation matrices can either be obtained through direct reasoning or as Jacobians of the final coordinates relative to the initial coordinates. The following examples will be square matrices of dimension 3, acting on sets of 3 parameters. These examples illustrate the procedure of obtaining the transformation matrices and are easily generalizable. To convert from a vector of control point values to a vector where the first element is the leftmost control point value and the other elements are the slopes, we use the following matrix:
\begin{equation}
\texttt{CtrlPts2Slopes} = \mqty(1&0&0\\-\Delta_{12}^{-1}&\Delta_{12}^{-1}&0\\0&-\Delta_{23}^{-1}&\Delta_{23}^{-1}),
\end{equation}
where $\Delta_{ij}$ is the chromosomal position (horizontal value) of the $j$-th control point minus the position of the $i$-th control point. We can thus relate the vector of slopes $\vec{\alpha}$ and the vector of (vertical) control point values $\vec{\theta}$:
\begin{align}
\vec{\alpha}&=\texttt{CtrlPts2Slopes}\cdot\vec{\theta}=\mqty(\theta_1\\ (\theta_2-\theta_1)/\Delta_{12}\\ (\theta_3-\theta_2)/\Delta_{23})
\end{align}
From the slopes, we can obtain the fork velocities with the following matrix:
\begin{equation}
\texttt{Slopes2Vels} = \mqty(1&0&0\\0&-k_G/\alpha_2^2&0\\0&0&-k_G/\alpha_3^2).
\end{equation}
The velocities $\vec{v}$ are related to $\vec{\alpha}$ like so:
\begin{equation}
\vec{v}=\texttt{Slopes2Vels}\cdot\vec{\alpha}=\mqty(\alpha_1\\-k_G/\alpha_2\\-k_G/\alpha_3).
\end{equation}

To properly transform the covariance matrix obtained in Sec.~\ref{sec:errorest}, we must use the transformation matrix on both sides, like so:
\begin{align}
\mathrm{cov}_\alpha &= \texttt{CtrlPts2Slopes}^T\cdot\mathrm{cov}_\theta\cdot\texttt{CtrlPts2Slopes},\\
\mathrm{cov}_v &= \texttt{Slopes2Vels}^T\cdot\mathrm{cov}_\alpha\cdot\texttt{Slopes2Vels}.
\end{align}
The error estimates for each parameter are the square roots of the corresponding diagonal elements.

\subsubsection{Behavior near the terminus}

\label{sec:ter}
Due to the stochastic nature of replisome progression, replication forks do not always meet at the terminus, which corresponds to the \textit{dif} site \cite{ref:Vibrio_Galli_data}. If one fork arrives first, it can continue past the \textit{dif} site, converting from antegrade to retrograde motion along the other arm. While there are Tus-\textit{Ter} traps in \textit{E. coli} and \textit{B. subtilis} to limit the amount of retrograde motion, they are not 100\% efficient and they are not present in \textit{V. cholerae} \cite{ref:Vibrio_Galli_data}. Thus, in a population, the true termination points along the chromosome would form a distribution around the \textit{dif site}. This means that in regions near the terminus, there can be antegrade and retrograde replication forks both contributing to the marker frequency, which contradicts one of the assumptions required for our analysis: replication proceeding only in one direction. Such retrograde motion would cause a flattening out around any fork convergence points observed in the marker-frequency data. Therefore, when doing a multi-segment analysis, we remove the two fork velocity measurements on either side of the marker-frequency minimum.


\subsubsection{Selection of the number of fitted knots}

\label{sec:modelsel}

First we bin the data into 1~kb regions, following the convention of previous marker frequency analysis studies. To determine the number of segments that would best fit the data, we use the Akaike Information Criterion (AIC). We compare a series of nested models with $2^k$ continuous piecewise-linear least squares fits, where $k=1,...,10$. The AIC-optimal model for fast growth (in LB) had 39 knots, spaced by 100~kb, generating 38 measurements of locus velocity across the two chromosomes of WT \textit{V. cholerae}. Other strains either have similar or smaller region sizes that minimized the AIC value, so for ease of comparison with other mutant strains, we take 100~kb to be the step size for determining fork velocity.

\section{Theoretical methods}
\label{sec:theoreticalmethods}

\subsection{Bilateral symmetry analysis}

\label{sec:sym}

To analyze whether an observed velocity profile was consistent with the predictions of the time-dependent mechanism, we test for bilateral symmetry between the left and right arms:
\begin{equation}
v(\ell) = v(-\ell), \label{eqn:bilat}
\end{equation}
where $\ell$ is the locus position relative to the origin. 

Assume the fork velocity has been computed over $2m$ equal-length genome segments, arranged symmetrically about the origin ($\ell=0$). The segments $i=-1...-m$ correspond to sequentially labeled segments along the left arm and the segments $i=1...m$ correspond to sequentially labeled segments along the right arm such that $\ell = \Delta \ell i$ corresponds to the end point of each segment where $\Delta \ell$ is the segment length. Let the mean fork velocity be defined:
\begin{equation}
\overline{v} \equiv \frac{1}{2m}\sum_{i=1}^mv_i+v_{-i}, 
\end{equation}
where $v_i$ is the velocity over segment $i$. 

To divide the variance into symmetric and antisymmetric contributions, we define symmetrized, $(i)$, and antisymmetrized velocities, $[i]$, for index pairs $\pm i$:
\begin{eqnarray}
\delta v_{(i)} &\equiv& {\textstyle\frac{1}{2}}(v_i+v_{-i}-2\overline{v}),\\
\delta v_{[i]} &\equiv& {\textstyle\frac{1}{2}}(v_i-v_{-i}).
\end{eqnarray}  
We define the symmetric and antisymmetric variances 
\begin{eqnarray}
\sigma^2_{S} &\equiv& \frac{1}{m} \sum_{i=1}^m \delta v_{(i)}^2 \\
\sigma^2_{A} &\equiv& \frac{1}{m} \sum_{i=1}^m \delta v_{[i]}^2,
\end{eqnarray}  
and the total variance is the sum of the symmetric and antisymmetric variances:
\begin{eqnarray}
\sigma^2  &\equiv& \frac{1}{2m}\sum_{i=1}^m(v_i-\overline{v})^2+(v_{-i}-\overline{v})^2\\
 &=& \sigma^2_{S} + \sigma^2_{A}.
\end{eqnarray}
Finally, we define the symmetric fraction of the variance:
\begin{equation}
f_S \equiv \sigma^2_{S} /\sigma^2.
\end{equation}
If the variation in the velocity obeys the bilateral symmetry (Eq.~\ref{eqn:bilat}), the variance is all symmetric (\textit{i.e.} $f_S = 1$). If on the other hand, the variation is randomly distributed along the genome, we expect equal symmetric and antisymmetric combinations (\textit{i.e.} $f_S = 1/2$).

\subsection{Estimation of average fork number per cell cycle}

The following analysis is only an estimation. There are a few key assumptions that are inconsistent with cell phenomenology, but are kept to make the analysis tractable. In particular, we take the assumption that neither arm of the chromosome has fork arrest, which is inconsistent with the results for \textit{B. subtilis}. This assumption allows us to obtain an estimate without prior knowledge of where fork arrest may occur. We also assume that the D period (time between end of chromosomal replication and cell division) is negligible, which is inconsistent when growth is extremely slow. This assumption allows us to use the copy number of the terminus as an approximation of the number of cells, which cannot be determined from MFA. 

The population number density of forks $n_f(\ell)$ at any position $\ell$ is given by the change in copy number between two consecutive positions along the chromosome:
\begin{equation}
n_f(\ell)=-\dv{}{\ell} N(\ell).
\end{equation}
This equation is a result of the following reasoning: If there are two copies at $\ell=x$ and one copy at $\ell=x+1$, then there must be a fork that has just replicated $\ell=x$. We need to integrate the number density over the entire chromosome to get the total number of forks for the population:
\begin{equation}
{N}_{fork,population}=2\int_{ori}^{ter} n_f(\ell)\,\dd \ell=2\int_{ori}^{ter}-\dv{}{\ell}N(\ell)\,\dd\ell=2(N_{ori}-N_{ter}),
\end{equation}
where the factor of 2 came from having two arms. To get the average number of forks per cell $\bar{N}_f$, we divide by the total number of cells, which is roughly equal to the number of termini $N_{ter}$:
\begin{equation}
\bar{N}_{f} = 2\frac{N_{ori}-N_{ter}}{N_{ter}}.
\end{equation}
This can be related to lag time:
\begin{equation}
\tau_\ell = -k_G^{-1}\ln\frac{N_{\ell}}{N_{ori}}\implies \frac{N_{ori}}{N_{ter}}=\exp(k_G\tau_{ter})=e^{\tau_{ter}\ln 2/T}=\qty(e^{\ln2})^{\tau_{ter}/T}=2^{\tau_{ter}/T},
\end{equation}
which can be substituted into the equation above to get:
\begin{equation}
\bar{N}_{f}=2*(2^{\tau_{ter}/T}-1).
\end{equation}
This is closely related to the result derived in \cite{Bremer:1977fv}. For two chromosomes, we add the number of forks for both, but divide everything just by the minimum $N_{ter,i}$, which corresponds to the terminus of Chr1 in this case:
\begin{equation}
\bar{N}_{f,2chr} = 2\frac{(N_{ori,1}-N_{ter,1})}{N_{ter,1}}+2\frac{(N_{ori,2}-N_{ter,2})}{N_{ter,1}}.
\end{equation}
In general, when there are multiple chromosomes:
\begin{equation}
\bar{N}_{fork,multichr} = \frac{2}{\min(N_{ter,i})}\sum_{i}(N_{ori,i}-N_{ter,i}),
\end{equation}
where $i$ denotes which chromosome. To get a relative lag time, we use:
\begin{equation}
\tau_{\ell\rightarrow m}=\tau_m-\tau_\ell=-k_G^{-1}\qty(\ln\frac{N_m}{N_{ori}}-\ln\frac{N_\ell}{N_{ori}})=k_G^{-1}\ln\frac{N_\ell}{N_m}.
\end{equation}
Letting the subscript $end$ denote the lag time corresponding to $\min(N_{ter,i})$. We thus have:
\begin{equation}
\bar{N}_{fork,multichr}=2\sum_{i}\qty(2^{(\tau_{end}-\tau_{ori,i})/T}-2^{(\tau_{end}-\tau_{ter,i})/T}),
\end{equation}
where $\tau_{end}$ is the maximum lag time measured. This can be simplified to:
\begin{equation}
\bar{N}_{fork,multichr}=2\times2^{\tau_{end}/T}\times\sum_{i}\qty(2^{-\tau_{ori,i}/T}-2^{-\tau_{ter,i}/T}),
\end{equation}
which enables a calculation of average fork number per cell cycle using lag times.

\subsection{Statistically significant deviations of local fork velocity from the global mean}
\label{sec:forkvelstatsig}
To test the statistical significance of the measured velocities, we use a null hypothesis test. The null hypothesis corresponds to a model with uniform fork velocity on the right and left arms. The alternative hypothesis corresponds to a model with a model-selected number of knots ($m$), corresponding to $m-1$ genomic region-specific velocities.


To calculate the p-value, we begin by obtaining the $\chi^2$-test statistic:
\begin{equation}
\chi^2=\sum_i\frac{(O_i-E_i)^2}{\sigma_i^2}
\end{equation}
where $O_i$ denotes the observed values, $E_i$ denotes the expected values, and $\sigma_i^2$ is the variance of the fork velocities measured using the fitter function (as described in Supplemental Material Sec.~\ref{sec:transformmatrices}). In this case, $E_i$ is the global mean fork velocity for all $i$, since that is the null hypothesis. Since there is only one degree of freedom for the model, we use the one-dimensional $\chi^2$ cumulative distribution function to determine the probability of measuring a $\chi^2$ statistic as extreme as the one we've measured, assuming the null hypothesis is true. This p-value is found to be $\ll 10^{-30}$ for all datasets except one with a p-value of $6\times 10^{-12}$. All of the p-values are much smaller than the standard threshold of 0.05, so we reject the null hypothesis. We thus conclude that there are statistically significant variations of the fork velocity relative to the mean.

\section{Supplemental theoretical results}

\subsection{Stochastic simulations}

\label{sec:model}

\subsubsection{Stochastic simulations method}

\begin{figure}[t]
\centering
\includegraphics[width=.65\linewidth]{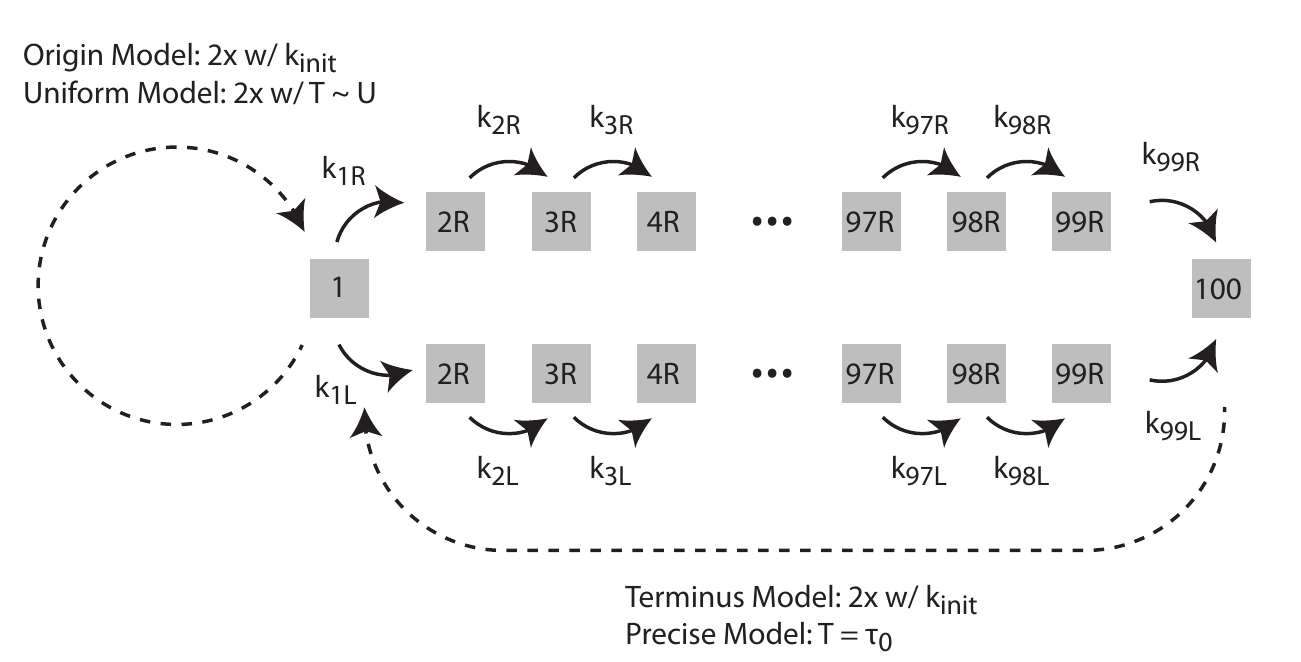}
\caption{\textbf{Stochastic simulation model schematics.} Replication initiates at the origin (state 1) and proceeds bi-directionally along the left (L) and right arms (R). Each transition between simulation bases (\textit{i.e.}~state number) represents the replication of roughly 25 kb. States have exponentially-distributed lifetimes with a rate equal to the fork velocity at that position: $k_X = v(\ell_X)$. A number of different initiation schemes were simulated. In the \textit{Origin Model}, the initiation rate was proportional to the number origins (per cell). In the \textit{Terminus Model}, the initiation rate was proportional to the number termini (per cell). In the \textit{Uniform Model}, there is a uniformly distributed wait time after initiation before the next initiation occurs.  In the \textit{Precise Model}, there is a fixed time interval after termination before the next initiation occurs.}
\label{fig:model}
\end{figure}

The purpose of the stochastic simulation was \textit{to investigate the role of stochasticity in determining the log-phase growth demography of the population}. It was \textit{not} to generate a mechanistically realistic and detailed model of the cell cycle.   

To simulate the cell cycle, we performed an exact stochastic simulation using the Gillespie Algorithm \cite{Gillespie1977}. We idealized the cell cycle as follows: {\textbf{Genome representation:}} The genome was divided into two equal length arms each consisting to 100 course-grained bases. An explicit realization of all 5 Mb would be too slow to rapidly explore different cell models. Finer coarse-grained basepairs were explored but made no difference to the results. \textbf{Replication initiation}: We experimented with a number of different models for initiation since we initially believed that the stochasticity of initiation would limit the ability to quantitate the fork velocity. (i) We initially investigate a \textit{Terminus Model} in which there was a constant rate of initiation $k_{\rm init}$ after the termination of the on-going replication. (ii) In order to simulate a more realistic model, we then considered an \textit{Origin Model} where there was constant rate $k_{\rm init}$ of initiation at each origin.  (iii) We explored a \textit{Precise Model} with precise timed B period length $T_B$. (iv) We explored a \textit{Uniform Model} with a uniform distribution of B periods. \textbf{Fork velocity:} We used the replication of the course-grained bases had exponentially distributed wait times $k_{\rm rep} = v/\Delta \ell$ (\textit{i.e.}~constant rate per unit time) where $v$ is the replication velocity and  $\Delta \ell$ is the length of the course-grained bases. We note that this assumption makes replication \textit{more stochastic} than a more realistic model in which wait times are exponentially distributed  at the single-base level. We experimented with a number of different types of locus-dependent fork velocities. In addition, we added exponentially-distributed pauses of various durations at a specific locus. \textbf{Termination:} We only allowed a single direction of fork propagation. When the fork reached the end of the arm, replication terminated. \textbf{Cell division:} We assumed cells divided immediately after termination of the slowest arm of the chromosome. \textbf{Initiation of the simulation:} All simulations initiated from a single new-born cell with an un-replicated chromosome. \textbf{Stop conditions:} Simulation were run until there were $10^5$ cells. \textbf{Determination of growth rate.} The number of cells was fit to an exponential to determine the growth rate $k_G$ between $N_{\rm cell}(t) = 10^4$ and $N_{\rm cell}(t) = 10^5$ cells. \textbf{Generation of simulated marker frequency.} The marker frequency was defined as the number of each genetic locus in the population at the termination of the simulation. In summary, the model was described by the parameters $k_{\rm init}$ and the fork velocity $v(\ell)$ (or pause time distribution) at each locus. The model output was the marker frequency at each locus $N(\ell)$ and the growth rate $k_G$.  See Fig.~\ref{fig:model}.

\subsubsection{Stochastic simulations match the predictions of the log-slope}

To test whether the log-slope law applies locally, irrespective of stochasticity in cell-cycle timing, we used a stochastic simulation to generate simulated marker-frequency data. The basic strategy was to simulate a stochastically timed cell cycle, including stochastically timed B periods as well as stochastic fork dynamics, and compare the observed population demographics to the prediction of non-stochastic models.

\idea{Log slope.} We define the log slope:
\begin{equation}
\alpha(\ell) \equiv \frac{\rm d}{\rm d\ell} \log N(\ell),
\end{equation}
where $N(\ell)$ is the population copy number of the locus at position $\ell$. Theoretically, the slope is predicted to be:
\begin{equation}
\alpha(\ell) = \frac{k_G}{v(\ell)},
\end{equation}
where $k_G$ is the growth rate and $v(\ell)$ is the locus-dependent fork velocity.

\idea{Log difference.} For long pauses, the dynamics are characterized by a step rather than a slope. The step is defined:
\begin{equation}
\Delta(\ell) \equiv \lim_{\delta\rightarrow 0^+} \log N(\ell+\delta)-  \log N(\ell),
\end{equation}
where $N(\ell)$ is the population copy number of the locus at position $\ell$. Theoretically, the log difference is predicted to be:
\begin{equation}
\Delta(\ell) \equiv k_G \delta \tau(\ell) = \log( 1+ \textstyle\frac{k_G}{k}),
\end{equation}
where $k_G$ is the growth rate, $\delta \tau(\ell)$ is the exponential mean of the step wait time and the last equality applies in the special case of an exponentially-distributed wait time with rate $k$.

\subsubsection{Simulation models}

The model descriptions and motivations are summarized below:

\idea{Terminus  Model 1:} In the terminus model, the replication initiation rate is proportional to the terminus number (per cell). In this model, there is a locus-independent fork velocity.

\idea{Terminus  Model 2:} Same as above, but with two fork velocities. On the last half of the right arm (region 1), the fork velocity is higher than the rest of the chromosome (region 0).

\idea{Terminus  Model 3:} Same as above, but with an exponentially-distributed pause close to the origin on the right arm. 

\idea{Terminus  Model 4:} Same as Terminus 1, but with a lower initiation rate. 

\idea{Origin Model:} In the origin model, the replication initiation rate is proportional to the origin number.

\idea{Uniform  Model:} The timing between initiations is uniformly distributed. This model was included to explore whether the observations were a special case of exponentially-distributed wait times.

\idea{Precise Model:} The timing between initiations is uniformly distributed. This model was included to explore whether the observations were a special case of exponentially-distributed wait times.

\subsubsection{Simulation results}

The results from the simulations are summarized in the table below. All numbers have at least precision of $1\%$ and are in simulation units: simulation bases (sb) and simulation time (st). 
\medskip

{
{\centering

\begin{tabular}{ | r || r | r || r | r || r |}
\hline
   {Model}  &  {Parameter}                      &  {Parameter}                &  {Observed}              &  {Observed}                         & {Predicted} \\
                      &  {Initiation}  &  {Fork dynamics}  &  {Growth rate: $k_G$ }  &  {Log slope/difference } & {Log slope/difference} \\
\hline
\hline
Terminus 1           &  $k_{\rm init} = 1$ st$^{-1}$  &     $v_0 = 20$ sb/st        &   $2.6\times 10^{-1}$   st$^{-1}$             &  $\alpha = 9.0\times 10^{-3}$ sb$^{-1}$    &   $\alpha = 9.0\times 10^{-3}$    sb$^{-1}$                               \\
\hline
Terminus 2           &  $k_{\rm init} = 1$ st$^{-1}$    &     $v_0 = 20$  sb/st       &         $2.7\times 10^{-1}$ st$^{-1}$         &  $\alpha_0 = 1.3\times 10^{-2}$   sb$^{-1}$   &    $\alpha_0 = 1.3\times 10^{-2}$  sb$^{-1}$                                \\
           &                      &     $v_1 = 30$  sb/st      &                                      &  $\alpha_1 = 8.8\times 10^{-3}$   sb$^{-1}$   &    $\alpha_1 = 8.8\times 10^{-3}$    sb$^{-1}$                              \\
\hline
Terminus  3          &  $k_{\rm init} = 1$ st$^{-1}$   &     $v_0 = 20$  sb/st       &     $2.6\times 10^{-1}$  st$^{-1}$            & $\alpha_0 = 1.3\times 10^{-2}$   sb$^{-1}$   &       $\alpha_0 = 1.3\times 10^{-2}$   sb$^{-1}$                            \\
               &                      &     $v_1 = 30$  sb/st       &                                     & $\alpha_1 = 8.8\times 10^{-3}$   sb$^{-1}$    &        $\alpha_1 = 8.8\times 10^{-3}$   sb$^{-1}$                           \\
                    &                      &     $k_3 = 2$   st$^{-1}$       &                                      &  $\Delta = 1.3\times 10^{-1}$                                 &     $\Delta = 1.3\times 10^{-1}$                                \\
                    \hline
Terminus 4            &  $k_{\rm init} = 0.3$ st$^{-1}$   &     $v_0 = 20$  sb/st       &     $1.4\times 10^{-1}$  st$^{-1}$            & $\alpha_0 = 7.3\times 10^{-3}$   sb$^{-1}$   &       $\alpha_0 = 7.3\times 10^{-3}$   sb$^{-1}$                            \\
                    \hline

Origin            &    $k_{\rm init} = 0.3$  st$^{-1}$   &     $v_0 = 20$  sb/st    &     $3.0\times 10^{-1}$ st$^{-1}$                     &      $\alpha = 1.5\times 10^{-2}$ sb$^{-1}$       &    $\alpha = 1.5\times 10^{-2}$ sb$^{-1}$                                \\
\hline
Uniform           &  $T\sim U([0,6])$    st      &     $v_0 = 20$  sb/st                       &    $2.7\times 10^{-1}$ st$^{-1}$                   &  $\alpha = 1.3\times 10^{-2}$ sb$^{-1}$   &    $\alpha = 1.3\times 10^{-2}$ sb$^{-1}$                                  \\
\hline
Precise           &  $T= 0.2$    st      &     $v_0 = 20$  sb/st                       &    $1.4\times 10^{-1}$ st$^{-1}$                   &  $\alpha = 6.8\times 10^{-3}$ sb$^{-1}$   &    $\alpha = 6.8\times 10^{-3}$ sb$^{-1}$                                  \\
\hline
  \end{tabular}

}}

\subsubsection{Re-scaling simulation units to compare to measured data}

The simulations were performed in simulation units. We call the coarse-grained bases simulation bases (sb) and time units simulation time (st). To compare these simulations to experimental data, the prediction need to be rescaled to place the observations in biological units. To model a 5 Mb bacterial genome, with typical fork velocity of 1 kb/s, the conversion factors are:
\begin{eqnarray}
\frac{5\times 10^6\ {\rm bp}}{200\ {\rm sb}} = 2.5\times 10^4\ \frac{{\rm bp}}{{\rm sb}},\\
\left(\frac{1\ {\rm s}}{10^3\ {\rm bp}}\right) \left(\frac{20\ {\rm sb}}{1\ {\rm st}}\right) \left(\frac{5\times 10^6\ {\rm bp}}{200\ {\rm sb}}\right) = 5\times 10^2\ \frac{{\rm s}}{{\rm st}},
\end{eqnarray}
since we simulated typical fork velocities of $20$ sb/st which we rescale the units to be equivalent to 1 kb/s. 

\idea{Aside:} Note that due to the smaller number of sb relative to bp, the stochasticity of fork dynamics should be exaggerated in the simulations, strengthening our argument that the stochasticity does not effect the model predictions.

\subsubsection{Movies of marker frequency dynamics approaching steady state growth}

We include two examples of movies tracking the marker frequency dynamics from a single cell to steady state growth. Movies \texttt{term1\_mov.mp4} and \texttt{term3\_mov.mp4} show the dynamics of the Terminus 1 and Terminus 3 models respectively. Movie time is shown in simulation units. The dynamics emphasizes the importance of performing the calculations of the marker frequency in the steady-state limit. Not  taking this limit correctly leads to anomalous results (\textit{e.g.}~\cite{Retkute2018}). 
\newpage

\section{Supplemental experimental results}

\subsection{Comparison with GC content}

\label{sec:gc}

\begin{figure}[h]
\centering
\includegraphics[width=.5\linewidth]{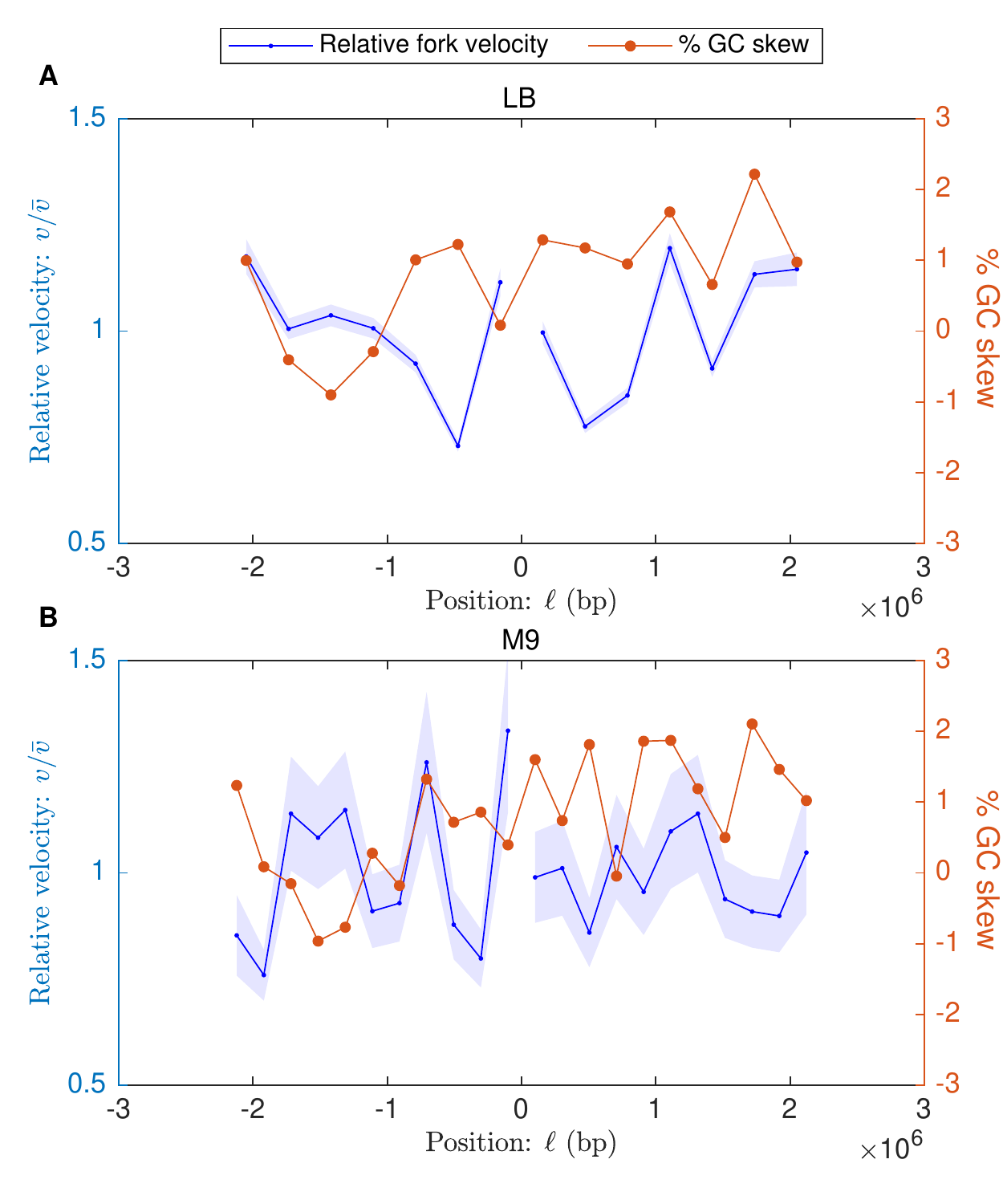}
\caption{\textbf{Relative fork velocity and \% GC skew as a function of position.} Results for wild-type \textit{E.~coli} strain MG1655. The blue curve represents the fork velocity divided by the mean fork velocity, with shaded error bars. The orange curve represents the \% GC skew. \textbf{Panel A:} Growth in LB medium. \textbf{Panel B:} Growth in M9 glucose minimal medium.}
\label{fig:EcGCContent} 
\end{figure}

One potential rate-limiting factor for replication is the distribution of GC content across the genome. However, at the current genomic resolution of 100 kb, we find no significant evidence of GC content determining the rate of replication. See Fig.~\ref{fig:EcGCContent} for the results of wild-type \textit{E.~coli} strain MG1655 in LB and M9.

\newpage

\subsection{\textit{V.~cholerae} WT on LB}

\begin{figure}[ht]
{
{\centering

\includegraphics[width=.5\linewidth]{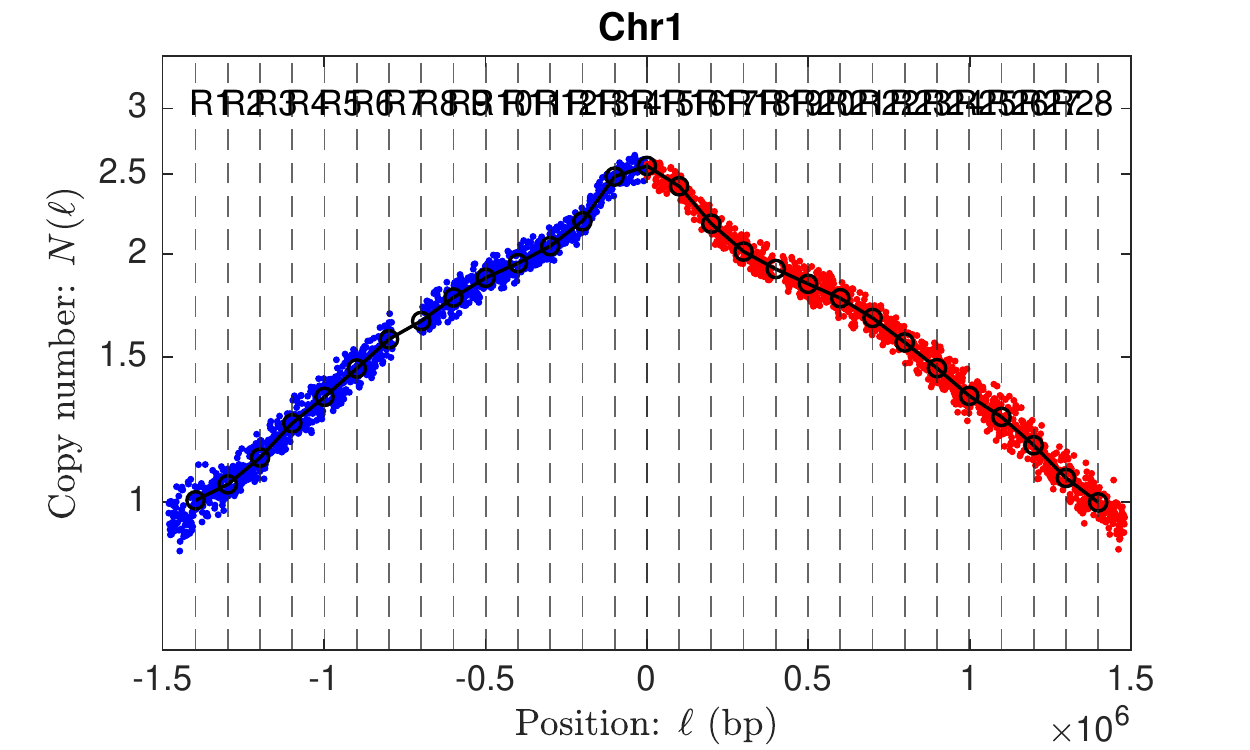}
\includegraphics[width=.5\linewidth]{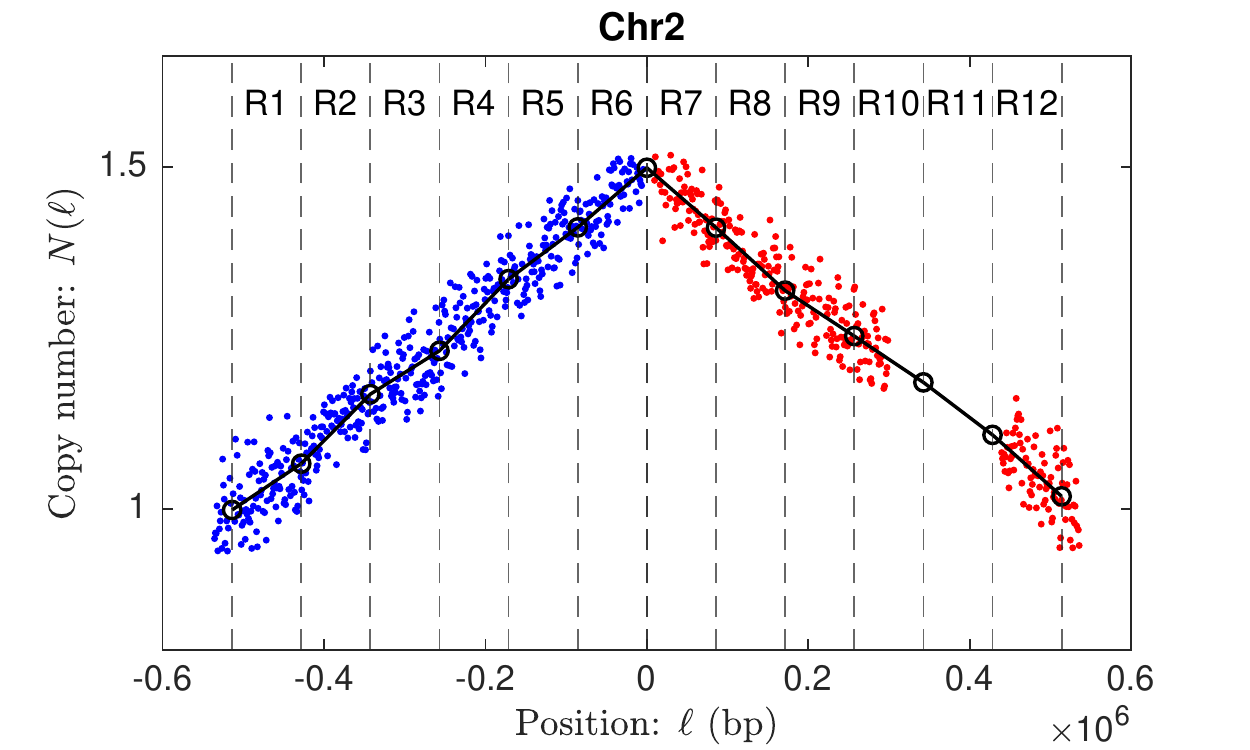}

Detailed results tabulated in \texttt{SuppData.xlsx}.

}}
\end{figure}

\newpage

\subsection{\textit{V.~cholerae} WT on M9 fructose}

\begin{figure}[ht]
{
{\centering

\includegraphics[width=.5\linewidth]{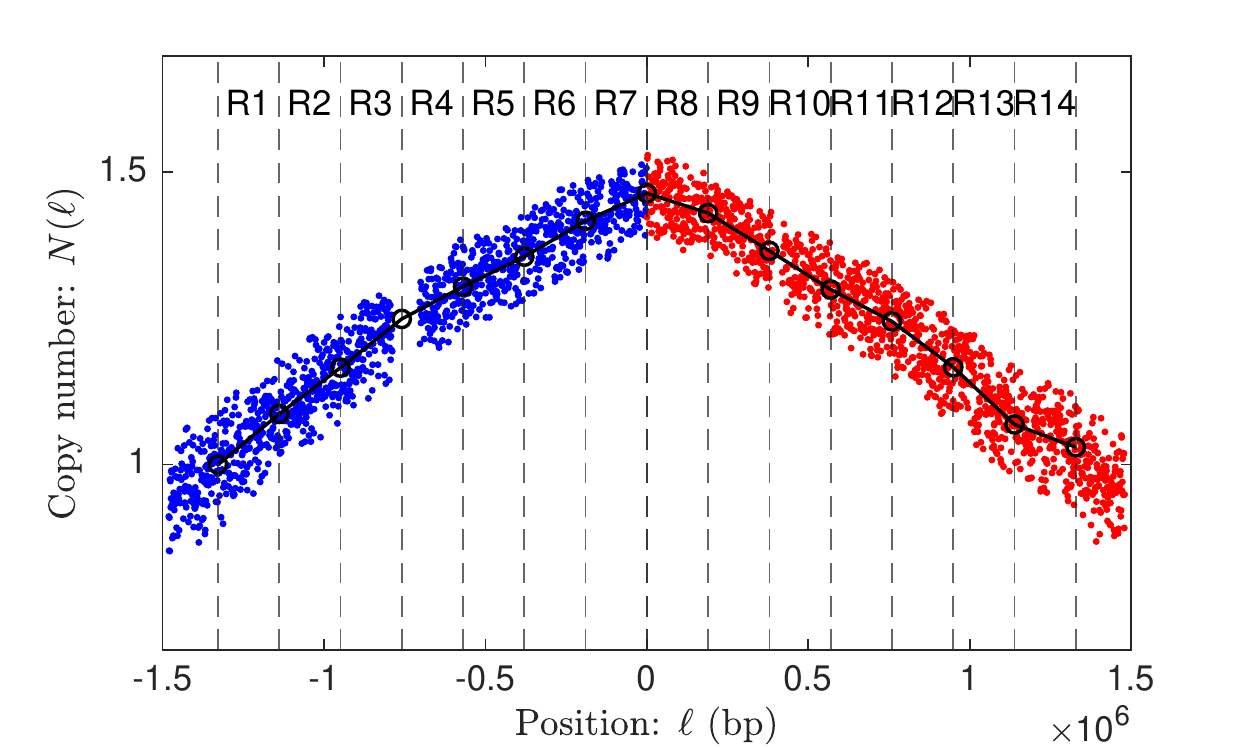}
\includegraphics[width=.5\linewidth]{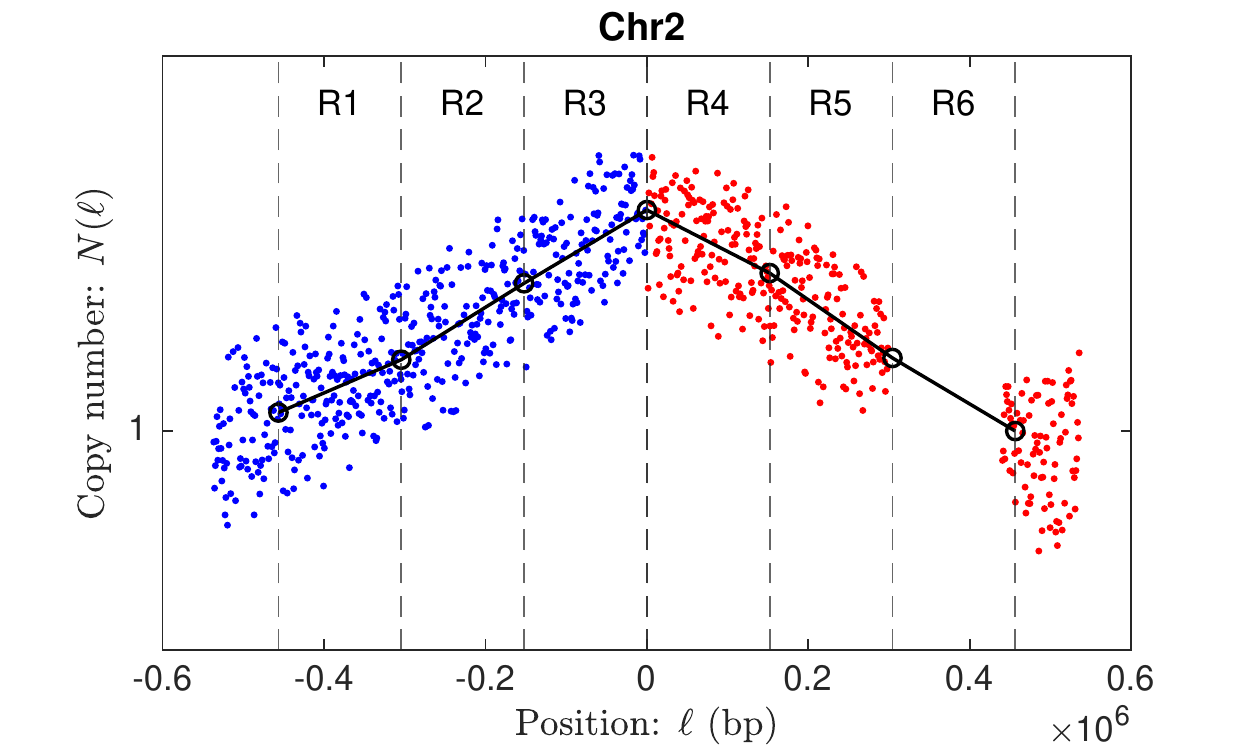}

Detailed results tabulated in \texttt{SuppData.xlsx}.

}}
\end{figure}

\newpage

\subsection{\textit{V.~cholerae} MCH1 on LB}

\begin{figure}[ht]
{
{\centering

\includegraphics[width=.5\linewidth]{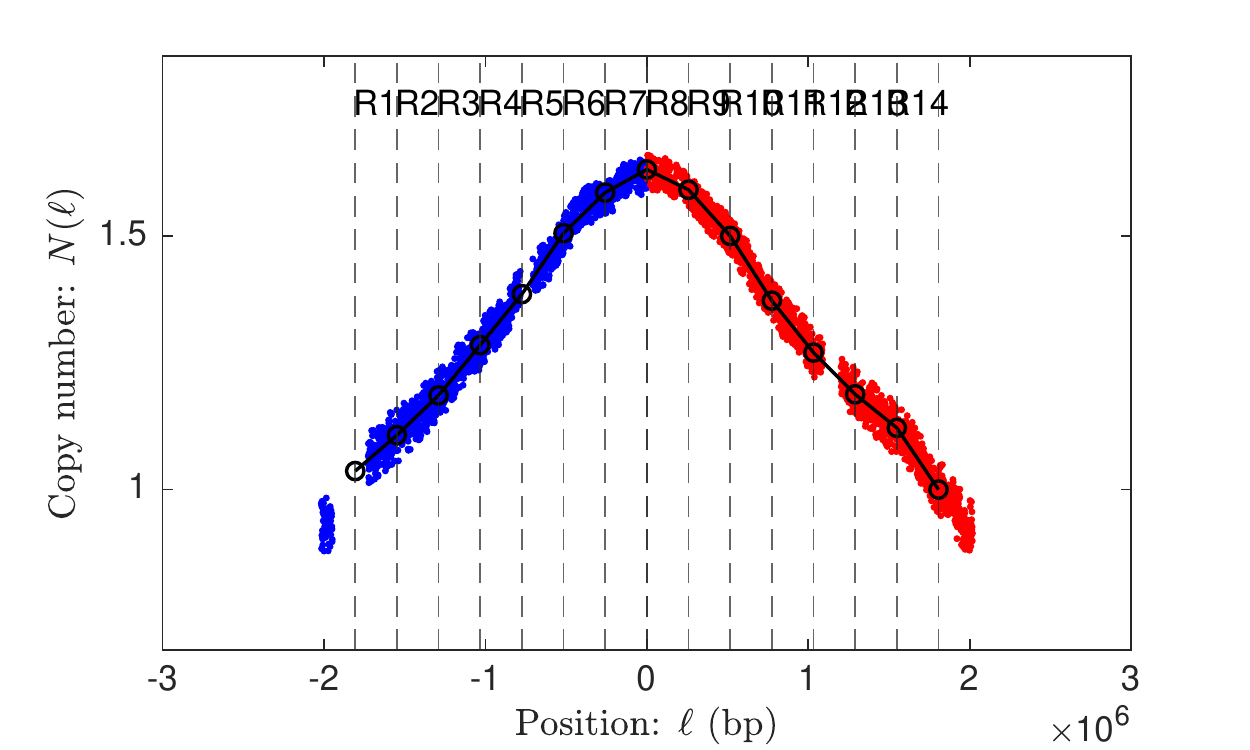}

Detailed results tabulated in \texttt{SuppData.xlsx}.

}}
\end{figure}

\subsection{\textit{V.~cholerae} MCH1 on M9 fructose}

{
{\centering

\includegraphics[width=.5\linewidth]{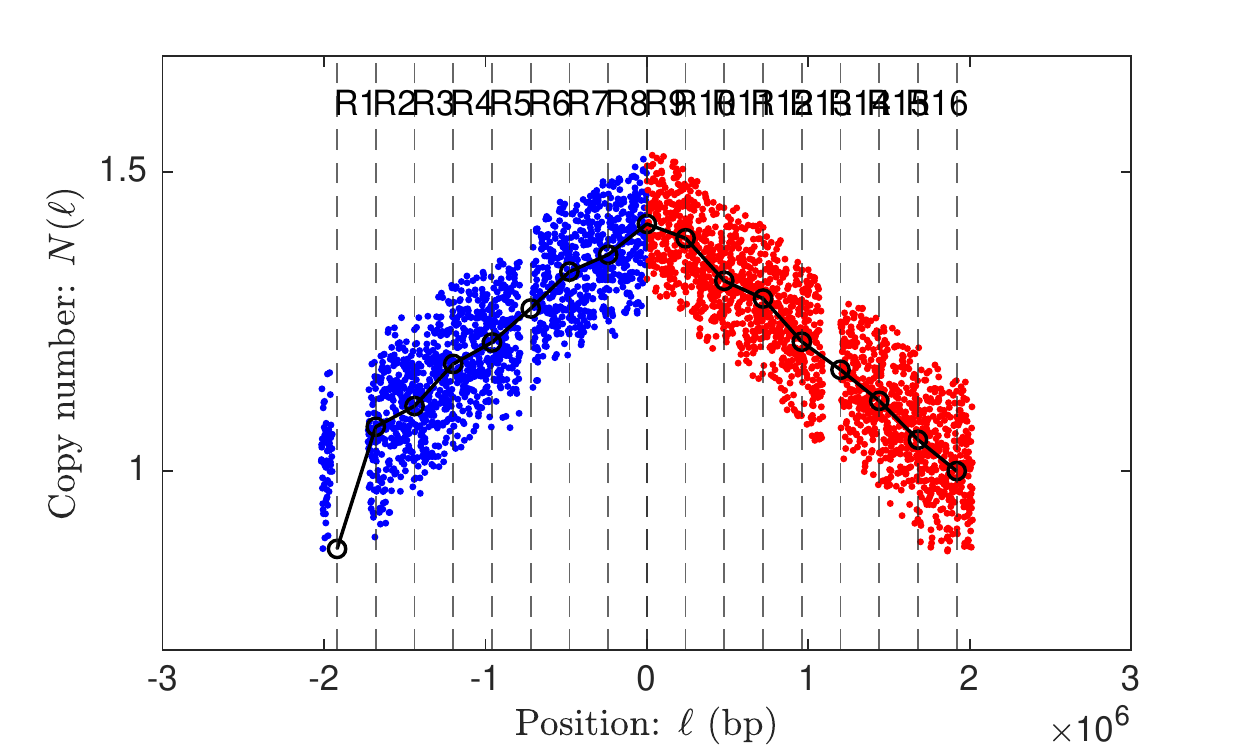}

Detailed results tabulated in \texttt{SuppData.xlsx}.

}}

\newpage

\subsection{\textit{V.~cholerae} \textit{oriR4} on M9 fructose}

{
{\centering

\includegraphics[width=.5\linewidth]{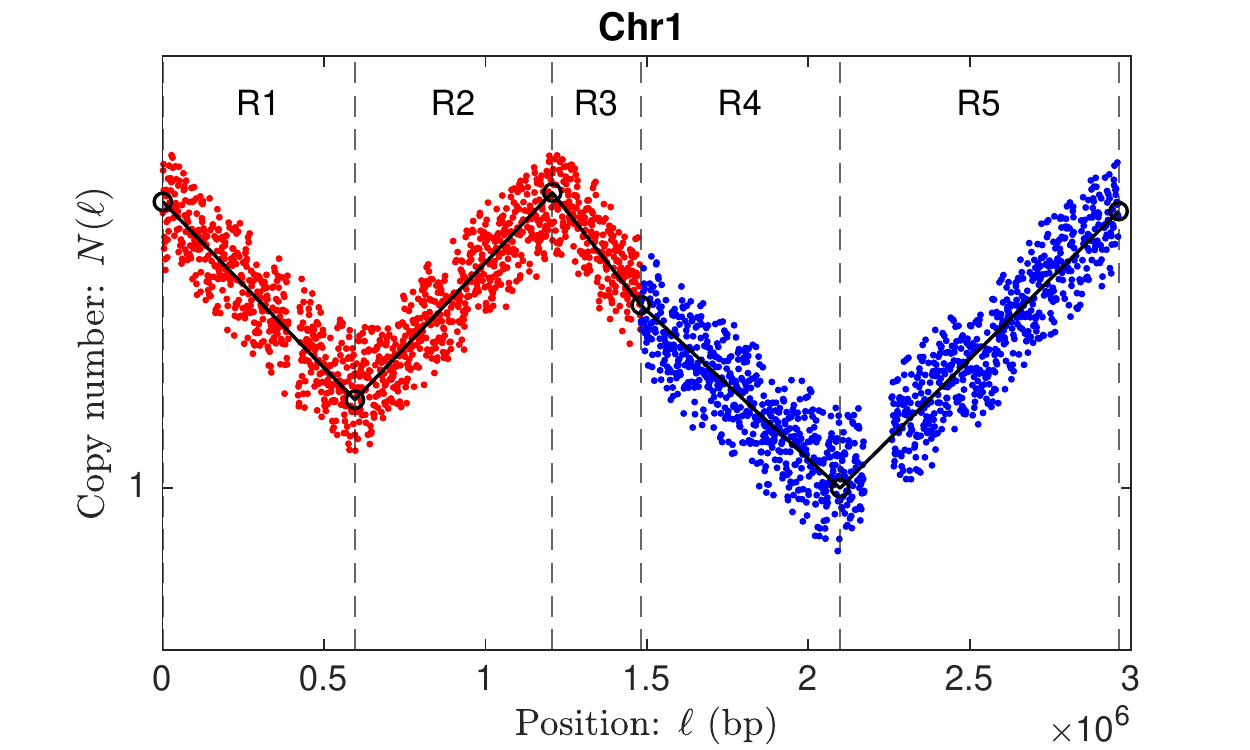}

\resizebox{\textwidth}{!}{\begin{tabular}{ | c | c | c | c | c | c | c | c | c | c |}
\hline
    Region   & Region start              & Region end                 &  Fork         & Replication   &  Slope:                 & Velocity:    & Log                      & Duration:   & Percent  \\
    name     &  position: $\ell_-$ (Mb)  & position: $\ell_+$ (Mb)    &  direction    & orientation   &  $\alpha$ (Mb$^{-1}$)   & $v$ (kb/s)   & difference: $\Delta$     & $\Delta \tau$ (min) & error \\
\hline
\hline
R1 & $0$ & $0.596$ & + & A & 0.308 & 0.750 & 0.183 & 13.2 & $\pm2.0\%$\\ 
\hline
R2 & $0.596$ & $1.21$ & - & R & 0.314 & 0.736 & 0.192 & 13.8 & $\pm1.7\%$\\ 
\hline
R3 & $1.21$ & $1.48$ & + & A & 0.380 & 0.608 & 0.104 & 7.51 & $\pm3.3\%$\\ 
\hline
R4 & $1.48$ & $2.10$ & + & R & 0.275 & 0.840 & 0.170 & 12.2 & $\pm1.9\%$\\ 
\hline
R5 & $2.10$ & $2.96$ & - & A & 0.297 & 0.777 & 0.256 & 18.5 & $\pm1.3\%$\\ 
\hline
  \end{tabular}
  }
  
\includegraphics[width=.5\linewidth]{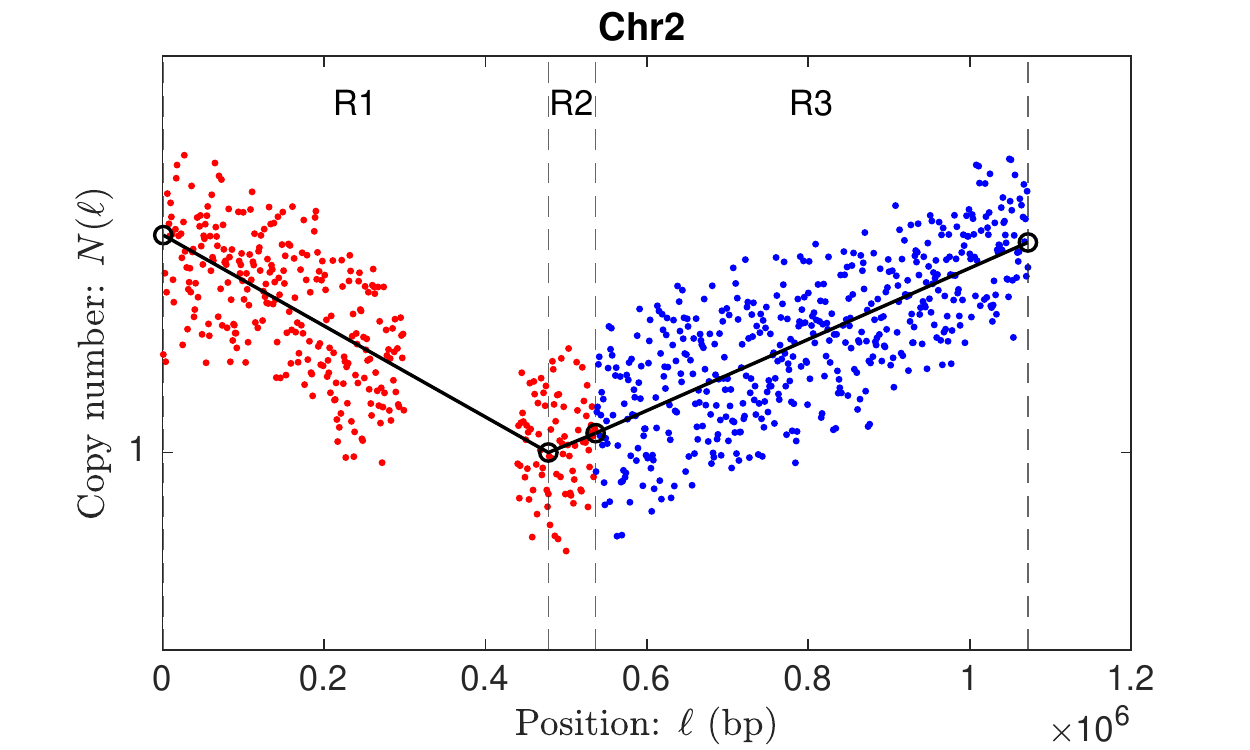}

\resizebox{\textwidth}{!}{\begin{tabular}{ | c | c | c | c | c | c | c | c | c | c |}
\hline
    Region   & Region start              & Region end                 &  Fork         & Replication   &  Slope:                 & Velocity:    & Log                      & Duration:   & Percent  \\
    name     &  position: $\ell_-$ (Mb)  & position: $\ell_+$ (Mb)    &  direction    & orientation   &  $\alpha$ (Mb$^{-1}$)   & $v$ (kb/s)   & difference: $\Delta$     & $\Delta \tau$ (min) & error \\
\hline
\hline
R1 & $0$ & $0.478$ & + & A & 0.257 & 0.898 & 0.123 & 8.85 & $\pm5.0\%$\\ 
\hline
R2 & $0.478$ & $0.537$ & - & R & 0.186 & 1.24 & 0.0109 & 0.787 & $\pm52\%$\\ 
\hline
R3 & $0.537$ & $1.07$ & - & A & 0.201 & 1.15 & 0.108 & 7.76 & $\pm4.4\%$\\ 
\hline
  \end{tabular}
  }}

\subsection{  \textit{B.~subtilis} \textit{oriC}-257$^\circ$ on S7 fumarate }

{\centering

\includegraphics[width=.5\linewidth]{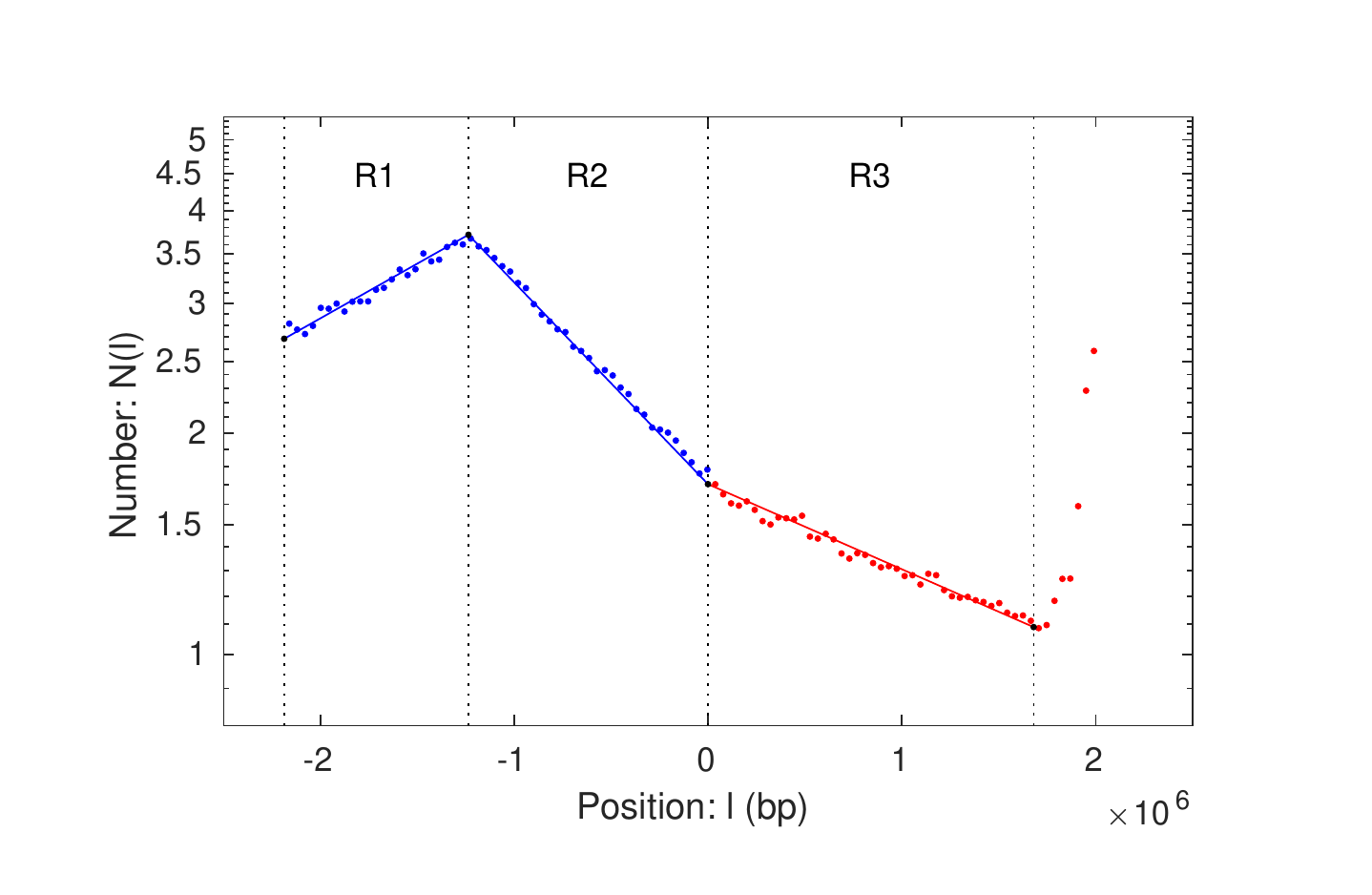}

\resizebox{\textwidth}{!}{\begin{tabular}{ | c | c | c | c | c | c | c | c | c |}
\hline
    Region   & Region start              & Region end                 &  Fork         & Replication   &  Slope:                 & Velocity:    & Log                      & Duration:     \\
    name     &  position: $\ell_-$ (Mb)  & position: $\ell_+$ (Mb)    &  direction    & orientation   &  $\alpha$ (Mb$^{-1}$)   & $v$ (kb/s)   & difference: $\Delta$     & $\Delta \tau$ (min) \\
\hline
\hline
R1 & $-2.19$ & $-1.24$ & - & A & $0.342 \pm 4.3\%$ & NA & $0.325 \pm 4.3\%$ & NA\\ 
R2 & $-1.24$ & $-8.72\times 10^{-4}$ & + & R & $0.631 \pm 1.3\%$ & NA & $0.779 \pm 1.3\%$ & NA\\ 
R3 & $-8.72\times 10^{-4}$ & $1.68$ & + & R & $0.266 \pm 2.6\%$ & NA & $0.447 \pm 2.6\%$ & NA\\ 
\hline
  \end{tabular}

}}

\subsection{  \textit{B.~subtilis} \textit{oriC}-94$^\circ$ on S7 fumarate }

\label{sec:othbs1}

{\centering

\includegraphics[width=.5\linewidth]{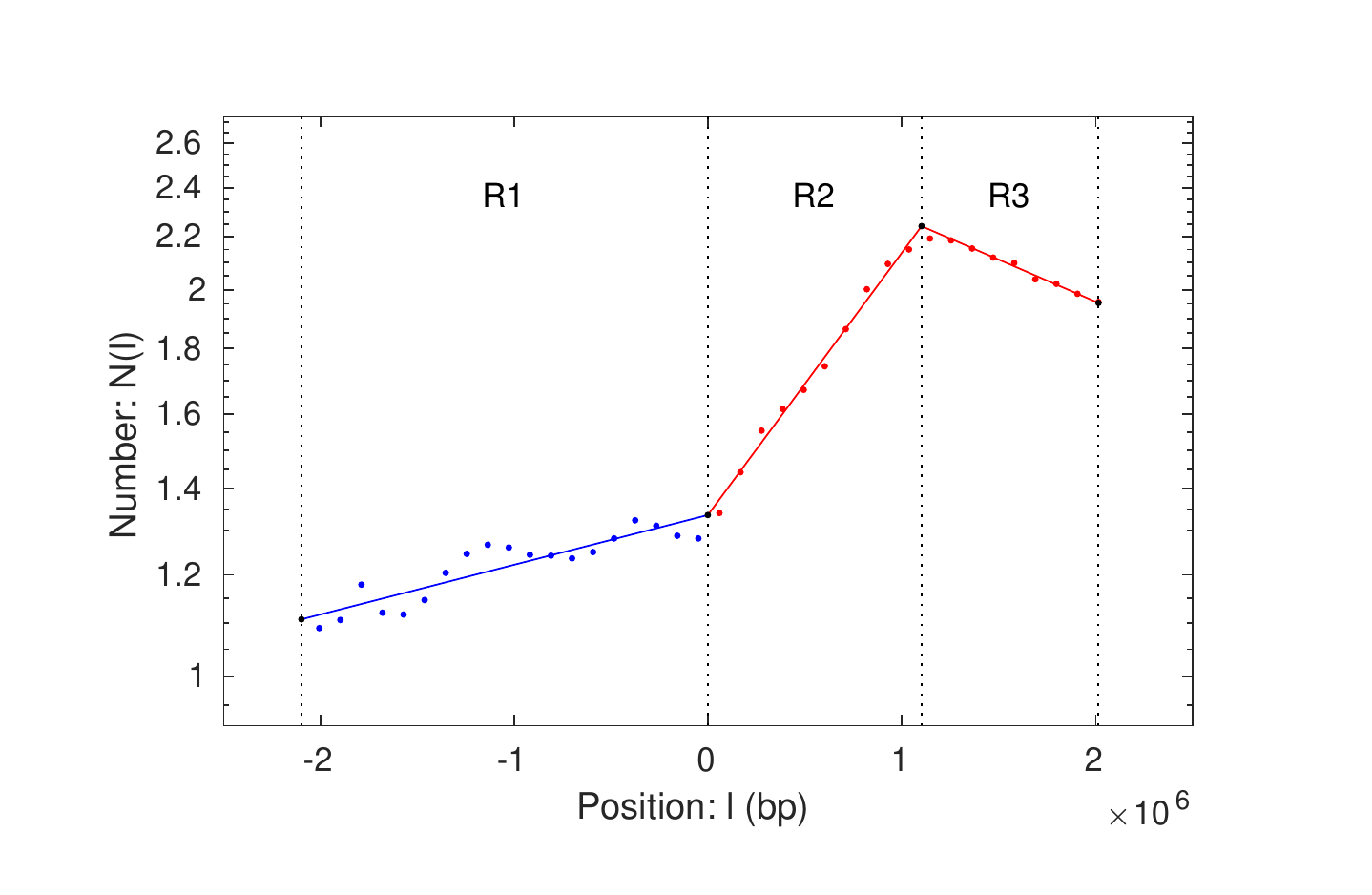}

\resizebox{\textwidth}{!}{\begin{tabular}{ | c | c | c | c | c | c | c | c | c |}
\hline
    Region   & Region start              & Region end                 &  Fork         & Replication   &  Slope:                 & Velocity:    & Log                      & Duration:     \\
    name     &  position: $\ell_-$ (Mb)  & position: $\ell_+$ (Mb)    &  direction    & orientation   &  $\alpha$ (Mb$^{-1}$)   & $v$ (kb/s)   & difference: $\Delta$     & $\Delta \tau$ (min) \\
\hline
\hline
R1 & $-2.10$ & $-1.44\times 10^{-3}$ & - & A & $8.91\times 10^{-2} \pm 10\%$ & NA & $0.187 \pm 10\%$ & NA\\ 
R2 & $-1.44\times 10^{-3}$ & $1.10$ & - & R& $0.469 \pm 3.5\%$ & NA & $0.518 \pm 3.5\%$ & NA\\ 
R3 & $1.10$ & $2.01$ & + & A & $0.151 \pm 18\%$ & NA & $0.138 \pm 18\%$ & NA\\ \hline
  \end{tabular}

}}

\subsection{  \textit{B.~subtilis} \textit{oriN} on S7 fumarate }

\subsubsection{ Two-slopes model}

{
{\centering

\includegraphics[width=.5\linewidth]{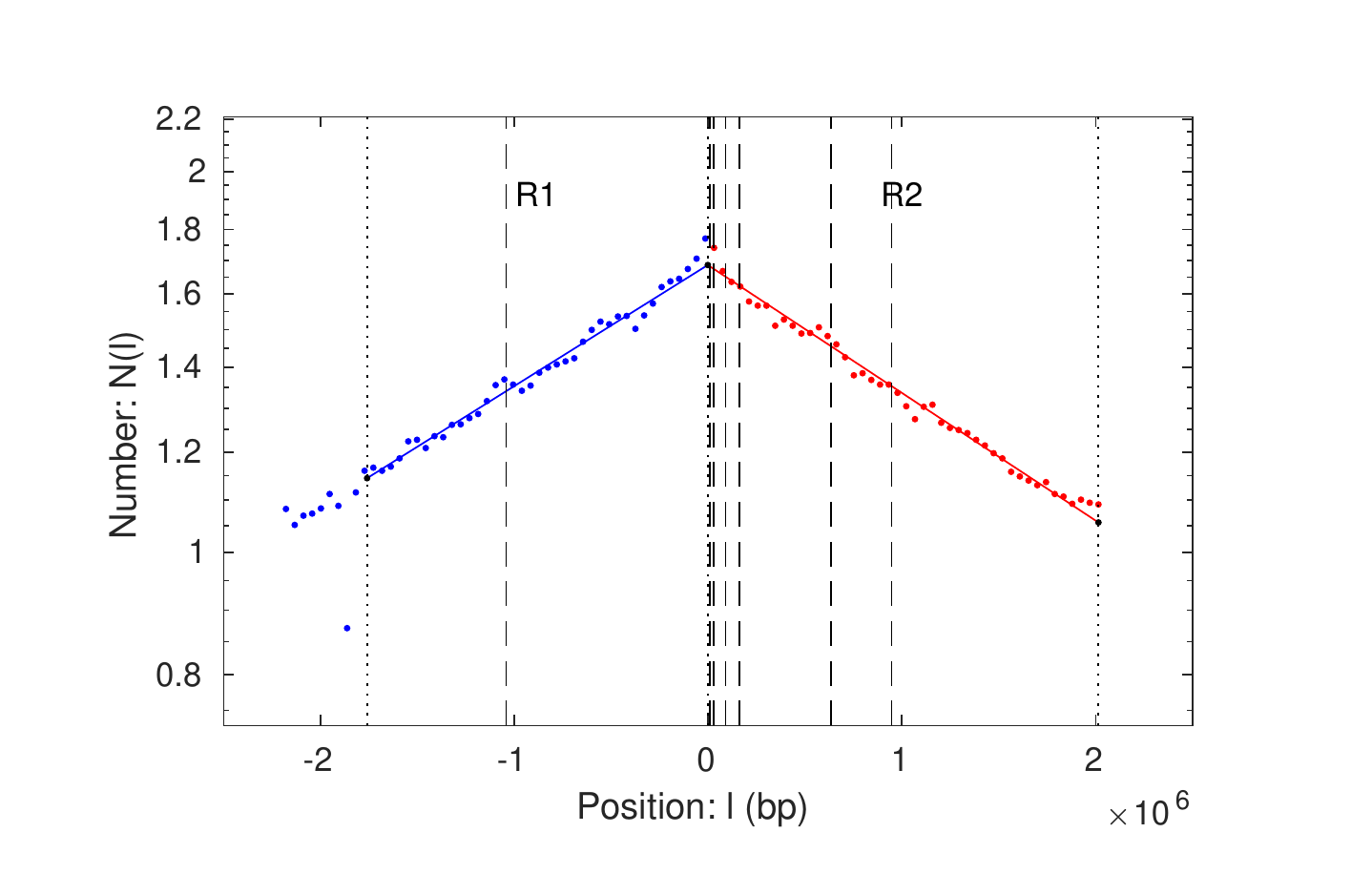}

\resizebox{\textwidth}{!}{\begin{tabular}{ | c | c | c | c | c | c | c | c | c |}
\hline
    Region   & Region start              & Region end                 &  Fork         & Replication   &  Slope:                 & Velocity:    & Log                      & Duration:     \\
    name     &  position: $\ell_-$ (Mb)  & position: $\ell_+$ (Mb)    &  direction    & orientation   &  $\alpha$ (Mb$^{-1}$)   & $v$ (kb/s)   & difference: $\Delta$     & $\Delta \tau$ (min) \\
\hline
\hline
R1 & $-1.76$ & $-1.59\times 10^{-3}$ & - & A & $0.221 \pm 2.3\%$ & $0.948 \pm 2.3\%$ & $0.388 \pm 2.3\%$ & $30.9 \pm 2.3\%$\\ 
R2 & $-1.59\times 10^{-3}$ & $2.01$ & + & A & $0.232 \pm 1.8\%$ & $0.900 \pm 1.8\%$ & $0.469 \pm 1.8\%$ & $37.3 \pm 1.8\%$\\ 
\hline
  \end{tabular}

}}}

\subsubsection{ Pause model}

{\centering

\includegraphics[width=.5\linewidth]{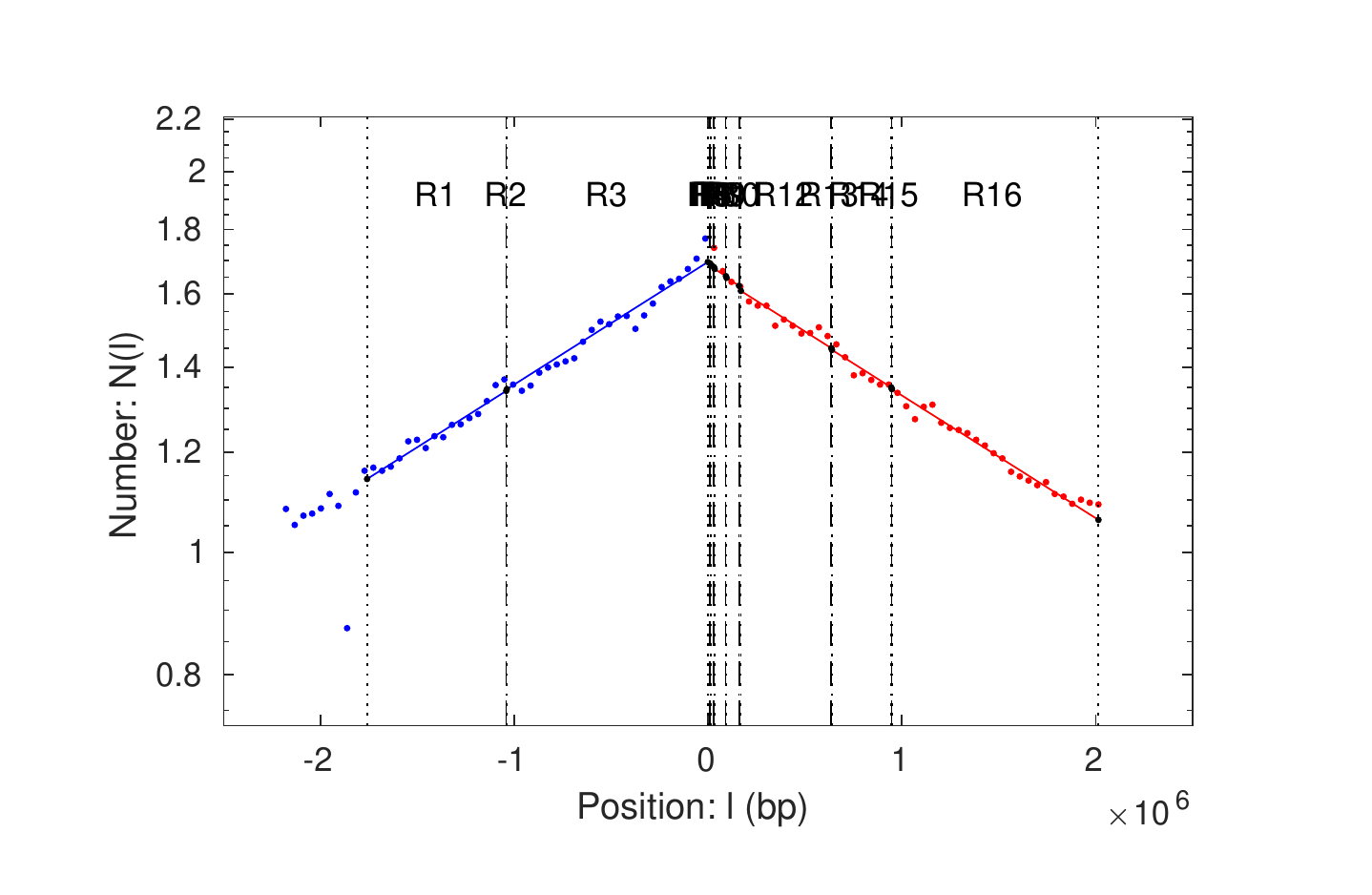}

\resizebox{\textwidth}{!}{\begin{tabular}{ | c | c | c | c | c | c | c | c | c |}
\hline
    Region   & Region start              & Region end                 &  Fork         & Replication   &  Slope:                 & Velocity:    & Log                      & Duration:     \\
    name     &  position: $\ell_-$ (Mb)  & position: $\ell_+$ (Mb)    &  direction    & orientation   &  $\alpha$ (Mb$^{-1}$)   & $v$ (kb/s)   & difference: $\Delta$     & $\Delta \tau$ (min) \\
\hline
\hline
R1 & $-1.76$ & $-1.04$ & - & A & $0.223 \pm 1.9\%$ & $0.937 \pm 1.9\%$ & $0.160 \pm 1.9\%$ & $12.8 \pm 1.9\%$\\ 
R2 & $-1.04$ & $-1.04$ & - & A & $0.973 \pm 21\%$ & $0.215 \pm 21\%$ & $3.13\times 10^{-3} \pm 21\%$ & $0.250 \pm 21\%$\\ 
R3 & $-1.04$ & $-1.59\times 10^{-3}$ & - & A & $0.223 \pm 1.9\%$ & $0.937 \pm 1.9\%$ & $0.232 \pm 1.9\%$ & $18.4 \pm 1.9\%$\\ 
R4 & $-1.59\times 10^{-3}$ & $1.13\times 10^{-2}$ & + & R & $0.223 \pm 1.9\%$ & $0.937 \pm 1.9\%$ & $2.88\times 10^{-3} \pm 1.9\%$ & $0.229 \pm 1.9\%$\\ 
R5 & $1.13\times 10^{-2}$ & $1.45\times 10^{-2}$ & + & A & $0.973 \pm 21\%$ & $0.215 \pm 21\%$ & $3.13\times 10^{-3} \pm 21\%$ & $0.250 \pm 21\%$\\ 
R6 & $1.45\times 10^{-2}$ & $3.06\times 10^{-2}$ & + & A & $0.223 \pm 1.9\%$ & $0.937 \pm 1.9\%$ & $3.59\times 10^{-3} \pm 1.9\%$ & $0.286 \pm 1.9\%$\\ 
R7 & $3.06\times 10^{-2}$ & $3.38\times 10^{-2}$ & + & A & $0.973 \pm 21\%$ & $0.215 \pm 21\%$ & $3.13\times 10^{-3} \pm 21\%$ & $0.250 \pm 21\%$\\ 
R8 & $3.38\times 10^{-2}$ & $9.18\times 10^{-2}$ & + & A & $0.223 \pm 1.9\%$ & $0.937 \pm 1.9\%$ & $1.29\times 10^{-2} \pm 1.9\%$ & $1.03 \pm 1.9\%$\\ 
R9 & $9.18\times 10^{-2}$ & $9.50\times 10^{-2}$ & + & A & $0.973 \pm 21\%$ & $0.215 \pm 21\%$ & $3.13\times 10^{-3} \pm 21\%$ & $0.250 \pm 21\%$\\ 
R10 & $9.50\times 10^{-2}$ & $0.159$ & + & A & $0.223 \pm 1.9\%$ & $0.937 \pm 1.9\%$ & $1.44\times 10^{-2} \pm 1.9\%$ & $1.15 \pm 1.9\%$\\ 
R11 & $0.159$ & $0.169$ & + & A & $0.973 \pm 21\%$ & $0.215 \pm 21\%$ & $9.40\times 10^{-3} \pm 21\%$ & $0.749 \pm 21\%$\\ 
R12 & $0.169$ & $0.636$ & + & A & $0.223 \pm 1.9\%$ & $0.937 \pm 1.9\%$ & $0.104 \pm 1.9\%$ & $8.30 \pm 1.9\%$\\ 
R13 & $0.636$ & $0.639$ & + & A & $0.973 \pm 21\%$ & $0.215 \pm 21\%$ & $3.13\times 10^{-3} \pm 21\%$ & $0.250 \pm 21\%$\\ 
R14 & $0.639$ & $0.945$ & + & A & $0.223 \pm 1.9\%$ & $0.937 \pm 1.9\%$ & $6.83\times 10^{-2} \pm 1.9\%$ & $5.44 \pm 1.9\%$\\ 
R15 & $0.945$ & $0.948$ & + & A & $0.973 \pm 21\%$ & $0.215 \pm 21\%$ & $3.13\times 10^{-3} \pm 21\%$ & $0.250 \pm 21\%$\\ 
R16 & $0.948$ & $2.01$ & + & A & $0.223 \pm 1.9\%$ & $0.937 \pm 1.9\%$ & $0.238 \pm 1.9\%$ & $19.0 \pm 1.9\%$\\ 
\hline
  \end{tabular}

}}

\subsection{  \textit{B.~subtilis} \textit{oriN}-257$^\circ$ on S7 fumarate }

\label{sec:othbs2}

{\centering

\includegraphics[width=.5\linewidth]{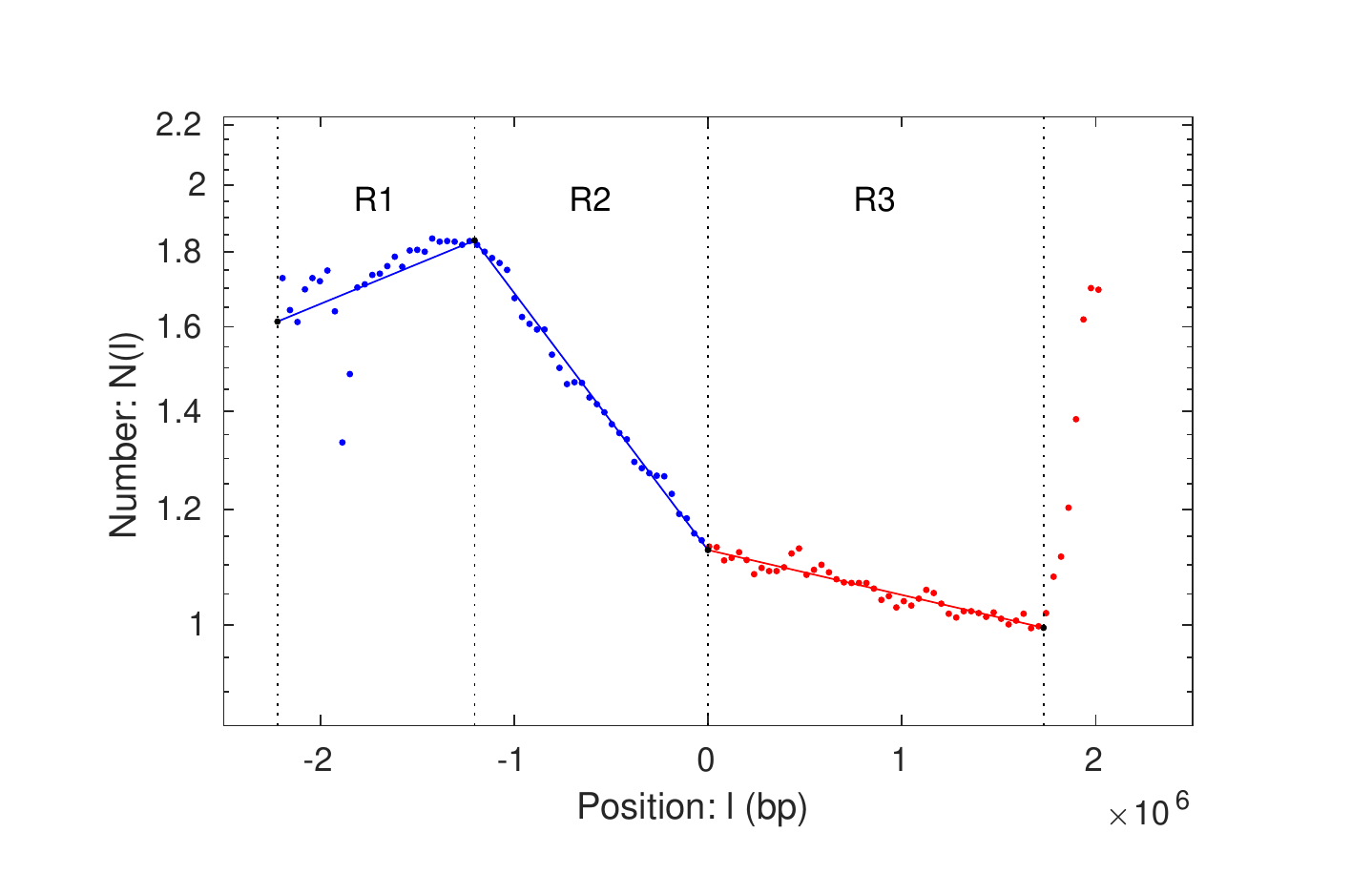}

\resizebox{\textwidth}{!}{\begin{tabular}{ | c | c | c | c | c | c | c | c | c |}
\hline
    Region   & Region start              & Region end                 &  Fork         & Replication   &  Slope:                 & Velocity:    & Log                      & Duration:     \\
    name     &  position: $\ell_-$ (Mb)  & position: $\ell_+$ (Mb)    &  direction    & orientation   &  $\alpha$ (Mb$^{-1}$)   & $v$ (kb/s)   & difference: $\Delta$     & $\Delta \tau$ (min) \\
\hline
\hline
R1 & $-2.22$ & $-1.20$ & - & A & $0.126 \pm 18\%$ & NA & $0.128 \pm 18\%$ & NA\\ 
R2 & $-1.20$ & $-9.18\times 10^{-4}$ & + & R & $0.405 \pm 3.5\%$ & NA & $0.488 \pm 3.5\%$ & NA\\ 
R3 & $-9.18\times 10^{-4}$ & $1.73$ & + & A & $7.08\times 10^{-2} \pm 16\%$ & NA & $0.123 \pm 16\%$ & NA\\ 
\hline
  \end{tabular}

}}

\subsection{ \textit{B.~subtilis} \textit{rrnIHG} inversion on MOPS glucose w/ Casamino Acids }

{
{\centering

\includegraphics[width=.5\linewidth]{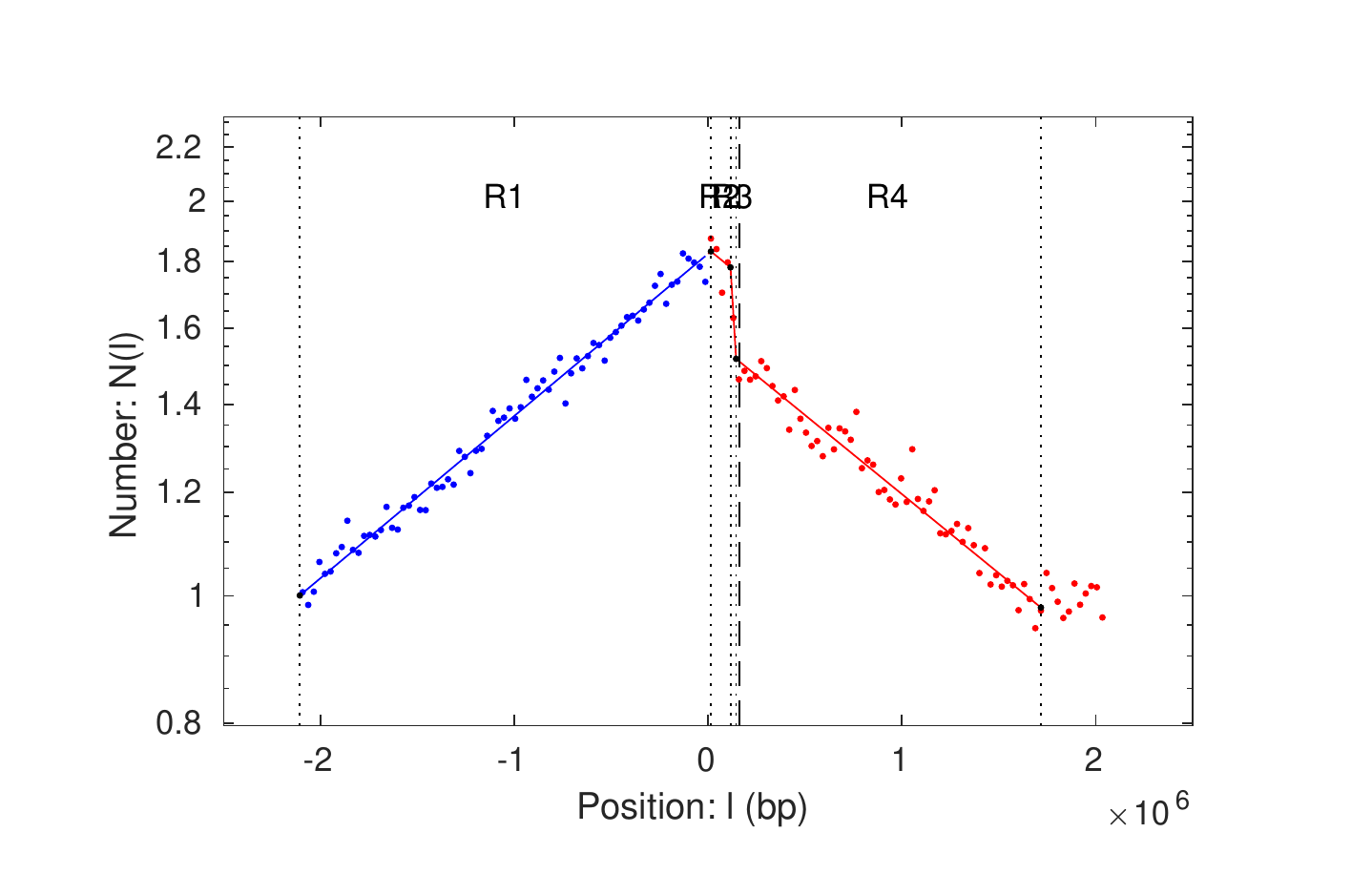}

\resizebox{\textwidth}{!}{\begin{tabular}{ | c | c | c | c | c | c | c | c | c |}
\hline
    Region   & Region start              & Region end                 &  Fork         & Replication   &  Slope:                 & Velocity:    & Log                      & Duration:     \\
    name     &  position: $\ell_-$ (Mb)  & position: $\ell_+$ (Mb)    &  direction    & orientation   &  $\alpha$ (Mb$^{-1}$)   & $v$ (kb/s)   & difference: $\Delta$     & $\Delta \tau$ (min) \\
\hline
\hline
R1 & $-2.13$ & $-6.09\times 10^{-3}$ & - & A & $0.285 \pm 2.3\%$ & $0.921 \pm 2.3\%$ & $0.605 \pm 2.3\%$ & $38.4 \pm 2.3\%$\\ 
R2 & $-6.09\times 10^{-3}$ & $9.49\times 10^{-2}$ & + & A & $0.278 \pm 4.0\%$ & $0.945 \pm 4.0\%$ & $2.81\times 10^{-2} \pm 4.0\%$ & $1.78 \pm 4.0\%$\\ 
R3 & $9.49\times 10^{-2}$ & $0.124$ & + & A & $5.57 \pm 8.4\%$ & $4.71\times 10^{-2} \pm 8.4\%$ & $0.161 \pm 8.4\%$ & $10.2 \pm 8.4\%$\\ 
R4 & $0.124$ & $1.70$ & + & A & $0.278 \pm 4.0\%$ & $0.945 \pm 4.0\%$ & $0.437 \pm 4.0\%$ & $27.7 \pm 4.0\%$\\ 

\hline
  \end{tabular}

}}}

\subsection{  \textit{B.~subtilis} \textit{rrnIHG} inversion on MOPS glucose -- Minimal }

{
{\centering

\includegraphics[width=.5\linewidth]{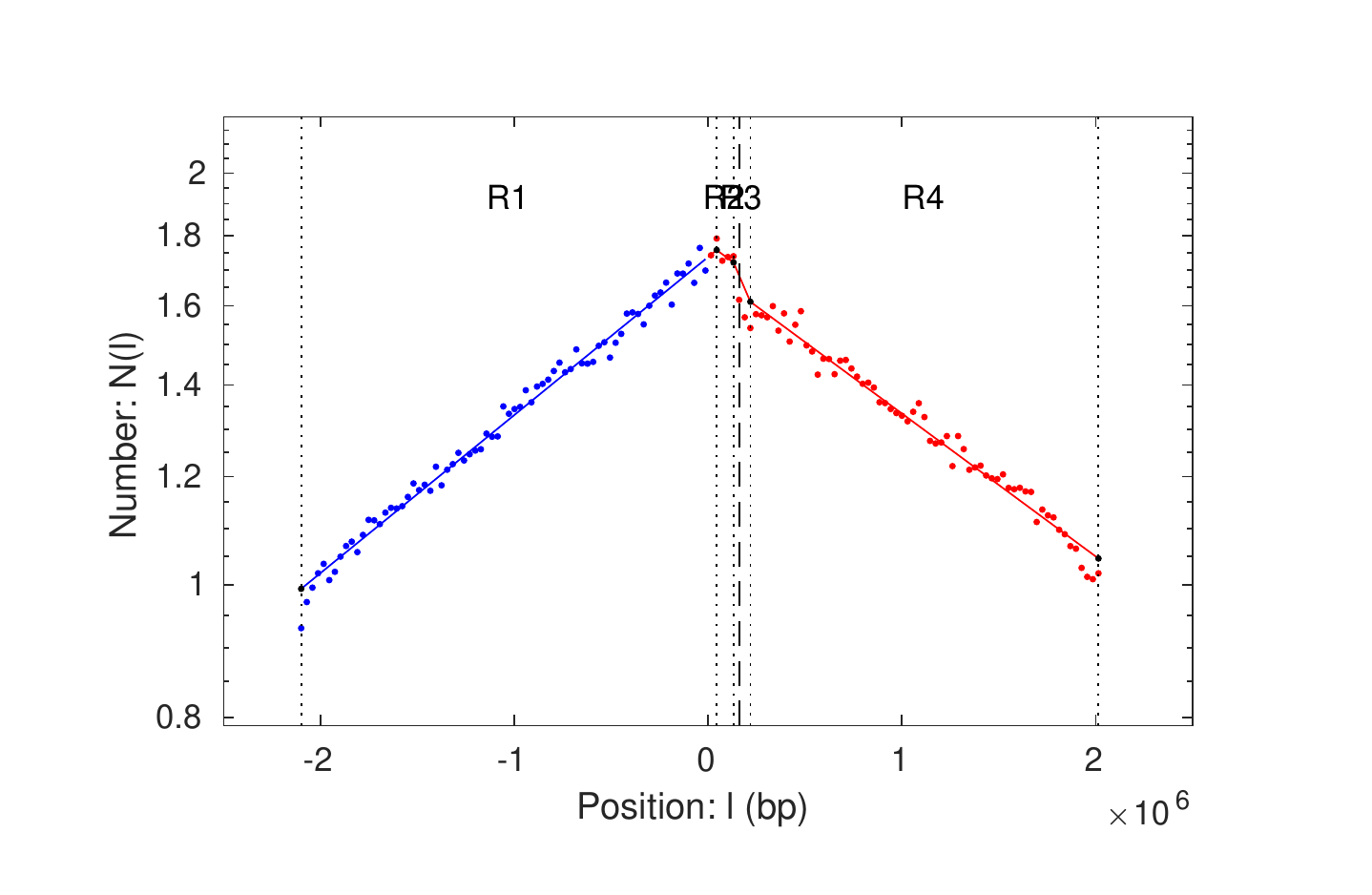}

\resizebox{\textwidth}{!}{\begin{tabular}{ | c | c | c | c | c | c | c | c | c |}
\hline
    Region   & Region start              & Region end                 &  Fork         & Replication   &  Slope:                 & Velocity:    & Log                      & Duration:     \\
    name     &  position: $\ell_-$ (Mb)  & position: $\ell_+$ (Mb)    &  direction    & orientation   &  $\alpha$ (Mb$^{-1}$)   & $v$ (kb/s)   & difference: $\Delta$     & $\Delta \tau$ (min) \\
\hline
\hline
R1 & $-2.10$ & $4.41\times 10^{-2}$ & - & A & $0.266 \pm 1.7\%$ & $0.869 \pm 1.7\%$ & $0.570 \pm 1.7\%$ & $41.1 \pm 1.7\%$\\ 
R2 & $4.41\times 10^{-2}$ & $0.131$ & + & A & $0.241 \pm 2.5\%$ & $0.961 \pm 2.5\%$ & $2.09\times 10^{-2} \pm 2.5\%$ & $1.51 \pm 2.5\%$\\ 
R3 & $0.131$ & $0.218$ & + & A & $0.761 \pm 13\%$ & $0.304 \pm 13\%$ & $6.61\times 10^{-2} \pm 13\%$ & $4.77 \pm 13\%$\\ 
R4 & $0.218$ & $2.01$ & + & A & $0.241 \pm 2.5\%$ & $0.961 \pm 2.5\%$ & $0.432 \pm 2.5\%$ & $31.2 \pm 2.5\%$\\ 
\hline
  \end{tabular}

}}}

\newpage

\subsection{  \textit{E.~coli} WT on LB  }

{
{\centering

\includegraphics[width=.5\linewidth]{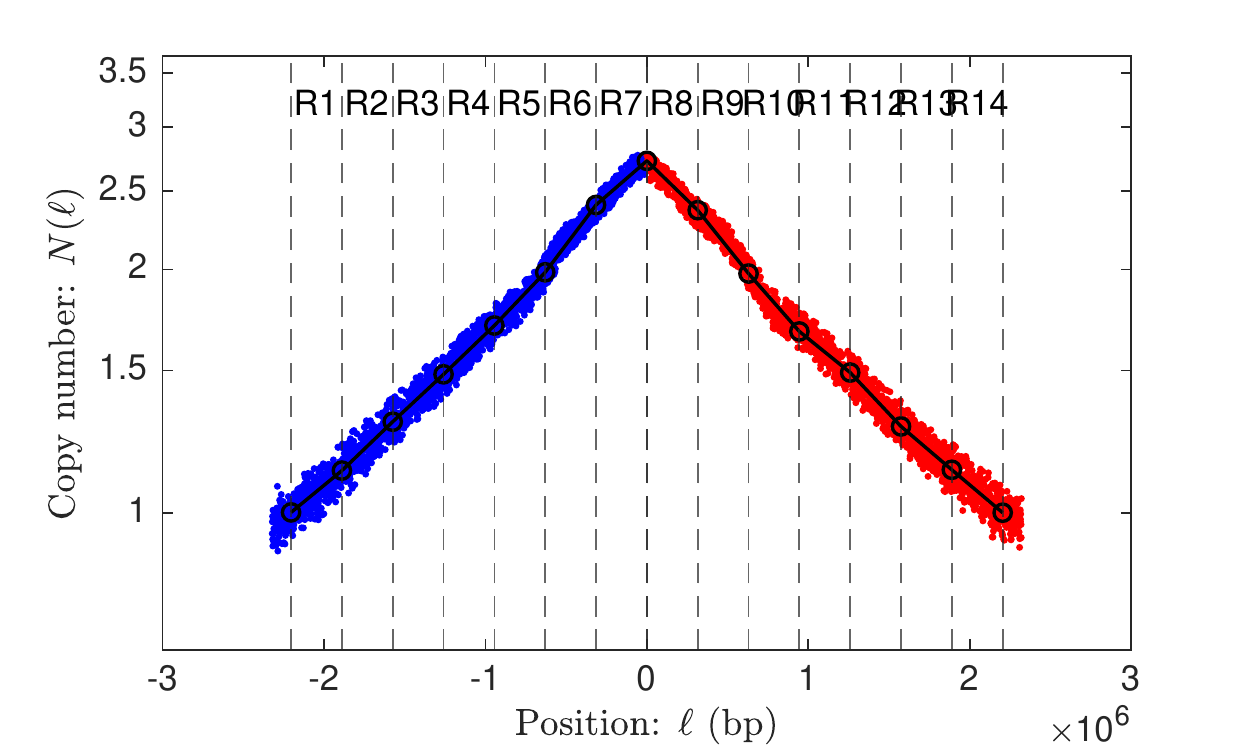}

Detailed results tabulated in \texttt{SuppData.xlsx}.

}}

\subsection{  \textit{E.~coli} WT on M9 glucose  }

{
{\centering

\includegraphics[width=.5\linewidth]{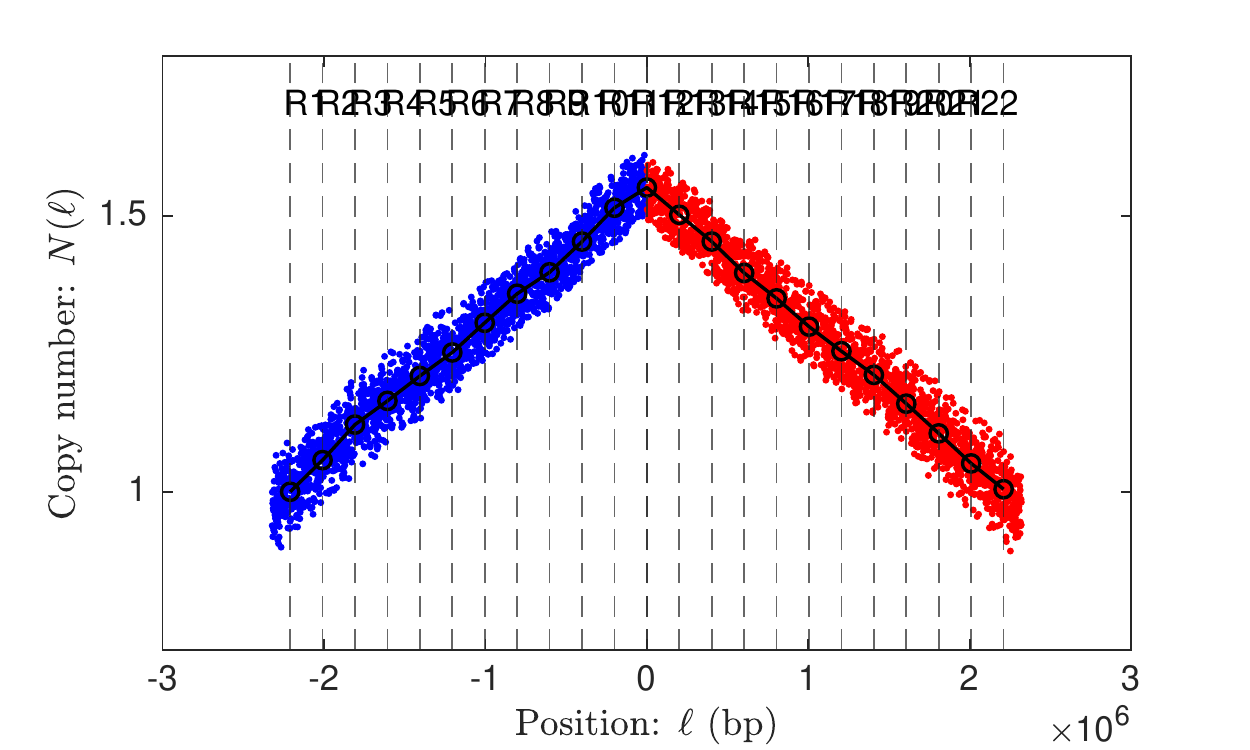}

Detailed results tabulated in \texttt{SuppData.xlsx}.

}}

%
%

\newpage

\bibliographystyle{apsrev4-1}
\bibliography{forkvel}